\documentclass[%
 reprint,
 cha,
 noshowpacs,
 superscriptaddress,
 amsmath,amssymb,
floatfix,
]{revtex4-1}
\usepackage[utf8]{inputenc}
\usepackage[T1]{fontenc}
\usepackage{amssymb}
\usepackage{bbm}
\usepackage{mathrsfs}
\usepackage{dsfont}

\usepackage{physics}
\usepackage{graphicx}
\usepackage{dcolumn}
\usepackage{bm}
\usepackage[x11names]{xcolor}

\usepackage{hyperref}
\usepackage{bookmark}

\hypersetup{%
  breaklinks=true,
  colorlinks=true,
  linkcolor=black,
  filecolor=black,
  urlcolor=black,
  citecolor=black,
  pdfborder=0 0 0,
}

\newcommand{\ii}{\text{i}}
\newcommand{\shortspace}{\enspace\enspace}
\newcommand{\midspace}{\enspace\enspace\enspace\enspace}
\newcommand{\longspace}{\enspace\enspace\enspace\enspace\enspace\enspace\enspace}
\newcommand{\height}{10pt}
\newcommand{\so}{\raisebox{-0.25\height}{\includegraphics[height=\height]{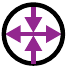}}}
\newcommand{\st}{\raisebox{-0.25\height}{\includegraphics[height=\height]{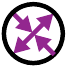}}}
\newcommand{\p}{\raisebox{-0.25\height}{\includegraphics[height=\height]{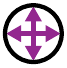}}}
\newcommand{\tor}{\raisebox{-0.25\height}{\includegraphics[height=\height]{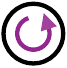}}}

\newcommand{\dtV}{\raisebox{-0.25\height}{\includegraphics[height=\height]{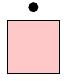}}}
\newcommand{\dtR}{\raisebox{-0.25\height}{\includegraphics[height=\height]{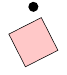}}}
\newcommand{\dtSo}{\raisebox{-0.25\height}{\includegraphics[height=\height]{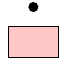}}}
\newcommand{\dtSt}{\raisebox{-0.25\height}{\includegraphics[height=\height]{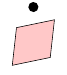}}}
\newcommand{\dtS}{\raisebox{-0.25\height}{\includegraphics[height=\height]{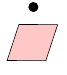}}}

\newcommand{\K}{\textbf{k}}
\newcommand{\V}{\textbf{v}}
\newcommand{\U}{\textbf{u}}
\newcommand{\R}{\textbf{r}}
\newcommand{\F}{\textbf{f}}

\newcommand{\s}{\hat{\bm{\sigma}}}

\newcommand{\KoR}{\textbf{k}\cdot\textbf{r}}

\newcommand{\Pk}{\mathcal{P}_{\textbf{k}}}
\newcommand{\Qk}{\mathcal{Q}_{\textbf{k}}}

\newcommand{\J}{\hat{\textbf{J}}}
\newcommand{\Jk}{\hat{\textbf{J}}_{\textbf{k}}}
\newcommand{\jka}{\hat{J}_{\textbf{k}, a}}
\newcommand{\jkb}{\hat{J}_{\textbf{k}, b}}
\newcommand{\jkc}{\hat{J}_{\textbf{k}, c}}

\newcommand{\Fp}{\textbf{F}_\textbf{k}^{\parallel}}
\newcommand{\Fo}{\textbf{F}_\textbf{k}^{\bot}}

\newcommand{\fo}{F^{\bot}}
\newcommand{\Fd}{\hat{\F}_{\,\K}^{\,\text{vel}}}

\newcommand{\iLt}{\ii \mathcal{L}t}
\newcommand{\iL}{\ii \mathcal{L}}

\newcommand{\G}{\bm{\Gamma}}

\newcommand{\e}{\text{e}}

\newcommand{\kb}{k_\text{B}}
\newcommand{\kT}{k_\text{B}T_\text{eff}}
\newcommand{\Sk}{\hat{\bm{\sigma}}_\textbf{k}}

\newcommand{\Sreg}{\hat{\bm{\sigma}}_{\textbf{k}}^{\text{reg}}}
\newcommand{\Skin}{\hat{\bm{\sigma}}_{\textbf{k}}^{\text{kin}}}
\newcommand{\Spos}{\hat{\bm{\sigma}}_{\textbf{k}}^{\text{pos}}}

\begin{document}

\title{Statistical mechanics of a chiral active fluid}

\author{Ming Han}
\affiliation{James Franck Institute,
  University of Chicago, Chicago, Illinois 60637, U.S.A.}
\affiliation{Pritzker School of Molecular Engineering,
  University of Chicago, Chicago, Illinois 60637, U.S.A.}
 
\author{Michel Fruchart}
\affiliation{James Franck Institute,
   University of Chicago, Chicago, Illinois 60637, U.S.A.}
\affiliation{Department of Physics, University of Chicago, Chicago, IL 60637, U.S.A.}

\author{Colin Scheibner}
\affiliation{James Franck Institute,
  University of Chicago, Chicago, Illinois 60637, U.S.A.}
\affiliation{Department of Physics, University of Chicago, Chicago, IL 60637, U.S.A.}

\author{Suriyanarayanan Vaikuntanathan}
\affiliation{James Franck Institute,
 University of Chicago, Chicago, Illinois 60637, U.S.A.}
\affiliation{Department of Chemistry, University of Chicago, Chicago, IL 60637, U.S.A.}


\author{William Irvine}
\affiliation{James Franck Institute,
  University of Chicago, Chicago, Illinois 60637, U.S.A.}
\affiliation{Department of Physics, University of Chicago, Chicago, IL 60637, U.S.A.}
  
\author{Juan de Pablo}
\affiliation{Pritzker School of Molecular Engineering,
  University of Chicago, Chicago, Illinois 60637, U.S.A.}
\affiliation{Center for Molecular Engineering, Argonne National Laboratory, Lemont, Illinois 60439, U.S.A.}

\author{Vincenzo Vitelli}
\affiliation{James Franck Institute,
  University of Chicago, Chicago, Illinois 60637, U.S.A.}
\affiliation{Department of Physics, University of Chicago, Chicago, IL 60637, U.S.A.}

\begin{abstract}
\color{black}
Statistical mechanics provides the foundation for describing complex materials using only a few thermodynamic variables. No such framework currently exists far from equilibrium.
In this Letter, we demonstrate how thermodynamics emerges far from equilibrium, using fluids composed of active spinners as a case study. 
Activity gives rise to a single effective temperature that parameterizes both the equation of state and the emergent Boltzmann statistics.
The same effective temperature, renormalized by velocity correlations, controls the linear response through canonical Green-Kubo relations for both the familiar shear viscosity and the odd (or Hall) viscosity observed in chiral fluids.
The full frequency dependence of these viscosities can be derived analytically by modelling the active-spinner fluid as a random walker undergoing cyclotron motion in shear-stress space.  
More generally, we provide a first-principles derivation of the Green-Kubo relations valid for a broader class of fluids far from equilibrium. Besides advancing non-equilibrium thermodynamics, our work demonstrates in silico a non-invasive microrheology of active fluids.
\end{abstract}

\maketitle

\begin{figure}[h]
\includegraphics[width=0.48\textwidth]{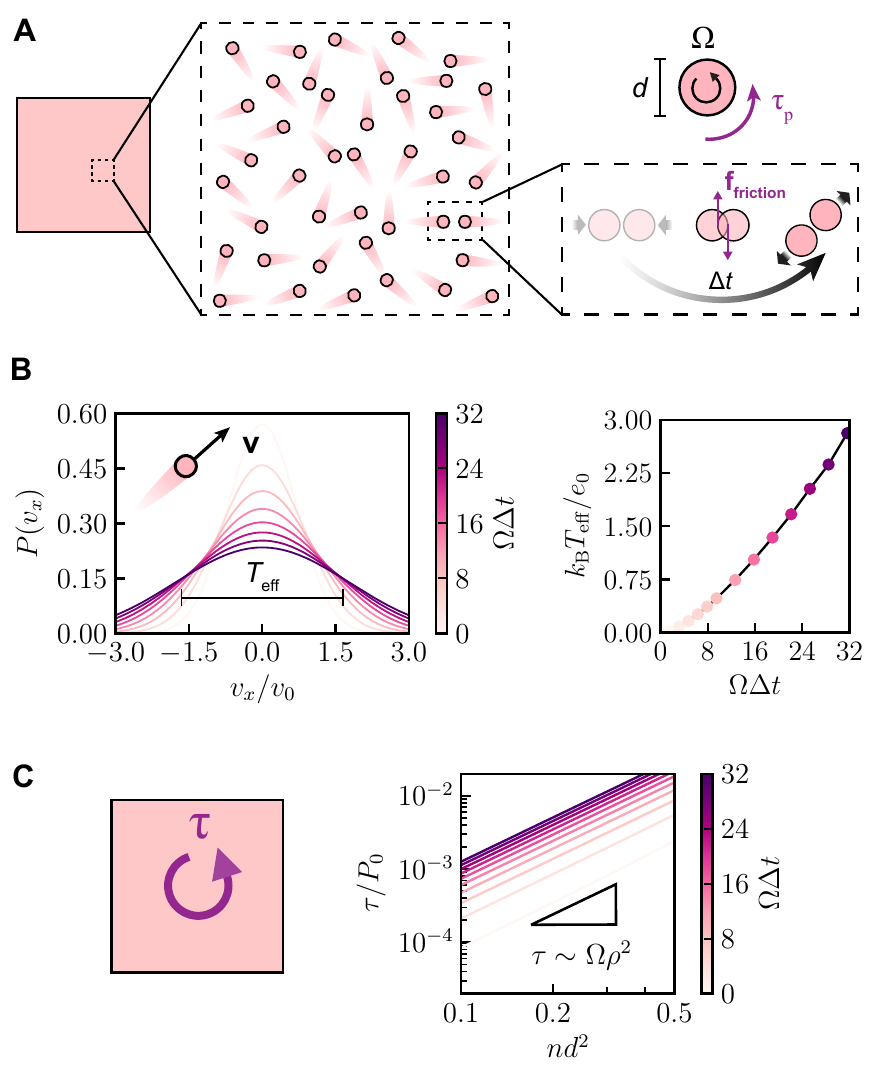}
\caption{\label{fig:equilibrium} \textbf{Thermodynamics of a chiral active fluid.} \textbf{A.} Schematic of the system setup. We simulate a 2D granular gas composed of frictional particles (diameter $d$), which are powered by an active torque $\tau_\text{p}$ to self-spin at a constant speed $\Omega$. During collision, two self-spinning particles slide respect to each other. The resultant interparticle friction causes transverse motion of the particles upon separation (Supplementary Mov.~S1). $\Delta t$ denotes the averaged collision duration. \textbf{B.} Velocity distribution. The $x$-component of translational velocity displays a Gaussian distribution $P(v_x)$ at various spinning speed $\Omega$. An effective temperature $T_\text{eff}$ is defined using the halfwidth of $P(v_x)$. Dependence of $T_\text{eff}$ on $\Omega$ is shown on the right. \textbf{C.} Anti-symmetric stress. At the steady state, the system acquires a nonvanishing torque density $\tau$. The dependence of $\tau$ on particle number density $n$ is shown on the right. Units: $v_0 = d/\Delta t$, $e_0 = m d^2/\Delta t^2$, $P_0 = m/d\Delta t^2$.}
\end{figure}

The fluctuation-dissipation relation is one of the most striking properties of thermodynamic equilibrium~\cite{kubo1966fluctuation}. 
It allows us to determine the response of a system without applying any perturbation. For example, the mobility of a Brownian particle can be extracted from its velocity fluctuations. Significant effort has gone into extending the fluctuation-dissipation relations to driven and active systems~\cite{Kurchan2005,Ciliberto2010,Cugliandolo2011,Seifert2012}.
For a single particle, experiments reveal that the mobility is related to velocity fluctuations via an effective temperature set by activity~\cite{Makse2002,DAnna2003,ojha2004statistical}. 
However, less is known about the collective response of an active fluid viewed as a whole. 
In equilibrium fluids, the fluctuation--dissipation theorem manifests as the Green--Kubo relation. This relation connects the fluid viscosities with fluctuations in the stress.
We ask, can such a relationship survive far from equilibrium?

In this Letter, we show that active fluids composed of spinning components~\cite{drescher2009dancing,furthauer2012active,nguyen2014emergent,petroff2015fast,kokot2018manipulation,tsai2005chiral,scholz2018rotating}
provide a case study of how Green-Kubo relations emerge in non-equilibrium steady states. 
In such fluids, broken detailed balance gives rise to additional viscosity coefficients, known as odd (or Hall) viscosities~\cite{avron1998odd,de2013non,banerjee2017odd,souslov2019topological,liao2019mechanism,epstein2019time,soni2019odd,alekseev2016negative,korving1966transverse,wiegmann2014anomalous,Berdyugin2019,pellegrino2017nonlocal,bradlyn2012kubo,offertaler2019viscoelastic,son2019chiral}, recently measured in fluids of spinning colloids~\cite{soni2019odd}.
We reveal that self-spinning and collisions generate a steady-state with a single effective temperature. This temperature enters both the Boltzmann distribution and the equation of state of the chiral fluid, in agreement with recent experiments~\cite{Farhadi2018}.
Crucially, the same effective temperature governs the linear response through canonical Green--Kubo relations that apply to both the shear and odd viscosities.

Generalized thermodynamic approaches have been successfully employed to describe non-equilibrium systems~\cite{Seifert2012,harada2005equality,Fodor2016,shankar2018hidden,Nardini2017,le2001motor,berthier2013non,palacci2010sedimentation,egolf2000equilibrium,prost2009generalized,GomezSolano2009,Seifert2010,Cengio2019,sarracino2019fluctuation,han2017effective}, but they are all subject to certain restrictions: (i) they lack a single effective temperature that governs distinct thermal properties~\cite{berthier2013non,palacci2010sedimentation}; (ii) they simply regain detailed balance at a coarse-grained level~\cite{egolf2000equilibrium}; (iii) they require drastic modifications of the fluctuation--dissipation relations~\cite{prost2009generalized,GomezSolano2009,Seifert2010,Cengio2019,sarracino2019fluctuation,han2017effective}. None of these restriction apply here.
We provide a first-principles derivation of the canonical Green--Kubo relations for the full viscosity tensor, including odd viscosities, in a broad class of fluids far from equilibrium.
Our findings are corroborated by large-scale numerical simulations. 


We start by demonstrating the emergence of equilibrium-like steady states from activity in the following microscopic model. Consider frictional granular particles, driven by large active torques, all spinning at a constant speed $\Omega$ (Fig.~\ref{fig:equilibrium}A). In this case, the angular degrees of freedom can be integrated out to find an effective Newton's equation for the centers of mass of the particles,
\begin{equation}
	\label{newton_eom}
	m \ddot{\textbf{x}}_i = \sum_{j \in N(i)} \textbf{f}^{\;\text{c}}_{ij} - \gamma\textbf{v}_{i j} +
	\gamma d\Omega \hat{ \boldsymbol{z}} \cross \hat{\textbf{r}}_{i j}
\end{equation}
where $\textbf{x}_i$ is the position of particle $i$ with  mass $m$ and diameter $d$. The right-hand side of Eq.~(\ref{newton_eom}) summarizes the interactions with the neighbors $N(i)$ of the particle $i$: $\textbf{f}^{\;\text{c}}_{ij}$ is a conservative soft repulsive force while the second and third terms are non-conservative interactions caused by interparticle friction, respectively describing the damping effects of head-to-head collision and the transverse interaction due to self-spinning. 

The system described by Eq.~(\ref{newton_eom}) is constantly randomized by collisions. We find that, as a result, it acquires equal-time ensemble properties typically associated with equilibrium:
(i) a Maxwell distribution of particle velocity (Fig.~\ref{fig:equilibrium}B) and (ii) a Boltzmann distribution of particle concentration in the presence of an external potential (Supplementary Fig.~S1). 
In all these cases, a single effective temperature $T_{\text{eff}}$ exists although no intrinsic thermal noise is included in our molecular dynamics simulations (see Supplementary Sec.~II and Fig.~S2). The effective temperature arises purely from activity. In the supplementary information, we derive that $T_{\text{eff}} \propto |\Omega|^{\alpha}$, where $\alpha$ is a non-universal exponent depending on $\textbf{f}^{\!\!\!\!\!\! \quad c}_{ij}$ and satisfying $4/3 \leq \alpha \leq 2$. Our simulations with a contact potential reveal a power-law behavior $T_{\text{eff}} \propto |\Omega|^{1.54\pm0.02}$ over two decades, consistent with our prediction.

We now consider the consequences of the effective temperature for the hydrodynamic description of a chiral active fluid. 
The stress tensor~$\bm{\sigma}$ determines the forces occurring at the boundary of a fluid, as well as the time evolution of its bulk velocity field~$\textbf{u}$ through the Navier-Stokes equation $\!\!\!\! \quad$ $\rho \text{D}_t \textbf{u} = \div{\bm{\sigma}} + \textbf{f},$ $\!\!\!\! \quad$ where $\rho = nm$ is the mass density of the fluid, $n$ is the number density, and $\textbf{f}$ denotes external body forces.
We performed hundreds of simulations varying particle density, self-spinning speed, and flow condition. 
In each, we measure the stress tensor, using the Irvine--Kirkwood formula~\cite{irving1950statistical} that expresses~$\bm{\sigma}$ in terms of the microscopic particle velocities and the forces between them. 

In the absence of any velocity gradient, the stress tensor is composed of only two components~\cite{de2013non}. 
First, an isotropic pressure $P$ that we find follows the ideal gas law $P = n k_\text{B}T_\text{eff}$, where $k_\text{B}$ is the Boltzmann constant (Supplementary Fig.~S3). 
Second, we find a non-vanishing anti-symmetric component of the stress that arises from the net torque density $\tau = \Gamma(n) \Omega$ with $\Gamma \sim n^2$ (Fig.~\ref{fig:equilibrium}C, Supplementary Fig.~S4). 

In the presence of small velocity gradients, surface forces appear between fluid layers generating the viscous stress $\sigma_{i j}^{\text{viscous}} = \eta_{i j k \ell} \partial_k u_{\ell}$, where $\eta_{i j k \ell}$ denotes the viscosity tensor. 
In order to keep track of all contributions to the constitutive relation between stress and strain-rate,
it is helpful to express the stress and the (unsymmetrized) strain-rate as the two vectors $\sigma_{\alpha}$ and $\dot{e}_{\beta}$ respectively, so that $\eta_{i j k \ell}$ can be represented as a matrix $\eta_{\alpha \beta}$ (see Supplementary Sec.~III and Refs.~\cite{scheibner2019odd,avron1998odd}). For an isotropic two-dimensional fluid, the constitutive relation reads
\begin{equation}
    \raisebox{-0.5\height}{\includegraphics[height=70pt]{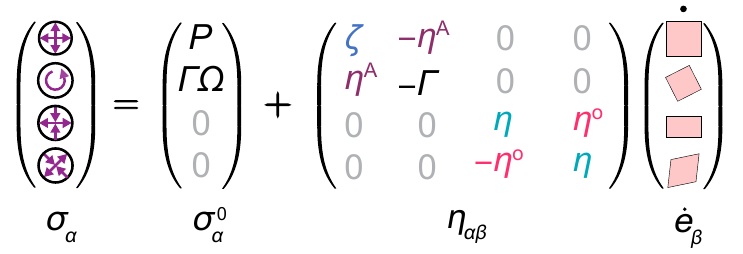}} \label{eq:matrix}
\end{equation}
where $\sigma_\alpha^0$ encodes the previously discussed contributions from the isostatic pressure $P$ and torque density $\Gamma \Omega$. 
The velocity gradients $\dot{e}_{\beta}$ are decomposed into dilation (\protect\dtV), rotation (\protect\dtR), and two pure shears rotated by $45^\circ$ (\protect\dtSo\; and \protect\dtSt) while the stress $\sigma_{\alpha}$ is decomposed into pressure (\protect\p), torque (\protect\tor), and two shear stresses  $s_1$~(\protect\so) and $s_2$~(\protect\st).

\begin{figure}[h!]
\includegraphics[width=0.48\textwidth]{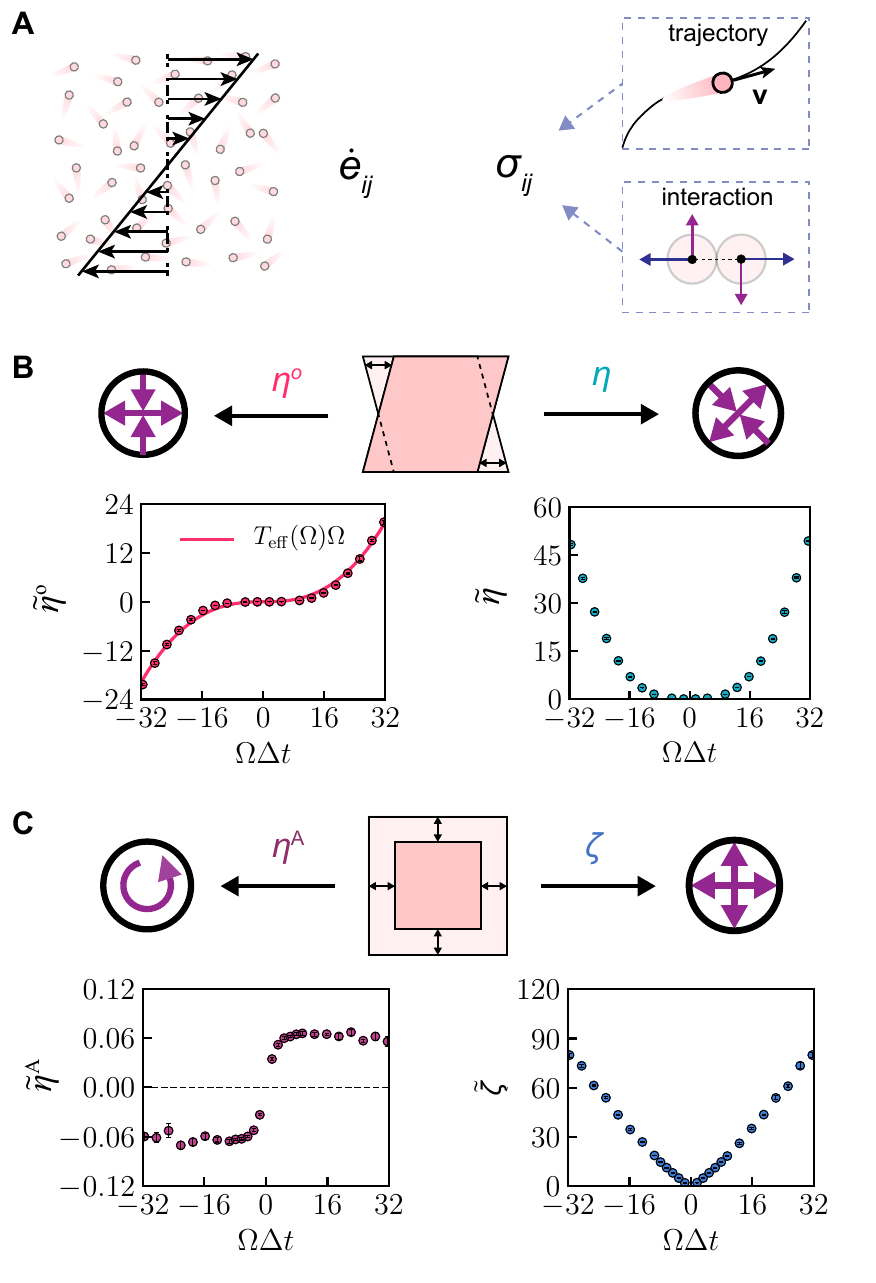}
\caption{\label{fig:response} \textbf{Rheology of chiral active fluid.} \textbf{A.} Schematic of rheological measurements. We perturb the simulation system with a velocity gradient $\dot{e}_{kl} = \partial_k u_l$ and then measure the linear response in stress $\sigma_{ij}$ to infer viscosity tensor $\eta_{ijkl}$. Rather than measuring the forces at the boundaries, the stress can be calculated in the bulk from particle trajectory and interactions using the Irvine--Kirkwood formula applied to either simulation or experimental data. In the dilute limit, the kinetic part of the Irvine--Kirkwood formula dominates so the stress tensor can be determined purely from movies of particle motion, without knowledge of microscopic interactions.
\textbf{B.} Odd and shear viscosities. A simple shear, which contains pure shear \protect\dtSt, induces shear stress $s_1$ (\protect\so) via odd viscosity $\eta^\text{o}$ and shear stress $s_2$ (\protect\st) via shear viscosity $\eta$. \textbf{C.} Compression-rotation and bulk viscosities. A dilation/compression alters $\tau$ (\protect\tor) via compression-rotation viscosity $\eta^\text{A}$ and pressure $P$ (\protect\p) via bulk viscosity $\zeta$. The dependencies of all the viscosities on spinning speed $\Omega$ are shown in \textbf{B-C}. They all obey the Onsager--Casimir reciprocal relation $\eta_{\alpha\beta}(\Omega) = \eta_{\alpha\beta}(-\Omega)$: $\eta$ and $\zeta$, the diagonal terms in Eq.~(\ref{eq:matrix}), are even in $\Omega$; $\eta^\text{o}$ and $\eta^\text{A}$, the anti-symmetric terms, are odd in $\Omega$. All the viscosities are in the unit of $\eta_0 = m/d\Delta t$.} 
\end{figure}

\begin{figure*}[p!]
\includegraphics[width=0.98\textwidth]{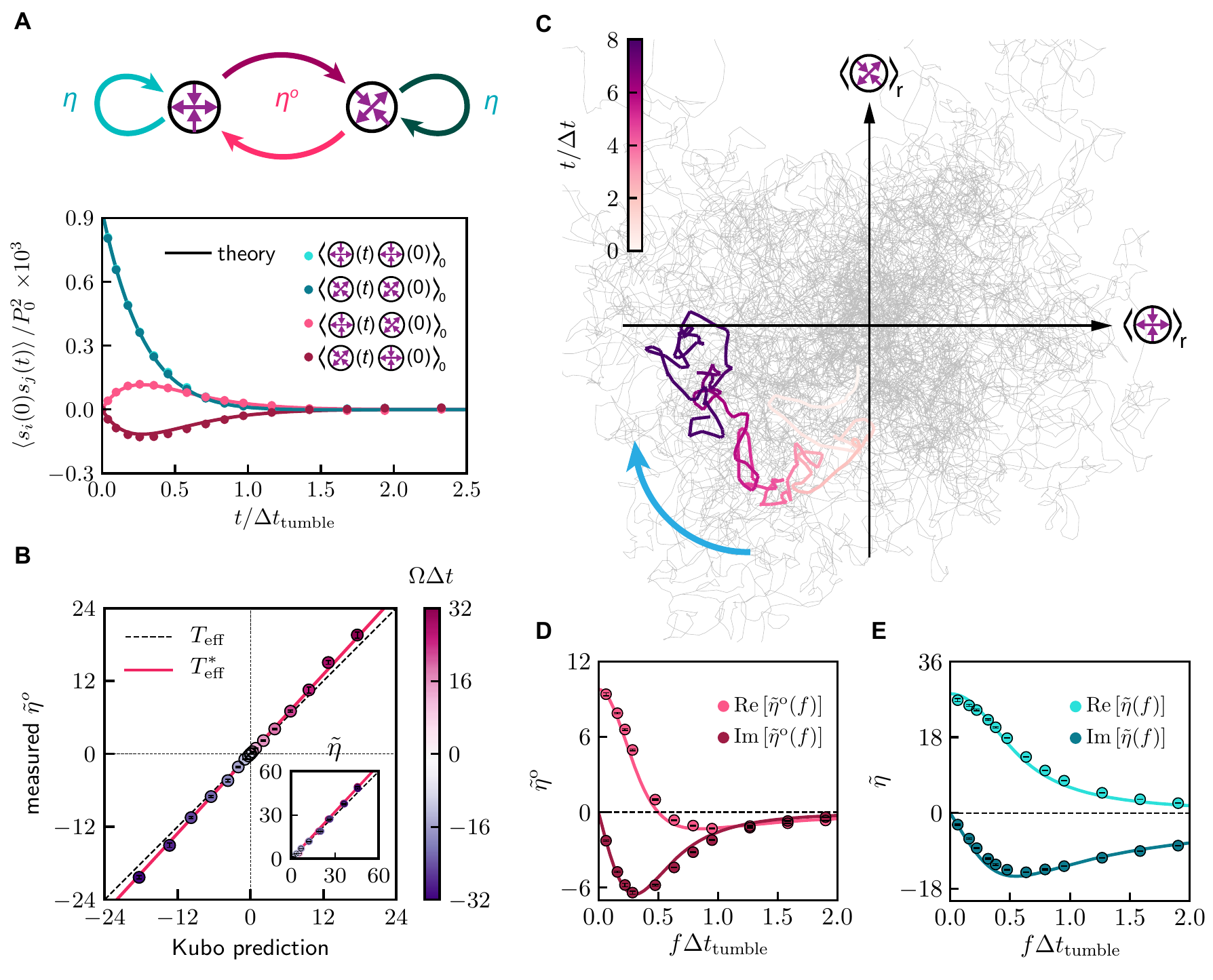}
\caption{\label{fig:fluctuations} \textbf{Green--Kubo relation.}
\textbf{A.} Stress--stress correlation functions. The time correlation functions of the two global shear stresses $\langle\protect\so\rangle_\text{r}$ and $\langle\protect\st\rangle_\text{r}$ are plotted. The shear viscosity $\eta$ leads to the auto-correlations (in green), whereas the odd viscosity $\eta^\text{o}$ gives rise to the cross-correlations (in red), also as summarized by the top schematic. The correlation functions predicted by our theory Eq.~(\ref{eq:langevin}) is compared with the values measured in simulations. The correlation time is set by the tumbling time of a particle $\Delta t_\text{tumble} = (\Delta t + \Delta t_\text{col}) \cdot \bar{v}/\overline{\Delta v}$, where $\Delta t$ is the collision duration, $\Delta t_\text{col}$ is the time between collisions, $\bar{v}$ is the mean velocity of the particle, and $\overline{\Delta v}$ is the average velocity change after a collision. We find that $\Delta t_\text{tumble} \approx 100\Delta t$ in this case. \textbf{B.} Green--Kubo relation. The coefficients of viscous response towards a steady shear can be predicted using the integral of the stress--stress correlation functions, known as the direct-current (d.c.) Green--Kubo relation. The predicted and measured odd viscosity $\eta^\text{o}$ is compared at a wide range of spinning speed $\Omega$. Inset: Comparison between the predicted and measured shear viscosity $\eta$. The Kubo predictions with $T_\text{eff}$ and renormalized $T_\text{eff}^*$ are marked as the dashed and solid lines, respectively.
\textbf{C.} Time evolution of the shear stress vector $\big(\langle\protect\so\rangle_\text{r}, \langle\protect\st\rangle_\text{r}\big)$  at spinning speed $\Omega = 26.7/\Delta t$. At the steady state of the chiral active fluid, the shear stress vector traces out a 2D random walk in the stress space (grey curve in background), which is loosely confined and rotates around the origin preferentially in a clockwise fashion over time (curve with gradient coloring). See Supplementary Mov.~S2. \textbf{D-E.} Green--Kubo relation in frequency domain. The frequency-dependent coefficients of viscous response towards an oscillatory shear can be estimated using the Fourier transform of the stress--stress correlation functions, known as the a.c. Green--Kubo relation. Comparisons between the Kubo prediction and the simulation measurement are presented for both odd viscosity (\textbf{D}) and shear viscosity (\textbf{E}) at various shear frequencies.}
\end{figure*}

\begin{figure*}[th]
\includegraphics[width=0.98\textwidth]{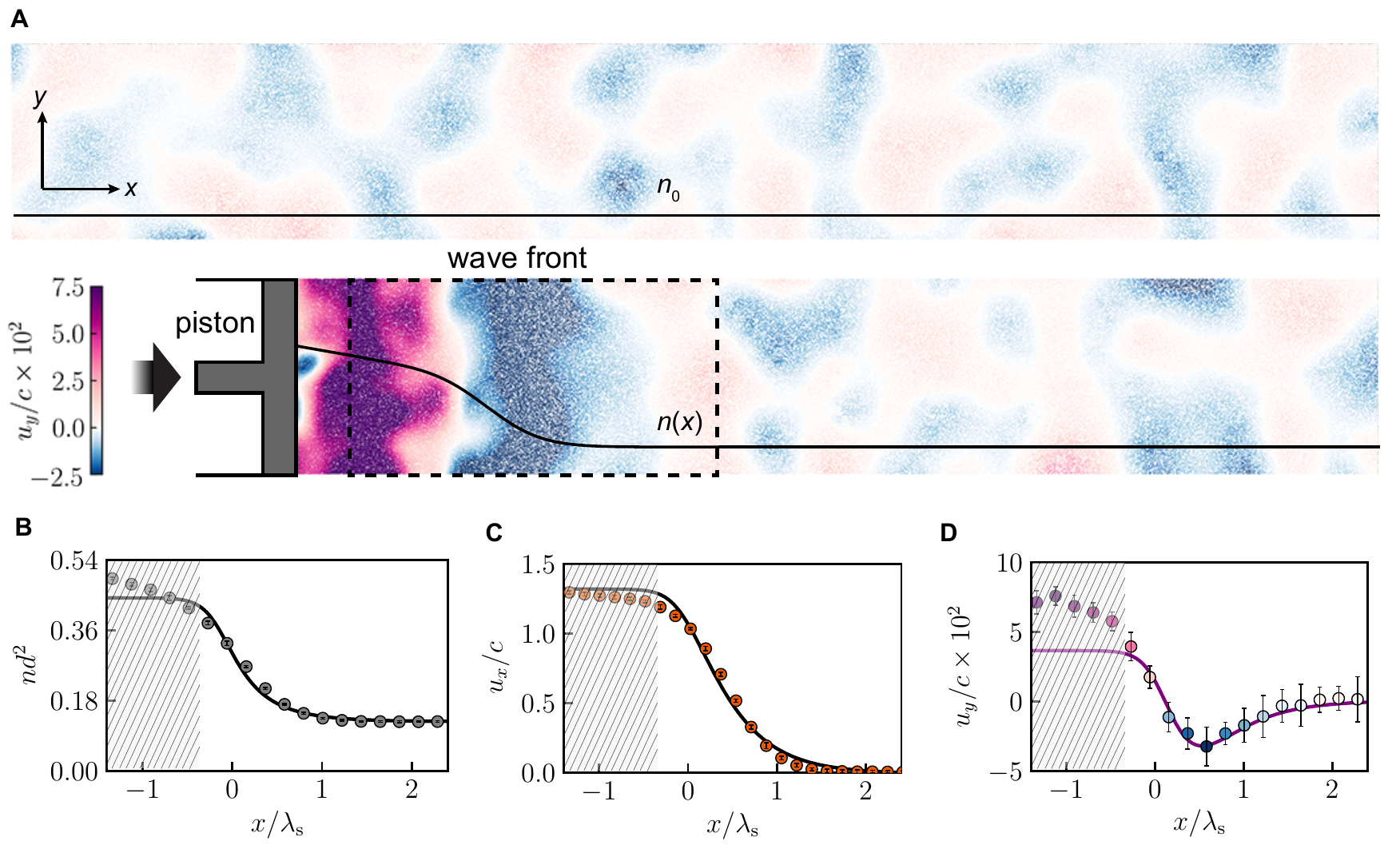}
\caption{\label{fig:flow} \textbf{Transverse mode in a  shock wave.} \textbf{A.} Shock wave. A piston moving at speed $U = 1.9d/\Delta t$ (faster than the speed of sound $c = 1.4d/\Delta t$) generates a shock wave accompanied with transverse flows, which is characterized by the vertical flow velocity $u_y$ (gradient coloring). See Supplementary Mov.~S3.
The particles self-spin counter-clockwise at speed $\Omega = 26.7/\Delta t$ and have an initial global density $n_0 = 0.125d^{-2}$.
According to the viscid Burgers' equation $\partial_t u + u\partial_x u = \nu \partial^2_x u$, the width of this shock is approximately $\lambda_\text{s} = 4\nu/U$, where $\nu = \eta/n_0m$ is the kinematic viscosity.
Hydrodynamic profiles are quantified near the wave front. \textbf{B.} Density profile $n(x)$. The simulation results are compared with continuum hydrodynamic theory (solid line), which employs parameters measured in a separate homogeneous microscopic systems of number density $n_0$ (dashed line). Thus, theoretical predictions would break down at extreme densities (shaded region). \textbf{C.} Horizontal flow velocity $u_x(x)$.  \textbf{D.} Vertical  flow velocity $u_y(x)$. The same color coding as panel A is applied here. Predictions using continuum hydrodynamic theory are plotted as solid lines. }
\end{figure*}

In order to determine the viscosities in Eq.~(\ref{eq:matrix}), we deform the simulation box at constant strain rates using the standard SLLOD algorithm (Fig.~\ref{fig:response}A, Supplementary Sec.~I and Fig.~S5).
We measure all the entries of the viscosity matrix and find, consistently with Eq.~(\ref{eq:matrix}), non-vanishing values only for $\xi$, $\Gamma$, $\eta^\text{A}$, $\eta^\text{B}$, $\eta$ and $\eta^\text{o}$.
Figure~\ref{fig:response} shows the dependence of the measured values on $\Omega$. 
The odd viscosity $\eta^\text{o}$ that couples the two shear stresses has magnitude comparable to the shear viscosity $\eta$.
The origin of the odd viscosity is traced to the microscopic breaking of time-reversal symmetry by interparticle collisions: the ratio $\eta^\text{o}/\eta$ is directly related to an angle characterizing the chirality of the collisions (Supplementary Figs.~S9-S10). By contrast, the other parity-violating viscosities $\eta^\text{A}$ and $\eta^\text{B}$ that couple compression and rotation have magnitude much smaller than the remaining viscosities. 

By comparing simulations with both clockwise and anticlockwise active torques, 
we test whether the viscosity matrix obeys the Onsager--Casimir reciprocity relation $\eta_{\alpha\beta}(\Omega) = \eta_{\beta\alpha}(-\Omega)$ ~\cite{casimir1945onsager}, a telltale sign of quasi-equilibrium  states. 
Consistently with Onsager--Casimir relations, we find that the antisymmetric term $\eta^\text{o}$ is an odd function of $\Omega$ while the symmetric terms $\eta$ and $\zeta$ are even (Fig.~\ref{fig:response}, Supplementary Fig.~S8).
We find that $\eta^\text{A}$ is an odd function of $\Omega$, but due to the numerical uncertainty in~$\eta^\text{B}$ (Supplementary Fig.~S6), we cannot determine the relation between~$\eta^\text{A}(\Omega)$ and~$\eta^\text{B}(-\Omega)$.
We find that $\eta^\text{o} \sim T_\text{eff} \, \Omega$ (Fig.~\ref{fig:response}B), which, under the substitutions $T_\text{eff} \to T$ and $\Omega \to B$, is similar to the odd viscosity of a thermal plasma at temperature $T$ in the limit of a weak magnetic field $B$~\cite{ChapmanCowling}.
The shear viscosity $\eta(T_{\text{eff}})$ depends on $\Omega$ only through the effective temperature.
When a thermostat with temperature $T_0$ is introduced in the simulation, we indeed observe that $\eta^\text{o} \sim (T_\text{eff} + T_0)\Omega$ (Supplementary Fig.~S8), further corroborating the effective-temperature concept. The shear viscosity of our chiral active fluids has an identical functional form $\eta(T_{\text{eff}})$ as the viscosity of the same fluid at equilibrium (i.e., without activity) as long as $T_\text{eff}$  is replaced $T$ (see Supplementary Fig.~S7).

A hallmark of equilibrium is that the response of a system at finite temperature $T$ can be simply determined from correlation functions of its thermal fluctuations. Can $T_\text{eff}$ play a similar role far-from equilibrium? To test this hypothesis we check the validity of the equilibrium form of the Kubo relations, with $T$ replaced by $T_\text{eff}$,
\begin{equation}
	\label{green_kubo}
	\eta_{\alpha \beta} = \frac{A}{k_\text{B} T_\text{eff}} \, \int_0^{\infty} \left< \sigma_{\alpha}(t) \sigma_{\beta}(0) \right>_0 \text{d}t,
\end{equation}
where $A$ is the area of the 2D system and $\left<\enspace \right>_0$ denotes an ensemble-average at the steady state. We numerically evaluate the right-hand side of Eq. (\ref{green_kubo}) focusing on the two fluctuating shear stresses \protect\so\, and \protect\st\, (Fig.~\ref{fig:fluctuations}). The auto-correlation function
$\left<\protect\so(t) \protect\so(0)\right>_0 = \left<\protect\st(t) \protect\st(0)\right>_0$ yields
the shear viscosity $\eta$ 
while the \textit{cross}-correlation function $\left<\protect\so(t) \protect\st(0)\right>_0 = -\left<\protect\st(t) \protect\so(0)\right>_0$  yields the odd viscosity $\eta^\text{o}$ (Fig.~\ref{fig:fluctuations}A).
Note that the latter relation manifestly violates time-reversal symmetry. As shown in Fig.~\ref{fig:fluctuations}B, the values of $\eta$ and $\eta^\text{o}$ computed from the Kubo formula agree well with the values we obtained using the direct hydrodynamic measurements reported in Fig.~\ref{fig:response}.~\footnote{We verified that the long-time tail associated with the breakdown of 2D hydrodynamics is too small to impact the viscosity prediction.}. 

Such a good agreement prompts us to seek a theoretical foundation for the Green--Kubo relation in the presence of activity and dissipative interactions. For thermal systems with conservative interactions, the Green--Kubo relation can be derived microscopically through the so-called Mori--Zwanzig formalism~\cite{Zwanzig2001,Mori1965,Nakajima1958,Zwanzig1960}. In Supplementary Sec.~IV, we extend this formalism to account for dissipative interactions in active fluids (where the Liouvillian can be non-Hermitian) and derive the Green--Kubo relation from first principles without assuming the Onsager regression hypothesis ~\footnote{The linear relation between stresses and velocity gradients holds only for the macroscopic, averaged (or on shell) quantities not the fluctuating ones.}.
We show that an equilibrium-like Green--Kubo relation holds near the steady-state of any isotropic active fluid with reciprocal dissipative interactions, as long as the ensemble-averaged  velocity--velocity correlation $c_{\textbf{v}\textbf{v}}(\textbf{r}) = \left<\textbf{v}(0) \cdot \textbf{v}(\textbf{r})\right>_0$ (see Supplementary Figs.~S11-S12) decays faster than $r^{-D}$ ($D$ the dimension of the system). We find that $T_\text{eff}$ in Eq.~\eqref{green_kubo} is, in general, renormalized by collective velocity fluctuations to $T_\text{eff} + nm\hat{c}_{\textbf{v}\textbf{v}} (\textbf{k} \to 0)/k_\text{B}D$.
For our chiral active fluids with a contact frictional interaction, $c_{\textbf{v}\textbf{v}}(\textbf{r})$ is both small and local, causing a small but detectable correction to $T_\text{eff}$ that matches our predictions (see the red line in Fig.~\ref{fig:fluctuations}C).
In wet active fluids, additional modifications of the Green-Kubo relation are required because the hydrodynamic interactions can be non-reciprocal (see Supplementary Sec.~IV).

Our Green--Kubo relation provides a powerful tool to extract the viscosities of active fluids from correlation functions of the stress. To gain insights into the time dependence of these correlations, we plot the \textit{spatially} averaged stresses 
$\langle\protect\so\rangle_\text{r}$ and $\langle\protect\st\rangle_\text{r}$ against each other as they evolve over time (Fig.~\ref{fig:fluctuations}C). The random trajectories of the collective variables $\langle\protect\so\rangle_\text{r}$ and $\langle\protect\st\rangle_\text{r}$ in shear-stress space are random, confined, and have a tendency towards rotation (Supplementary Mov.~S2).  
Inspired by this observation, we introduce a minimal model based on the following Langevin equation (see discussions in Supplementary Sec.~V and Fig.~S13)
\begin{equation}
    \raisebox{-0.5\height}{\includegraphics[height=42pt]{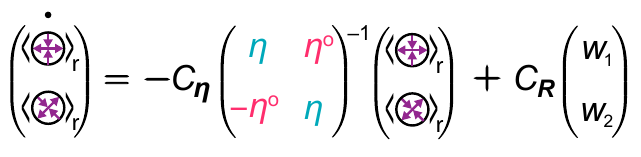}}, \label{eq:langevin}
\end{equation}
where $w_1$ and $w_2$ are two independent white-noise components, the prefactors $C_{\bm{\eta}} =  \big<\langle\protect\so\rangle_\text{r}^2(0)\big>_0\, A/k_\text{B}T_\text{eff}$ and $C_\textbf{R} = \big<\langle\protect\so\rangle_\text{r}^2(0)\big>_0 \sqrt{A/k_\text{B}T_\text{eff} \cdot \eta/(\eta^2 + \eta^{\text{o}2})}$. When the odd viscosity $\eta^\text{o}$ vanishes,
Eq.~\eqref{eq:langevin} simply describes the evolution of an overdamped random walker with Cartesian coordinates $(\langle\protect\so\rangle_\text{r}, \langle\protect\st\rangle_\text{r})$ moving in a harmonic trap. In the presence of a non-vanishing $\eta^\text{o}$, the random walker experiences an additional  {\it azimuthal} nonconservative force proportional to its distance from the origin~\cite{scheibner2019odd} that makes it rotate as shown in Fig.~\ref{fig:fluctuations}C.
In Supplementary Sec.~V, we solve Eq.~(\ref{eq:langevin}) analytically and find closed-form expressions for the stress--stress correlation functions (plotted as continuous lines in Fig.~\ref{fig:fluctuations}A) that match very well with the molecular dynamics simulation measurements (plotted as dots in Fig.~\ref{fig:fluctuations}A).
By Fourier transforming these analytically derived correlation functions, we can predict how the viscous coefficients $\eta(f)$ and $\eta^\text{o}(f)$ depend on the shear frequency $f$, in excellent agreement with numerical data shown in Fig.~\ref{fig:fluctuations}D-E. 
Since viscosities originate from momentum transfer among interacting particles, their characteristic frequencies are controlled by the tumbling time $\Delta t_\text{tumble}$ required for a particle to randomize its direction.

\medskip

We have shown that the viscous coefficients $\eta$, $\eta^\text{o}$ and $\Gamma$ can be obtained from microscopic measurements of the stress fluctuations.
To validate this approach, it is important to test whether  a hydrodynamic description of our chiral fluid with these viscosity coefficients, can accurately describe the resulting macroscopic fluid dynamics.
To do so, we perform large-scale molecular dynamics simulations in which we compress our gas of spinners with a piston as illustrated in Fig.~\ref{fig:flow}A.
The result is a non-linear compression shock: a sharp change in the density profile $n(x)$ (Fig.~\ref{fig:flow}B) moving at constant velocity, with a finite width. 
The hydrodynamic theory of such a compression shock in a chiral fluid~\cite{banerjee2017odd} predicts that  all the transport coefficients contribute to the fluid flow, making it an ideal testing ground for our measured transport coefficients, see Supplementary Sec.~VI. In particular the width of the shock is set by $\eta$, and the shock  is accompanied by a localized shear flow (color map in Fig.~\ref{fig:flow}A, Supplementary Mov.~S3, Supplementary Fig.~S14) controlled by the interplay of $\eta^\text{o}$ and $\Gamma$.
We find that the  velocity profiles $u_x(x)$ and $u_y(x)$ determined from molecular dynamics simulations (dots in Figs.~\ref{fig:flow}C and D) match with the numerical solution of the hydrodynamic equations (solid lines) without any fitting parameters, supporting the validity of our approach. 
Conversely, we find that an imposed steady-state shear flow induces density modulation in the transverse direction (Supplementary Fig.~S15), a phenomenon that allows us to measure $\eta^\text{o}$ in agreement with both direct rheological probes and with the Green-Kubo formula.

Our demonstration of the validity of the Green-Kubo formula far from equilibrium provides a theoretical foundation for a non-invasive rheology of active fluids. This approach enables the measurement of viscous coefficients from movies of steady-state particle motion combined with knowledge of their interactions.

\vspace{5mm}

\noindent\textbf{Acknowledgments} \enspace We thank S. Atis, A. G. Abanov, D. T. Son and H. C. \"{O}ttinger for valuable discussions. 
S.V., W.T.M.I., J.J.d.P. and V.V. acknowledge
primary support through the Chicago MRSEC,
funded by the NSF through grant No.~DMR-1420709. 
S.V. acknowledges support from the National Science Foundation under Grant No.~DMR-1848306. V.V. acknowledges support from the Complex Dynamics and Systems Program of the Army Research Office under grant No.~W911NF-19-1-0268.
M.H. and M.F. acknowledge support from the University of Chicago MRSEC through Kadanoff-Rice postdoctoral fellowships.
C.S. acknowledges support by the National Science Foundation
Graduate Research Fellowship under grant No.~1746045. 
M.H. acknowledges use of the GM4 cluster supported by the National Science Foundation’s Division of Materials Research under the Major Research Instrumentation (MRI) program award No. 1828629.

\onecolumngrid
\clearpage
\newpage

\onecolumngrid
\begin{titlepage}
\centering
{\Large \textbf{Supplementary information}}
\vspace{20mm}
\end{titlepage}

\onecolumngrid
\tableofcontents
\clearpage
\newpage
\twocolumngrid

\setcounter{figure}{0}
\setcounter{equation}{0}
\renewcommand{\thefigure}{S\arabic{figure}}
\renewcommand{\theequation}{S.\arabic{equation}}

\section{Simulation procedure}\label{simulation}

We study the behavior of a chiral active fluid by performing particle-based simulations using customized LAMMPS package. In particular, we consider a two-dimensional (2D) granular gas in which individual particles are powered by active torques
\begin{equation}
    \tau_{i} = \gamma_\text{rot} (\Omega - \Omega_i), 
\end{equation}
to self-spin at a targeted speed $\Omega$. A large coefficient $\gamma_\text{rot}$ is chosen to enforce a homogeneous $\Omega$-field across the system.

In the system we study, the particles interact via excluded-volume effects and interparticle friction.
We employ a Hookean repulsion to model the excluded-volume effects:
\begin{equation}
    \textbf{f}^\text{ c}_{ij} = 
    \begin{cases}
    -k(r_{ij})h_{ij}\hat{\textbf{r}}_{ij}, &\enspace\enspace r_{ij} < d \\
    0, & \enspace\enspace r_{ij} \ge d
    \end{cases}
\end{equation}
where $d$ is the particle diameter, $\R_{ij} = \R_i - \R_j$ is the center-to-center vector between particles $i$ and $j$, and $h_{ij} = d-r_{ij}$ is their radial overlapping depth. 
To prevent complete interpenetration between the particles, a nonlinearity is added to the repulsion, with a distance-dependent Hookean coefficient that diverges at  $r_{ij} = 0$:
\begin{equation}
    k(r_{ij}) = k\left(1+\alpha\frac{d}{r_{ij}}\right).
\end{equation}
In addition to such conservative force, the particles also experience a dissipative force due to interparticle friction, 
\begin{equation}
    \textbf{f}^\text{ d}_{ij} = \begin{cases} - \gamma\left(\textbf{v}_{i j} -
	 \Omega_{ij} \hat{\textbf{z}} \cross d\hat{\textbf{r}}_{i j}\right), \enspace& r_{ij} < d, \\
	 0 & r_{ij} \geq d,\end{cases}
\end{equation}
where $\gamma$ is the friction coefficient. This frictional force is linear with the relative surface velocity between the two particles at contact, $\V_{ij}^\text{sf} = \V_{ij} - \Omega_{ij} \hat{\textbf{z}} \cross d\hat{\textbf{r}}_{i j}$, where $\textbf{v}_{ij} = \textbf{v}_{i} - \textbf{v}_{j}$ denotes the relative velocity of their center of mass (COMs) and $\Omega_{ij} = (\Omega_i + \Omega_j)/2$ denotes their average self-spinning speed.

Without loss of generality, we choose the area fraction of the system $\phi = 0.2$ and set the aforementioned parameters as $\gamma_\text{rot} = 3md^2/\Delta t$, $\gamma = 0.015m/\Delta t$,  $k = m/\Delta t^2$, $\alpha = 0.3$, where $m$ is the particle mass and $\Delta t$ denotes the timescale of the interactions. To focus on the viscous effects emergent from particle interactions, a frictionless background is used. All the simulations are initialized with a random velocity distribution.
The results are collected after the system reaches a steady state. Below, we detail specific procedures for investigating the thermodynamics, kinetics, linear response, Green--Kubo relation and hydrodynamics.

\vspace{10mm}
\noindent\textbf{Thermodynamics} \enspace
To examine the thermodynamic properties of this chiral active fluid in a nonequilibrium steady state, we perform the following analysis on a square system of side length $L = 36d$ with periodic boundary conditions.

\emph{Boltzmann statistics.} 
We first measure the distribution of the particle velocities at 15 different spinning speeds $\Omega \in [1/\Delta t, \, 30 /\Delta t]$. 
We confirm that the particle velocities follow from a Maxwell-Boltzmann distribution and then extract an effective temperature $T_\text{eff}$ from the velocity variance. 
We further examine the Boltzmann statistics in spatial arrangement of the particles by introducing potential bias, i.e. a potential barrier or well of magnitude $|U| < \kT$, into the system. The results are discussed in Section \ref{SSE} below.


\emph{Equation of the state.} We study the density dependence of the hydrodynamic stresses at 8 different particle-number densities $n \in [0.01/d^2, 0.4 /d^2]$. The hydrodynamic stress of the entire system is measured using the Irvine--Kirkwood formula~\cite{irving1950statistical},
\begin{equation}\label{eq:irvstress}
\begin{split}
    \bm{\sigma} &= -\frac{1}{A} \left[\sum_{i}^{N} m\textbf{v}_i\textbf{v}_i + \frac{1}{2}\sum_{ij, \; i \neq j}^{N(N-1)} \textbf{f}_{ij}\textbf{r}_{ij}\right], \\
\end{split}
\end{equation}
where $A$ denotes the total area of the system and $N$ denotes the total number of the particles. In Section \ref{SSE}, we show that the pressure $P \triangleq -(\sigma_{xx} + \sigma_{yy})/2$ follows the ideal-gas law $P = n \kT$. Furthermore, we measure the anti-symmetric stress $\tau \triangleq (\sigma_{xy} - \sigma_{yx})/2$ and determine its density dependency $\tau = \Gamma(n) \Omega$. 

\vspace{10mm}

\noindent\textbf{Kinetics} \enspace To study the microscopic origin of the anti-symmetric stress $\tau$ and odd viscosity $\eta^\text{odd}$, we analyze two-particle scattering simulations. In these simulations, two particles undergo a head-to-head collision. The incident relative velocities between the particles are sampled from a Maxwell--Boltzmann distribution with reduced mass $m/2$ and temperature $T_\text{eff}(\Omega)$. The impact parameter $b$ is sampled from a uniform distribution $\mathcal{U}[-d, d]$. To reveal the origin of $\tau$, we measure the angular momentum change $\Delta L$ of the particle pair caused by the collision. To investigate the cause of $\eta^\text{odd}$, we quantify the change in their relative velocity $\Delta \V$ instead. Here, at each of 20 different $\Omega \in [-30/\Delta t, \, 30/\Delta t]$, we simulate over $10,000$ independent collisions to reduce statistical errors. The results are discussed in Section \ref{linResp}.

\vspace{10mm}

\noindent\textbf{Linear responses} \enspace
We study the linear response of our chiral active fluid by imposing uniform deformations and measuring the stress response. The deformation is implemented via the SLLOD algorithm~\cite{evans1984nonlinear,daivis2006simple,evans2008statistical} with periodic boundary conditions. We measure bulk viscosity $\xi$ and compression-rotation viscosity $\eta^\text{A}$ by imposing compression/dilation (\protect\dtV). We measure shear viscosity $\eta$, odd viscosity $\eta^\text{o}$ and compression-rotation viscosity $\eta^\text{B}$ under simple shear (\protect\dtS). Furthermore, we confirm our measurements of $\eta$ and $\eta^\text{o}$ by performing additional simulations under pure shear (\protect\dtSo). The results are discussed in Section \ref{linResp}.

To avoid artifacts caused by dramatic changes in system size, we apply oscillatory deformations with a time-modulated strain rate,
\begin{equation}
  \dot{e}_{\beta}(t) =
    \begin{cases}
      \epsilon &  nT < t \leq (n+\frac{1}{2})T\\[3pt]
      -\epsilon & (n+\frac{1}{2})T < t \leq (n+1)T \\
    \end{cases}       
\end{equation}
a square wave of small magnitude $\epsilon < 0.25\%/\Delta t$ and long period $T = 1000\Delta t$. We calculate the time-weighted averages of both the strain rate and the resultant stress:
\begin{align}
    \dot{e}_{\beta}(\epsilon) &= \left<\dot{e}_{\beta}(t) \cdot \text{sgn}[\dot{e}_\beta(t)]\right>_\text{t},\\[6pt]
    \sigma_{\alpha}(\epsilon) &= \left<\sigma_{\alpha}(t) \cdot \text{sgn}[\dot{e}_\beta(t)]\right>_\text{t}
\end{align}
where the function $\text{sgn}(x)$ extracts the sign of $x$.
Note that the strain rate $\dot{e}_{\beta}(\epsilon) = \epsilon$. Regarding the stress, the oscillatory nature of $\text{sgn}[\dot{e}_\beta(t)]$ naturally removes the steady-state stresses, such as pressure $P$ and antisymmetric stress $\tau$, which are invariant under deformation. It also avoids the influences from the normal stress difference~\cite{sierou2002rheology,weissenberg1947continuum,campbell1989stress} caused by micro-structure formation of the particles under shear, which is quadratic with the strain rate. The viscous coefficients are extracted from the linear response:
\begin{equation}
    \eta_{\alpha\beta} = \left<\frac{\sigma_\alpha(\epsilon)}{\dot{e}_\beta(\epsilon)}\right>_{\epsilon},
\end{equation}
which is an average over 15 different deformation magnitudes $\epsilon \in [0.08\%/\Delta t, \, 0.25\%/\Delta t]$.

We repeat such viscosity measurement at 30 different spinning speeds $\Omega \in [-30/\Delta t, \, 30/\Delta t]$ to evaluate the Onsager--Casimir relation $\eta_{\alpha\beta}(\Omega) = \eta_{\beta\alpha}(-\Omega)$. Lastly, we also  investigate the temperature dependence of $\eta^\text{o}$ by introducing an intrinsic temperature $T_0$ via added random forces.

\vspace{5mm}

\noindent\textbf{Green--Kubo relation} \enspace 
To study the Green--Kubo relation, we investigate the dynamics of the fluctuating stresses at the steady state. In particular, we measure the correlation functions between the two shear stresses (\protect\so\, and \protect\st) of the entire system, and use them to estimate the shear and odd viscosities via the Green--Kubo formula (Eq.~(3) in the main text). Such Kubo predictions are compared with direct measurements via linear responses for 30 different spinning speeds $\Omega \in [-30/\Delta t, \, 30/\Delta t]$.

To further verify the Green--Kubo relation in frequency domain, we perform linear-response analysis at $\Omega = 26.7/\Delta t$ under a sinusoidal simple shear,
\begin{equation}
    \protect\dtS (t) = \epsilon \text{cos}(2\pi ft),
\end{equation}
with frequency $f$ and magnitude $\epsilon$. The frequency-dependent strain rate and stresses are calculated as follow:
\begin{align}
    \hat{\protect\dtS} (f, \epsilon) &= \left<\protect\dtS (t) \, \e^{-i 2\pi ft}\right>_\text{t},\\[3pt]
    \hat{\protect\so} (f, \epsilon) &= \left<\protect\so (t) \, \e^{-i 2\pi ft}\right>_\text{t}, \\[3pt]
    \hat{\protect\st} (f, \epsilon) &= \left<\protect\st (t) \, \e^{-i 2\pi ft}\right>_\text{t},
\end{align}
where $\left<\enspace\right>_t$ denotes time average.
From the frequency dependent strain rate and stresses, we extract the frequency-dependent shear and odd viscosities
\begin{align}
    \eta(f) &= \left<\frac{\hat{\protect\st} (f, \epsilon)}{\hat{\protect\dtS} (f, \epsilon)}\right>_\epsilon,\\[3pt]
    \eta^\text{o}(f) &= \left<\frac{\hat{\protect\so} (f, \epsilon)}{\hat{\protect\dtS} (f, \epsilon)}\right>_\epsilon,
\end{align}
which are averages over 15 different shear magnitudes $\epsilon \in [0.08\%/\Delta t, \, 0.25\%/\Delta t]$. These measured viscosities are then compared with the Fourier transform of the stress--stress correlation functions. The results are discussed in the main text and theoretical derivations are provided in Section \ref{sec:Kubo}.

\vspace{10mm}

\noindent\textbf{Hydrodynamics} \enspace 
We first study the steady-state flow of our chiral active fluid. 
We employ a large system of length $L_x = 600d$ and width $L_y = 300d$, with periodic boundary conditions applied in both dimensions.  
To investigate the effects of odd viscosity $\eta^\text{o}$, we generate a shear flow by imposing a $y$-directional force field $\textbf{F} = (0, F_0 \text{sin}(2\pi x/L_x))$ onto the particles and measure the variation of particle density in the $x$-direction. The results are discussed in Section \ref{secHydro}.

Secondly, we study a shock wave as an example of a nonlinear hydrodynamic phenomenon. To do so, we employ an even larger system size $L_x = 6000d$ and $L_y = 300d$. Harmonic walls with Hookean constant $k_\text{wall} = 100m/\Delta t^2$ are placed at the boundaries in the $x$-direction, while periodic boundary conditions are applied in the $y$-direction. Right next to the walls, we add a thin buffer zone (with thickness $w = d$) where particles experience a strong damping towards their $y$-directional velocity via the force $\F = (0, -\gamma_\text{buf} v_y)$, where $\gamma_\text{buf} = m/\Delta t$ is the local drag coefficient.
The buffer zone gives an effective stick boundary condition for our chiral active fluid. To generate a density shock wave in $x$-direction, we move the left wall and its buffer zone at speed $v = 1.4c$, where $c = 1.4 d/\Delta t$ is the speed of sound of this fluid. We record the time-dependent flow and density profiles, $u_x(x, t)$, $u_y(x, t)$ and $n(x, t)$, which are averaged over 500 independent simulations to reduce statistical errors. The results are discussed in Section \ref{secHydro}.

All the particle-based hydrodynamic simulations are compared with the predictions of the continuum theory with hydrodynamic parameters determined by the measurements in the previous sections.  

\onecolumngrid
\clearpage
\newpage
\twocolumngrid

\section{Steady-state ensemble} \label{SSE}

\begin{figure}[h]
\includegraphics[width=0.48\textwidth]{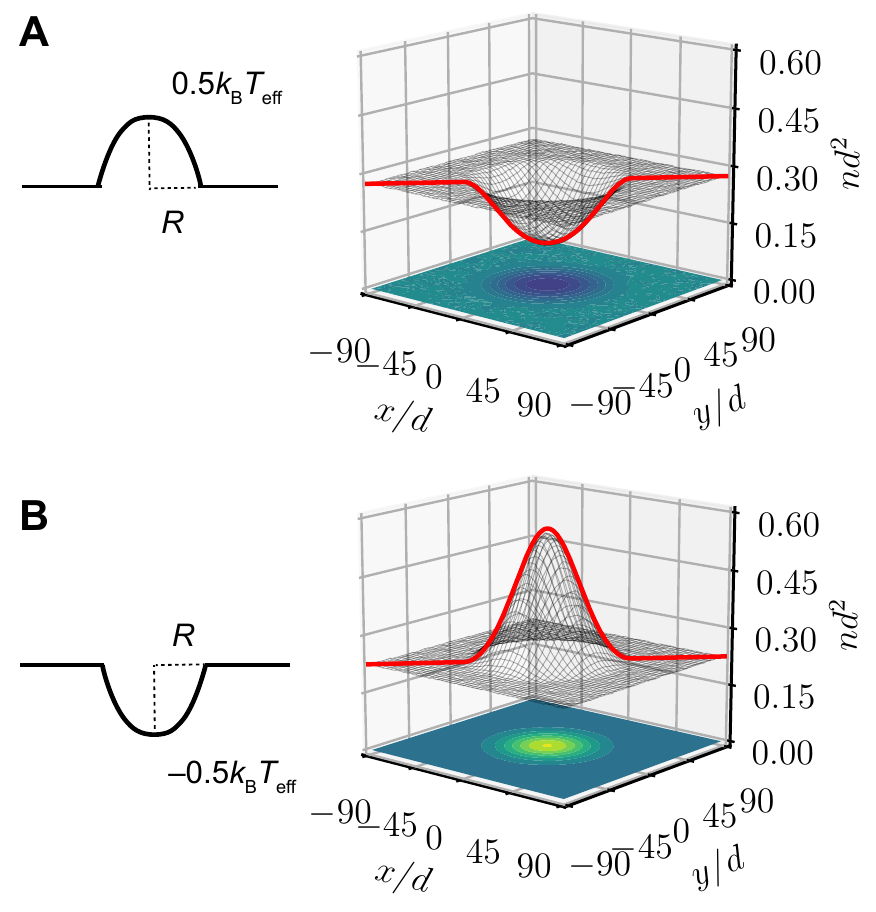}
\caption{\label{fig:Boltzmann} \textbf{Boltzmann statistics.} We investigate the density distribution $n(\R)$ of the particles when a potential bias is applied. Here the particles self-spin at speed $\Omega = 26.7/\Delta t$. \textbf{A.} Potential barrier $U(\R) = 0.5\kT\,\text{cos}(\pi 
r/2R)$ for  $r < R$, where $r$ denotes the distance from the center of the system. \textbf{B.} Potential well $U(\R) = -0.5\kT\,\text{cos}(\pi 
r/2R)$ for $r < R$. The measured density profiles (2D mesh) match well with the theoretical predictions (red lines) using Boltzmann statistics $n(-r) \propto \text{exp}\left[-U(r)/\kT\right]$.}
\end{figure}

\noindent\textbf{Boltzmann statistics} \enspace In the main text, we have shown the Maxwell--Boltzmann distribution of particle velocity. Here we verify that Boltzmann statistics parameterized by the same effective temperature $T_\text{eff}$ govern the spatial arrangement of the partiles as well (Fig.~\ref{fig:Boltzmann}).

\vspace{10mm}

\noindent\textbf{Effective thermodynamics} \enspace 
Here we use a mean-field approximation to derive an effective Langevin dynamics for the system. This effective Langevin dynamics justifies the equilibrium-like ensemble properties of the system's steady state.
For an arbitrary particle $i$, each collision causes a drag effect via the dissipative interaction $-\gamma\V_{ij}$. At the steady state, the particles are uniformly distributed and acquires random velocities. Thus the collisions with neighboring particles provide a background drag with drag coefficient
\begin{equation} \label{eq:drag}
    \gamma_\text{eff} = p_\text{col} \gamma,
\end{equation}
where $p_\text{col} = n \pi d^2$ accounts for the probability of a particle colliding with another at a given time. In addition, since the interparticle vector $\R_{ij}$ is random, the active part of the interaction $\gamma d\Omega \hat{z} \cross \hat{\textbf{r}}_{i j}$ due to self-spinning acts as an effective random force $\bm\xi(t)$. By replacing those two interactions with the effective drag and random forces, we can rewrite Eq.~(1) in the main text as an effective Langevin equation:
\begin{equation} \label{eq:lang}
    m \ddot{\textbf{x}}_i = \sum_{j \in N(i)} \textbf{f}^{\;\text{c}}_{ij} - \gamma_\text{eff}\textbf{v}_{i} + 
	\bm\xi(t),
\end{equation}
where the random force $\bm\xi(t)$  satisfies $\expval{\xi_a(t) \xi_b(t')} = 2 \gamma_\text{eff} k_{\text{B}} T_{\text{eff}} \delta_{ab} \delta(t - t')$ to produce the effective temperature. 
\newline

\begin{figure}[h]
\includegraphics[width=0.48\textwidth]{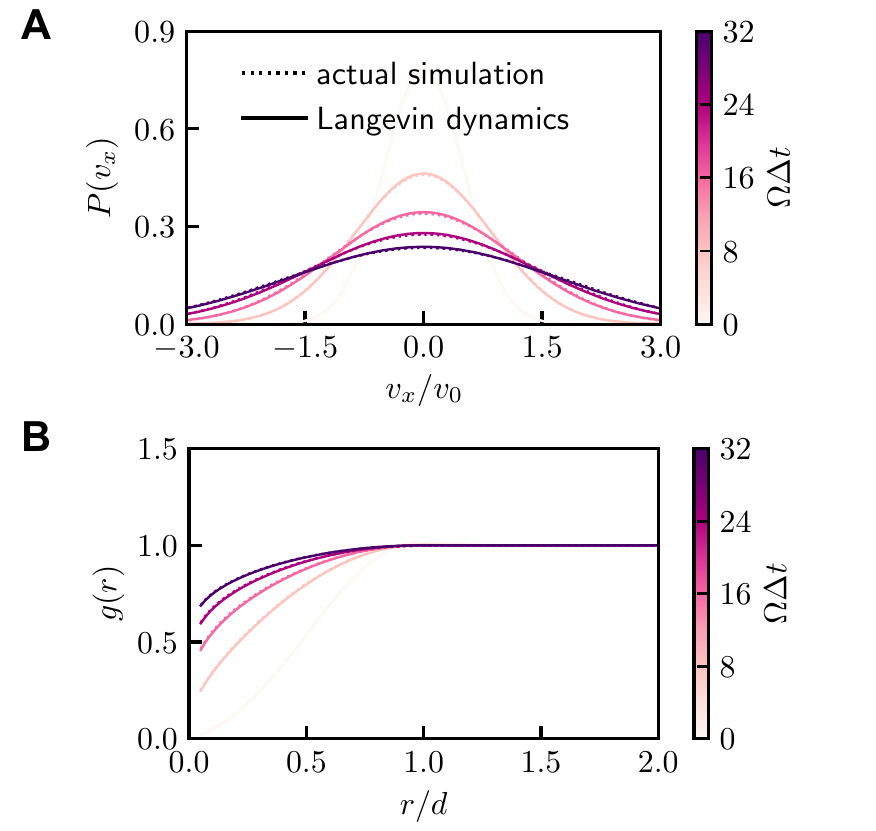} 
\caption{\label{fig:thermal} \textbf{Validating effective Langevin dynamics.} \textbf{A.} Velocity distribution. \textbf{B.} Radial distribution function. The results generated by the effective Langevin dynamics Eq.~(\ref{eq:lang}) (solid lines) are almost identical those with actual simulations (dashed lines).}
\end{figure}

By evaluating the particle velocity distribution and radial distribution function, we confirm that this effective Langevin dynamics produces the Boltzmann statistics observed in the actual simulation (Fig.~\ref{fig:thermal}). 
While this effective theory explains the equilibrium-like properties of the steady state at any single point in time, the substitution of interparticle friction $\textbf{f}^\text{ d}_{ij}$ with the single-particle forces $\gamma_\text{eff}\V_i$ and $\bm\xi_i$ cannot capture the right dynamics. 
In particular, the random noise $\bm\xi_i$ lacks the chiral nature of the active interaction $\gamma d\Omega \hat{z} \cross \hat{\textbf{r}}_{i j}$, which gives rise to odd viscosity $\eta^\text{o}$ (later shown by Fig.~\ref{fig:odd-viscosity-dV}) and breaks time-reversal symmetry~\cite{avron1998odd,banerjee2017odd}. 
Therefore, the time-correlated properties of our chiral active fluid are not entirely captured by this effective Langevin model. For instance the relation $\left<\protect\so(t) \protect\st(0)\right>_0 = -\left<\protect\st(t) \protect\so(0)\right>_0$ implied by the presence of $\eta^\text{o}$ (see Fig.~3A in the main text) does not follow from the Langevin model.
\newline

\vspace{5mm}

\noindent\textbf{Effective temperature} \enspace 
In addition to justifying the equilibrium-like behavior of the equal-time statistics, the Langevin-dynamics model provides an estimate of the effective temperature. Each collision makes a contribution to the random force of magnitude $\gamma d\Omega$. Given that $\R_{ij}$ is random, we argue that different components of the random force are independent of each other. Thus, 
\begin{equation}
    \left<\xi_a(0)\xi_b(0)\right> = p_\text{col} \frac{(\gamma d\Omega)^2}{2} \delta_{ab},
\end{equation}
where the one half accounts for the equipartition over each dimension of this 2D system. Furthermore, the collision duration $\Delta t_\text{col}$ sets the minimum time scale of the Langevin dynamics. By assuming different collisions are independent of each other, we can rewrite the requirement $\expval{\xi_a(t) \xi_b(t')} = 2 \gamma_\text{eff} \kT \delta(t - t')\delta_{ab}$ as
\begin{equation}
    p_\text{col}\, \frac{(\gamma d\Omega)^2}{2}= 2 \gamma_\text{eff} \kT \frac{1}{\Delta t_\text{col}}.
\end{equation}
As $\gamma_\text{eff} = p_\text{col}\gamma$, the effective temperature becomes
\begin{equation}\label{eq:Teff}
    \kT = \frac{1}{4} \gamma d^2 \Omega^2 \Delta t_\text{col}.
\end{equation}
The collision duration $\Delta t_\text{col}$ depends on the repulsive force. In the extreme of a strong Hookean interaction, a collision between two particles corresponds to a half-cycle harmonic oscillator with effective mass $m/2$ and spring constant $k$. Thus, the duration time $\Delta t_\text{col} = \pi\sqrt{m/2k}$ is a constant independent of particle velocity. However, in the other extreme, when the repulsion is too weak to consider, the two particles simply penetrate each other, with a duration time  $\Delta t_\text{col} = d/\bar{v}_\text{rel}$, where $\bar{v}_\text{rel} =  m/4\pi\kT \cdot \int_0^\infty 2\pi v^2 \text{exp}[-mv^2/4\kT]\,dv = \sqrt{\pi\kT/m}$ is the averaged relative speed. By plugging these two extreme cases of $\Delta t_\text{col}$ into Eq.~(\ref{eq:Teff}), we find that the effective temperature follows a power-law behavior
\begin{equation}\label{eq:Teff}
    \kT \propto |\Omega|^{\alpha},
\end{equation}
with exponent $\alpha$ constrained in the range of $4/3 \leq \alpha \leq 2$. This is consistent with the power-law behavior $\kT \propto |\Omega|^{1.54 \pm 0.02}$ observed in our case.
 
\vspace{10mm}

\begin{figure}[h]
\includegraphics[width=0.46\textwidth]{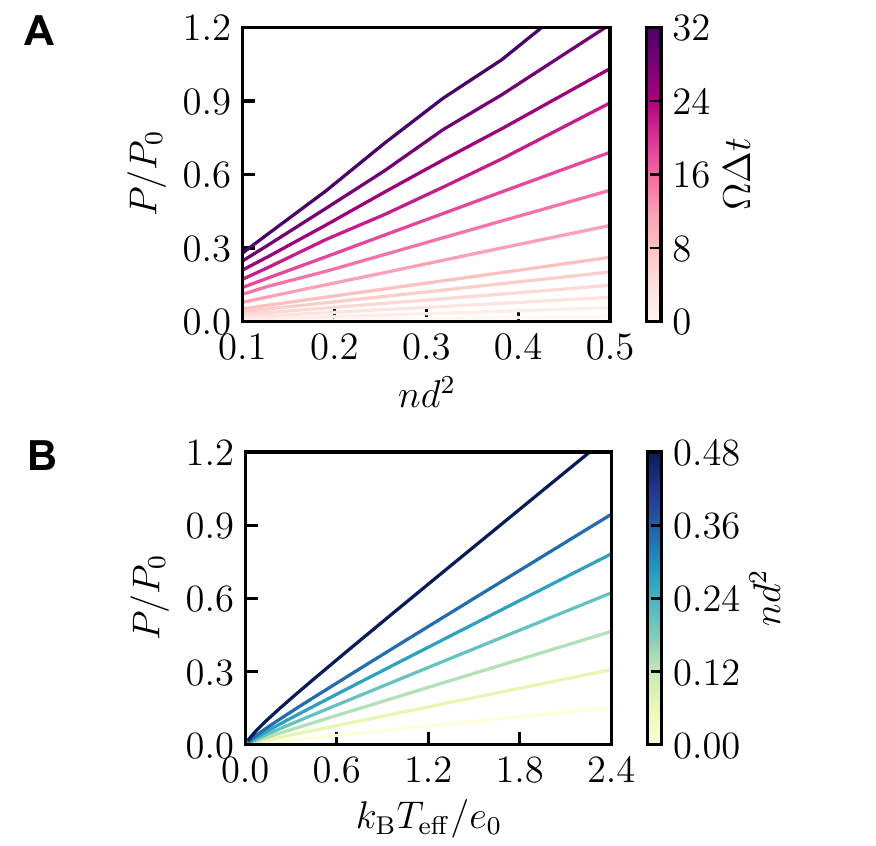}
\caption{\label{fig:pressure} \textbf{Ideal-gas law.} A determination of the equation of state for the chiral active gas. The linear relationship is consistent with the ideal-gas law  $P = n\kT$. \textbf{A.} Dependence of pressure $P$ on particle density $n$.  \textbf{B.} Dependence of $P$ on effective temperature $T_\text{eff}$.}
\end{figure}

\noindent\textbf{Ideal-gas law} \enspace  Fig.~\ref{fig:pressure} shows the result of measuring the system pressure via the Irvine--Kirkwood stress at various particle density $n$ and spinning speed $\Omega$. We confirm that the fluid does follow the ideal-gas law $P = n\kT$ (Fig.~\ref{fig:pressure}). Hence, the microscopic Boltzmann statistics gives rise to a macroscopic equation of state resembling a thermal gas. 

\vspace{10mm}

\begin{figure}[ht]
\includegraphics[width=0.54\textwidth]{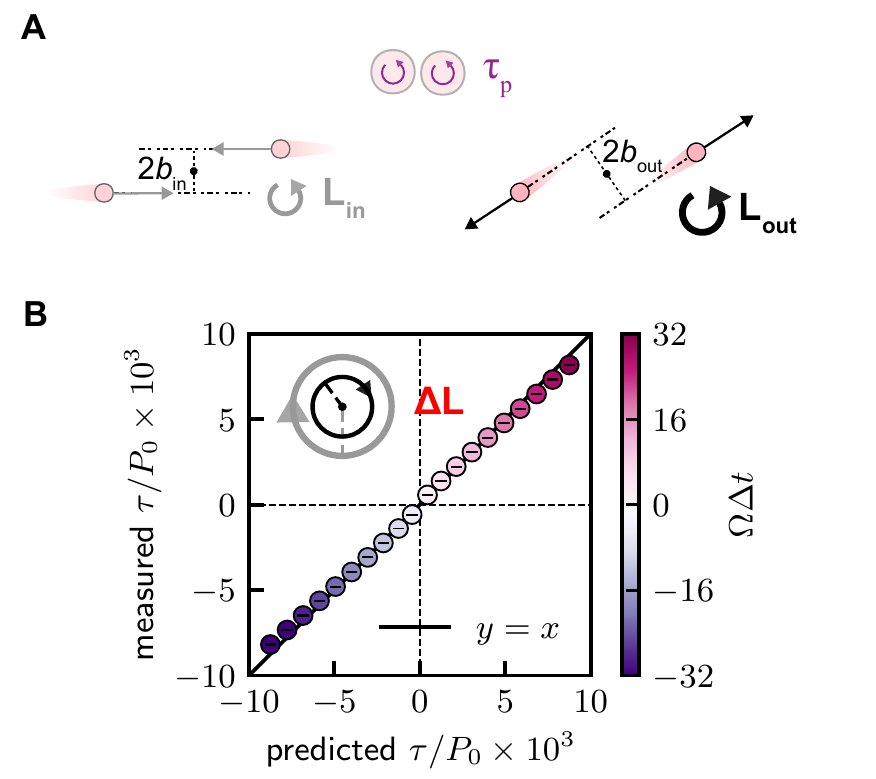} 
\caption{\label{fig:odd-stress} \textbf{Microscopic origin of antisymmetric stress.} \textbf{A.} Schematic of an interparticle collision. When two particles collide, the self-spinning is converted into the co-rotation of the particles around their center-of-mass, leading to an angular momentum change $\Delta L = L_\text{out} - L_\text{in}$, where $L = mv_\text{rel}b$ is the angular momentum of the co-rotation ($b$ the impact parameter). On average, this gives rise to the anti-symmetric stress $\tau$. \textbf{B.} Comparison between the theoretical prediction and simulation measurement of $\tau$. We measure the average angular momentum change upon collision $\overline{\Delta L}$ by performing scattering simulations, and then use it to predict $\tau$ based upon the kinetic theory Eq.~\eqref{eq:tau}. Such predictions agree well with the steady-state measurement of the many-body system. }
\end{figure}

\noindent\textbf{Anti-symmetric stress} \enspace Unlike a common fluid at thermal equilibrium, our chiral active fluid has the tendency to rotate even at the steady state due to a non-vanishing anti-symmetric stress $\tau$. The antisymmetric stress arises from the active torques which constantly inject angular momentum into the system via the self-spinning of the particles. When two particles collide, the angular momentum of self-spinning is partially converted into the angular momentum of the co-rotation of the two particles around their center of mass, $L = mv_\text{rel}b$, where  $b$ is the impact parameter of the collision (Fig.~\ref{fig:odd-stress}A). The angular momentum change $\Delta L = L_\text{out} - L_\text{in}$ caused by the collision gives rise to the anti-symmetric stress $\tau$ at the macroscopic level, leading to an additional equation of state.
\newline

In fact, we can analytically derive the anti-symmetric stress using a simple kinetic theory. In 2D, a particle moves across the system with a collisional cross-section $2d$. The frequency of it colliding with another particle is 
\begin{equation}
    f_\text{col}^\text{p} = 2d \cdot \bar{v}_\text{rel} \cdot n.
\end{equation}
Thus the total collision frequency of the entire system is 
\begin{equation}
    f_\text{col}^\text{s} = \frac{1}{2} \cdot f_\text{col}^\text{p} \cdot N =  n^2 Ad \bar{v}_\text{rel}
\end{equation}
where the one half accounts for the double counting of the collision pairs, and $N = n A$ is the total number of particles. Let us denote the average angular momentum change due to a single collision as $\overline{\Delta L}$. Then we can derive the anti-symmetric stress:
\begin{equation}\label{eq:tau}
    \tau = \frac{f_\text{col}^\text{s} \cdot \overline{\Delta L}}{A} = \sqrt{\frac{\pi\kT}{m}} n^2 d \overline{\Delta L}.
\end{equation}

To confirm our kinetic theory, we perform molecular dynamics simulations of the particle kinetics. We numerically measure the averaged angular momentum change $\Delta L$ during interparticle collision and plug it into Eq.~\eqref{eq:tau} to predict $\tau$. In Fig.~\ref{fig:odd-stress}B, we show that the predicted value agrees well with the simulation measurement at the steady state.

\onecolumngrid
\clearpage
\newpage

\twocolumngrid

\section{Linear response}

\noindent \textbf{Linear response} \label{linResp} \enspace 
Linear hydrodynamics relies on the existence of a linear relationship between stresses and velocity gradients. This linear relationship is summarized by the following equation: 
\begin{align}
    \sigma^{(v)}_{ab} = \eta_{abcd} \dot e_{cd}, 
    \label{lin}
\end{align}
where $\sigma_{ab}^{(v)}$ is the viscous stress tensor, $\dot e_{cd} =\partial_d u_c$ is the (unsymmetrized) velocity gradient tensor, and $\eta_{abcd}$ is the viscosity tensor. Here, we introduce the notation used in Eq.~(2) of the main text and discuss how various physical symmetries restrict the form of $\eta_{abcd}$. \newline

Following the example in Ref.~\cite{scheibner2019odd}, we introduce the following basis for rank-2 tensors in two dimensions:
\begin{align}
    &\tau^{0}_{ab} = \mqty(1 & 0 \\ 0 & 1) 
    &&\tau^{1}_{ab} = \mqty(0 & -1 \\ 1 & 0) \\
    &\tau^{2}_{ab} = \mqty(1 & 0 \\ 0 & -1)
    &&\tau^{3}_{ab} = \mqty(0 & 1 \\ 1 & 0).
\end{align}
We note that $\tau_{ab}^0$ transforms as a scalar under rotations while $\tau_{ab}^1$ as a pseudoscalar. The matrices $\tau^2_{ab}$ and $\tau^3_{ab}$ form a basis for symmetric traceless tensors, and transform together as bivectors under rotations. We use the $\tau^\alpha_{ab}$ to decompose the stress and velocity gradient tensors into irreducible representations of $SO(2)$ via the following definitions: 
\begin{align}
    \shortspace \protect\p & \triangleq \frac12\tau^0_{ab} \sigma_{ab} = (\sigma_{xx} + \sigma_{yy})/2, \\[3pt]
    \protect\tor & \triangleq \frac12\tau^1_{ab} \sigma_{ab} = (\sigma_{yx} - \sigma_{xy})/2, \\[3pt]
    \protect\so & \triangleq \frac12\tau^2_{ab} \sigma_{ab} = (\sigma_{xx} - \sigma_{yy})/2,  \\[3pt]
    \protect\st & \triangleq  \frac12\tau^3_{ab} \sigma_{ab} = (\sigma_{xy} + \sigma_{yx})/2,
\end{align}
and 
\begin{align}
    \protect\dtV & \triangleq  \tau^0_{ab} \dot e_{ab} = \dot{e}_{xx} + \dot{e}_{yy},\midspace\enspace\\[6pt]
     \protect\dtR & \triangleq  \tau^1_{ab} \dot e_{ab} = \dot{e}_{yx} - \dot{e}_{xy}, \\[6pt]
     \protect\dtSo & \triangleq  \tau^2_{ab} \dot e_{ab} =\dot{e}_{xx} - \dot{e}_{yy}, \\[6pt]
     \protect\dtSt & \triangleq  \tau^3_{ab} \dot e_{ab} = \dot{e}_{xy} + \dot{e}_{yx}.
\end{align}
Furthermore, we define the four-by-four matrix $\eta^{\alpha \beta} = \tau^\alpha_{ab} \eta_{abcd} \tau^{\beta}_{cd}$. With these definitions, Eq.~(\ref{lin}) can be written as:
\begin{align}
  \mqty(\protect \p \\ \protect \tor \\ \protect \so \\ \protect \st)^{(v)} =
  \mqty(
  \eta^{00} & \eta^{01} & \eta^{02} & \eta^{03} \\
  \eta^{10} & \eta^{11} & \eta^{12} & \eta^{13} \\
  \eta^{20} & \eta^{21} & \eta^{22} & \eta^{23} \\
  \eta^{30} & \eta^{31} & \eta^{32} & \eta^{33} 
  )
  \mqty( \protect\dtV \\ \protect\dtR \\ \protect \dtSo \\ \protect \dtSt),
\end{align}
where the superscript $(v)$ denotes the viscous stresses. Certain basic physical assumptions restrict the form of $\eta^{\alpha \beta}$. For example, under the assumption of isotropy alone, $\eta^{\alpha \beta}$ takes the form~\cite{scheibner2019odd}:
\begin{align}
    \eta^{\alpha \beta} = 
    \mqty( 
    \xi & \eta^\text{B} & 0 & 0 \\
    \eta^\text{A} & -\Gamma & 0 & 0 \\
    0 & 0 & \eta & \eta^0 \\
    0 & 0 & -\eta^0 & \eta \\
    ).\label{gen}
\end{align}
We note that the Onsager--Casimir reciprocity relations imply that the antisymmetric contribution to $\eta^{\alpha \beta}$ must be odd under microscopic time-reversal symmetry, while the symmetric portion must be even under microscopic time reversal symmetry~\cite{DGM,deGroot1954}. In standard tensor notation, Eq.~(\ref{gen}) may be written as:
\begin{align}
    \eta_{abcd} =& \;\xi \delta_{ab} \delta_{cd} -  \eta^\text{A} \epsilon_{ab} \delta_{cd} - \eta^\text{B} \delta_{ab} \epsilon_{cd}  - \Gamma \epsilon_{ab} \epsilon_{cd}\notag \\[6pt]
    & + \eta (\delta_{ac} \delta_{bd} + \delta_{ad} \delta_{bc} - \delta_{ab} \delta_{cd}) + \eta^\text{o} E_{abcd} ,
\end{align}
where $\delta_{ab}$ and $\epsilon_{cd}$ denote the Kroneker delta and Levi-Civita tensors, respectively, and 
\begin{equation}
    E_{abcd} = \frac12 (\epsilon_{ac} \delta_{bd} + \epsilon_{ad} \delta_{bc} + \epsilon_{bd} \delta_{ac} + \epsilon_{bc} \delta_{ad}).
\end{equation}
\vspace{1mm}

When the viscosity coefficients do not depend on space, we have the general form of the Navier-Stokes equation $\rho D_t \bm{u} = \div{\bm{\sigma}}$ for chiral active fluids:
\begin{align}
	\label{explicit_navier_stokes}
    \rho D_t \bm{u} = &\; \div{\bm{\sigma}_\text{ss}} + \xi \, \bm\nabla \, (\div{\bm{u}}) + \eta^\text{A} \mathcal{R} \cdot \bm\nabla \, (\div{\bm{u}}) \notag\\[6pt]
    &-\Gamma \, \bm\nabla \times (\bm\nabla \times \bm{u}) - \eta^\text{B} \mathcal{R} \cdot \bm\nabla \times (\bm\nabla \times \bm{u}) \notag \\[6pt]
    &+\, \eta \, \Delta \bm{u}
    \,+\, \eta^\text{o} \, \mathcal{R} \cdot \Delta\bm{u},
\end{align}
where 
\begin{align}
    \mathcal{R} = \begin{pmatrix} 0 & 1 \\ -1 & 0\end{pmatrix}
\end{align}
is the rotation matrix by $-\pi/2$.
\newline

Equation~\eqref{explicit_navier_stokes} can be compared, for instance, with Eq.~(37) of Ref.~\cite{DGM} (CH. XII, \S~1, p. 310).
In this reference, our coefficient $\Gamma$ is called the rotational viscosity $\eta_r$. 
Both the pressure and the anti-symmetric stress $\eta_r \text{rot}(2 \omega)$ in Eq.~(37) of Ref.~\cite{DGM} (in this reference, $\omega$ is the mean angular velocity of the fluid) are included in the term $\div{\bm{\sigma}_\text{ss}}$ of our Eq.~\eqref{explicit_navier_stokes}.
The coefficient $\xi$ in Eq.~\eqref{explicit_navier_stokes} is related to the volume viscosity $\eta_{\text{v}}$ in Ref.~\cite{DGM} through $\xi = \eta/D + \eta_{\text{v}}$, where $D$ is the spatial dimension. The terms involving $\eta^\text{A}$ and $\eta^\text{B}$ are additional contributions, which are generally allowed in a chiral active fluid.
\newline

\medskip

\begin{figure}[h]
\includegraphics[width=0.52\textwidth]{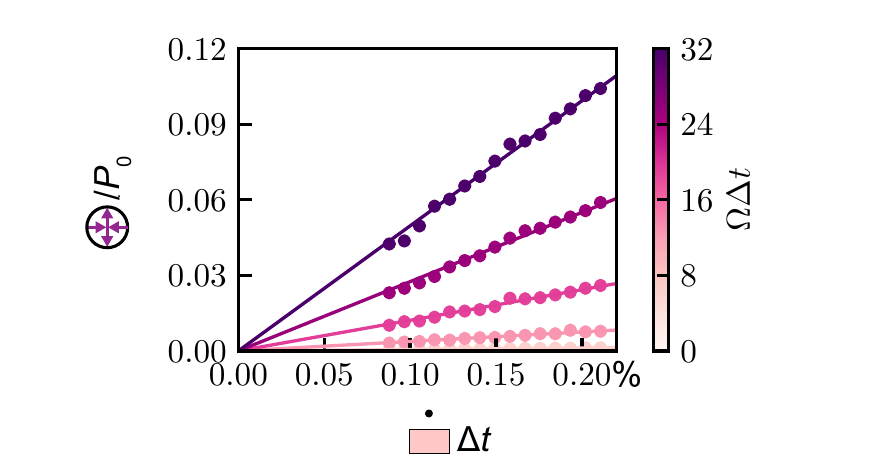}
\caption{\label{fig:linear-response} \textbf{Linear response.}  As an illustrative example of linear response, we show the dependence of the pure shear stress \protect\so\, on the corresponding strain state \protect\dtSo\, at different spinning speeds $\Omega$. Here, the strain rate is chosen to be large enough ($\dot{e}> 0.08\%/\Delta t$) to achieve a good signal-noise ratio but at the same time small enough ($\dot{e} < 0.25\%/\Delta t$) to keep the system in the linear regime.}
\end{figure}

Figure~\ref{fig:linear-response} shows the linear response of our system when a small strain rate is applied. This allows us to directly measure all the viscous coefficients and further study their dependence on particle spinning speed $\Omega$ to evaluate the Onsager--Casimir relation.
\newline

\vspace{10mm}

\begin{figure}[h]
\includegraphics[width=0.52\textwidth]{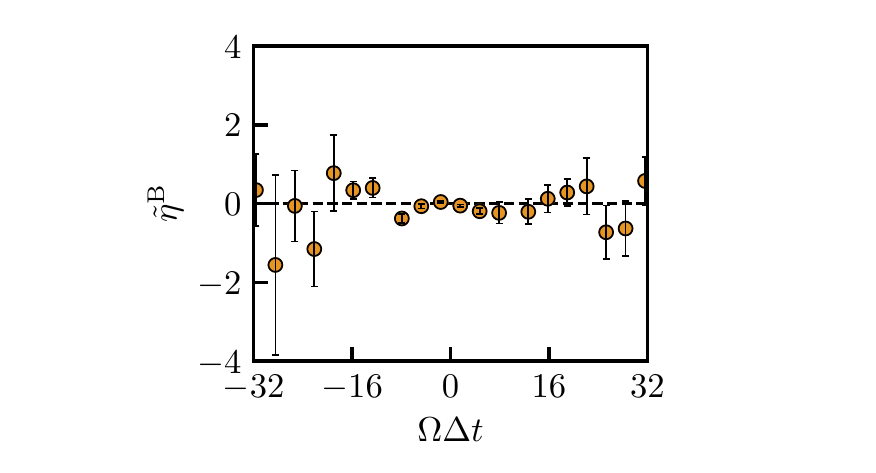}
\caption{\label{fig:rotB-viscosity} \textbf{Rotation-compression viscosity.} We measure $\eta^\text{B}$ by evaluating the linear response of \protect\p\, towards \protect\dtR. However, the statistical uncertainty is too large to draw the conclusion $\eta^\text{A} (\Omega) = \eta^\text{B} (-\Omega)$.}
\end{figure}

\noindent \textbf{Rotation-compression viscosities} \enspace The viscous coefficients $\eta^\text{A}$ and $\eta^\text{B}$ determine the coupling between compression and rotation. In the main text, we have shown that $\eta^\text{A}$ is an odd function of $\Omega$. However, statistical uncertainty in simulations (Fig.~\ref{fig:rotB-viscosity}) precludes the determination of the symmetry of $\eta^\text{B}(\Omega)$. Note that compared to the large steady-state pressure $P/P_0 \sim \mathcal{O}(1)$, where $P_0 = m/d\Delta t^2$, the linear response via $\eta^\text{B}$ is an undetectable correction. This is different from the linear response via $\eta^\text{A}$, which is still measurable in the presence of the small anti-symmetric stress $\tau/P_0 \sim \mathcal{O}(10^{-2})$.

\vspace{10mm}

\begin{figure}[h]
\includegraphics[width=0.52\textwidth]{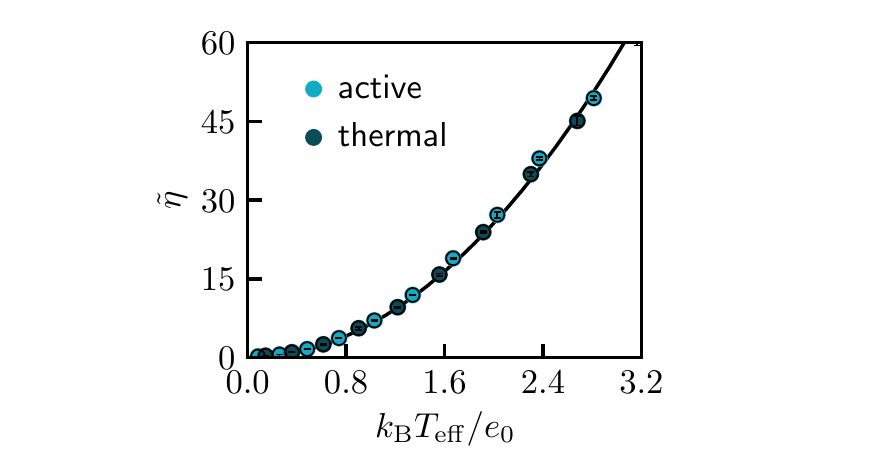}
\caption{\label{fig:shear-viscosity-T} \textbf{Shear viscosity.} Here we compare the temperature dependence of shear viscosity in our chiral active fluid versus its thermal counterpart, where rotational activity is replaced with a thermostat. The shear viscosity $\eta(T_\text{eff})$ obtained from the chiral active fluid and the shear viscosity $\eta(T)$ obtained from the thermal system share the same functional form, $\eta \sim T^2$ (black curve).}
\end{figure}

\noindent \textbf{Shear viscosity $\bm\eta$} \enspace For a thermal fluid, the shear viscosity is a function of temperature $\eta(T)$. Here we investigate whether $T_\text{eff}$ also plays the role of temperature in determine the value of $\eta$ in our system. To do so, we replace rotational activity with a thermostat and create a thermal counterpart of our chiral active fluid. We find that the shear viscosity of this thermal system displays the same temperature dependence as what we measured before (Fig.~\ref{fig:shear-viscosity-T}).

\vspace{10mm}

\begin{figure*}[ht]
\includegraphics[width=0.98\textwidth]{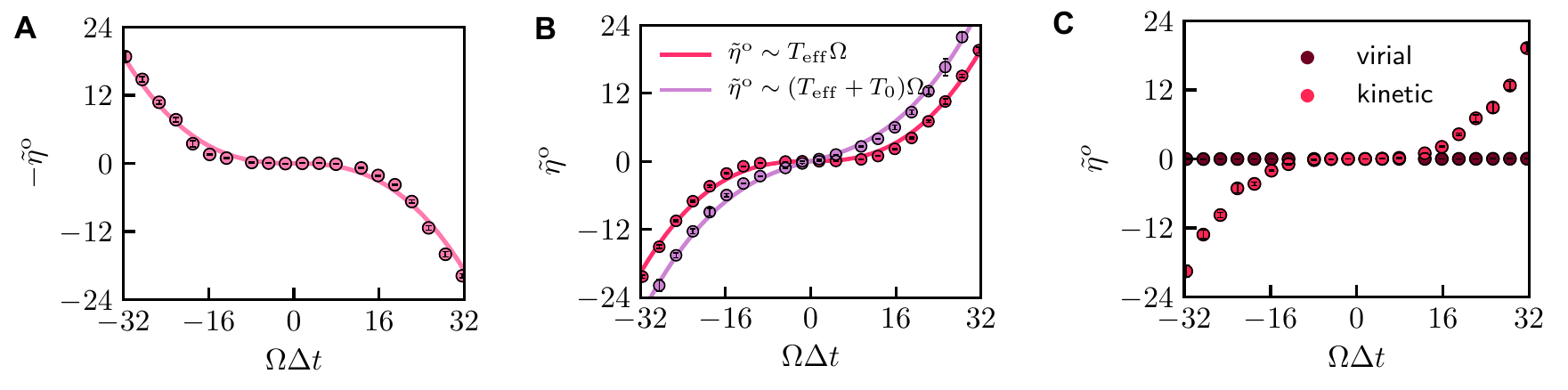} 
\caption{\label{fig:odd-viscosity} \textbf{Odd viscosity.} \textbf{A.} Measurement of odd viscosity $\eta^\text{o}$ under the pure shear \protect\dtSo\, at various spinning speeds $\Omega$. The curve is opposite to $\eta^\text{o}(\Omega)$ shown in Fig.~1B in the main text, confirming the anti-symmetric nature of odd viscosity. \textbf{B.} Temperature dependence of $\eta^\text{o}$. We introduce an intrinsic temperature $T_0$ into the system by adding random force to each particle. The intrinsic temperature changes the odd viscosity to $\eta^\text{o} \sim (T_\text{eff}+ T_0)\Omega$, which confirms the linear dependence of odd viscosity on both temperature and spinning speed. \textbf{C.} Contributions of the kinetic and virial stresses to $\eta^\text{o}$. We evaluate the linear responses in the kinetic component $\bm\sigma^\text{kin} = -\sum_i^N m\V_i\V_i/A$ and the virial component $\bm\sigma^\text{vir} = -\sum_{ij}^{N^2} \F_{ij}\R_{ij}/2A$ of the stress,
and extract their contributions to $\eta^\text{o}$ for a wide range of $\Omega$.}
\end{figure*}

\noindent \textbf{Odd viscosity $\eta^\text{o}$} \enspace According to Eq.~(2) in the main text, odd viscosity governs the interplay between the two pure shears (\protect\dtSo\, and \protect\dtSt). 
We have measured $\eta^\text{o}$ by evaluating the linear response of the fluid towards a simple shear that contains \protect\dtSt\, (Fig.~2B in the main text). 
To confirm the anti-symmetric nature of $\eta^\text{o}$, here we evaluate the linear response towards \protect\dtSo\,  and find that the viscosity does become the opposite to what we measured before (Fig.~\ref{fig:odd-viscosity}A).  
To verify the temperature dependence $\eta^\text{o} \sim T_\text{eff}\Omega$, we increase the system temperature by $T_0$ through a thermostat and find that the resulting odd viscosity then follows $\eta^\text{o}  \sim (T_\text{eff} + T_0)\Omega$  (Fig.~\ref{fig:odd-viscosity}B). 
By decomposing the Irivine--Kirkwood stress into the kinetic and virial parts, we find that $\eta^\text{o}$ is dominantly contributed by particle kinetics (Fig.~\ref{fig:odd-viscosity}C).

\vspace{10mm}

\begin{figure*}[htb]
\includegraphics[width=0.94\textwidth]{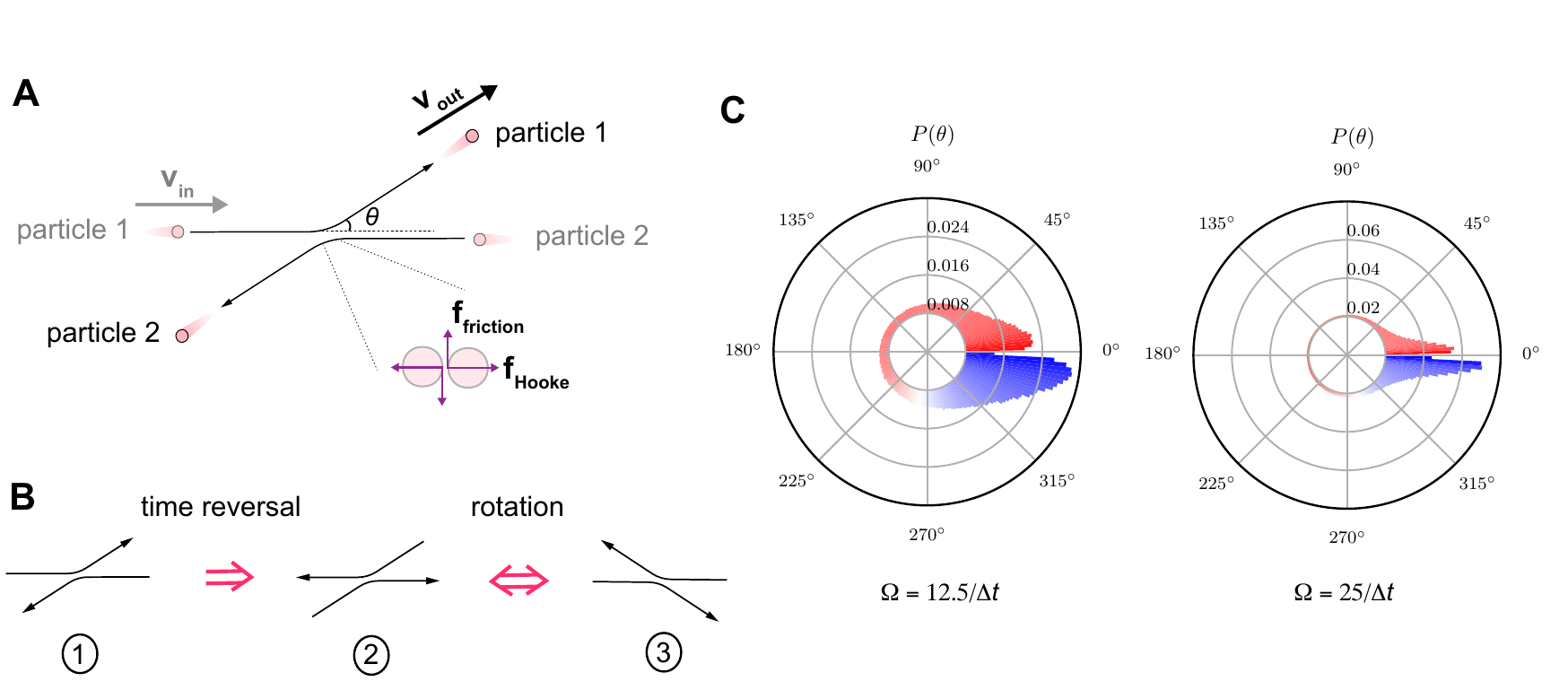}
\caption{\label{fig:odd-viscosity-P} \textbf{Time-reversal symmetry breaking in the interparticle collision.} \textbf{A.} Schematic of interparticle collisions. The surface friction between two spinners generates a transverse interaction, which can cause a preferential bias in the turning angle of either particle (also see Supplementary Mov.~S1). Here we define the turning angle $\theta$ as the angle between the incoming velocity $\V_\text{in}$ and outgoing velocity  $\V_\text{out}$ of particle 1. \textbf{B.} Chirality in the collision. In 2D, a collision with a preferential turning angle $\theta$ is chiral ({\large \textcircled{\small 1}}). Such chirality leads to the breaking of time-reversal symmetry in the collision. One can show this by sequentially applying time reversal ($\V \to -\V$, see {\large \textcircled{\small 2}}) and rotation (allowed by the isotropy of the system). These two operations result in a configuration {\large \textcircled{\small 3}} that differs from the original configuration {\large \textcircled{\small 1}}.
\textbf{C.} Probability distribution of $\theta$. We find that the collisions in our chiral active fluid display a preferential bias in $P(\theta)$. However, such bias becomes weaker at a larger spinning speed $\Omega$. The plots are colorcoded by the impact parameter $b$. Blue means that particle 1 is initially below particle 2, whereas red means particle 1 is initially above particle 2.} 
\end{figure*}

\noindent \textbf{Microscopic origin of odd viscosity $\eta^\text{o}$} \enspace To investigate the microscopic origin of odd viscosity, we perform molecular dynamic simulations of the particle kinetics. We quantify the effects of interparticle collision as the turning angle $\theta$ between the incoming and outgoing velocities of a given particle (Fig.~\ref{fig:odd-viscosity-P}A).
The active part of the interparticle friction 
$\gamma d\Omega \hat{z} \cross \hat{\textbf{r}}_{i j}$ drives the particles in the transverse direction. 
This gives rise to the chirality of the collision, which is characterized by a preferential bias in $\theta$.
As illustrated in Fig.~\ref{fig:odd-viscosity-P}B, such chirality is associated with broken time-reversal symmetry, the key ingredient of odd viscosity~\cite{avron1998odd,banerjee2017odd}.
We find that the collisions in our system are indeed chiral (Fig.~\ref{fig:odd-viscosity-P}C).  
However, the chirality becomes weaker at larger $\Omega$. At first sight, this is contradictory to the observation that $\eta^\text{o}$ monotonically increases with $\Omega$ (see Fig.~\ref{fig:odd-viscosity}B). 

\vspace{10mm}

\begin{figure*}[ht]
\includegraphics[width=0.96\textwidth]{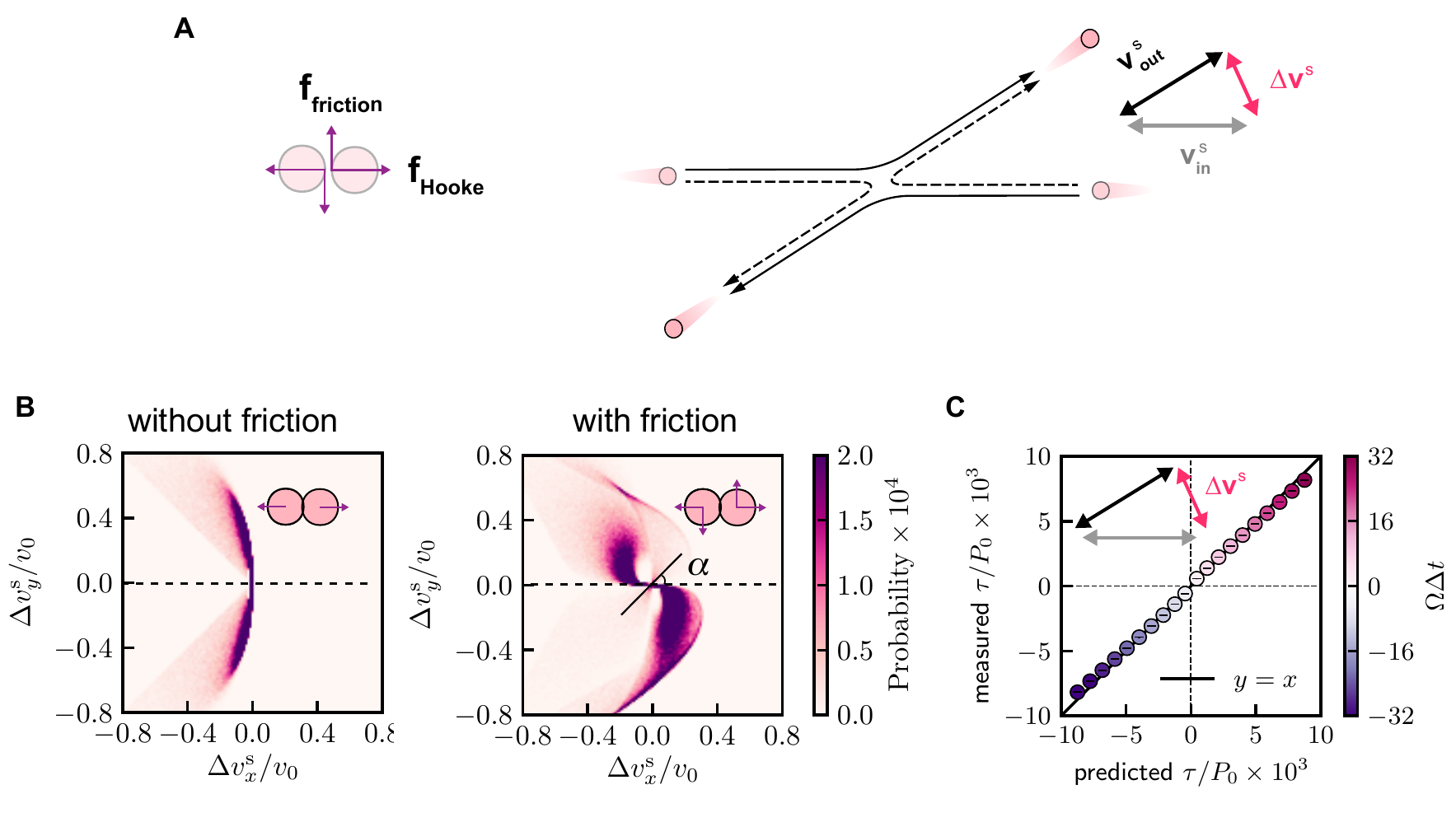} 
\caption{\label{fig:odd-viscosity-dV} \textbf{Microscopic origin of odd viscosity.} \textbf{A.} Schematic of an interparticle collision. Viscosity is the consequence of multiple collisions. 
Since the participating particles are identical, we parameterize the collisions in terms of the 
the symmetrized velocities, as shown above. \textbf{B.} Distribution of the symmetrized velocity change $\bm\Delta \V^\text{s}$. Without friction, $\bm\Delta \V^\text{s}$ is symmetric about the $x$-axis. With friction, $\bm\Delta \V^\text{s}$ displays a chiral distortion with a characteristic twisting angle $\alpha$. \textbf{C.} Linear relation between the viscosity ratio $\eta^\text{o}/\eta$ and the twisting angle $\alpha$.}
\end{figure*}

We notice that the effects of interparticle collision can be also quantified as the resultant velocity change. Since viscosity is a consequence of multiple collisions, it should not depend on particle identity. 
Without distinguishing the particles, for any given initial and final state of a collision, there are two possible pathways (Fig.~\ref{fig:odd-viscosity-dV}A). 
To account for this, we consider the symmetrized velocity change 
\begin{equation}
   \bm\Delta \V^\text{s} =
    \begin{cases}
      \V_\text{out} - \V_\text{in} & \shortspace \V_\text{in}\cdot\V_\text{out} \geq 0 \\[6pt]
      -\V_\text{out} - \V_\text{in} & \shortspace \V_\text{in}\cdot\V_\text{out} < 0  \\
    \end{cases}.       
\end{equation}
See the illustration of $\bm\Delta \V^\text{s}$ in Fig.~\ref{fig:odd-viscosity-dV}A.
We find that due to the interparticle friction between the spinners, the distribution of $\bm\Delta \V^\text{s}$ becomes chiral.
We characterize the chirality of $P(\bm\Delta \V^\text{s})$ by means of a twisting angle $\alpha$.
Similar to the aforementioned turning angle $\theta$, $\alpha$ also suggests the existence of odd viscosity. 
More precisely, $\alpha$ encodes the competition between
the onset of the transverse motion and the reduction of the longitudinal motion due to the collision. 
The former is driven by the chiral active interaction $\gamma d\Omega \hat{z} \cross \hat{\textbf{r}}_{i j}$ and leads to odd viscosity, whereas the latter is driven by the remaining passive interactions and gives rise to shear viscosity.
Remarkably, the viscosity ratio $\eta^\text{o}/\eta$ is indeed linear with $\alpha$, making $\alpha$ a reliable predictor for odd viscosity.

\onecolumngrid
\clearpage
\newpage

\twocolumngrid

\section{Derivation of the Kubo relation}\label{sec:Kubo}

Here we provide a first-principle derivation of the Green--Kubo relation using the Mori--Zwanzig formalism~\cite{Nakajima1958,Zwanzig1960,Mori1965,Zwanzig2001}, a systematic coarse-graining procedure to study the dynamics of a many-body system. We show that the equilibrium-like Green--Kubo relation
\begin{equation}
	\eta_{\alpha \beta} = \frac{A}{\kT^*} \, \int_0^{\infty} \left< \sigma_{\alpha}(t) \sigma_{\beta}(0) \right> dt,
\end{equation}
holds near the steady-state of an isotropic active fluid, as long as the steady state is stable and displays fast-decaying velocity--velocity corrections $|\left<\V(\R) \cdot \V(0)\right>_0 | \leq \mathcal{O}(r^{-D})$, where $D$ is the dimension of the system. 
Our analysis focuses on systems with pairwise interactions that are arbitrary functions of the relative coordinate and are at most linear in the relative particle velocity. 


In the derivation, we choose the momentum current densities $\Jk$ of the fluid as the slow variables, to characterize momentum transfer at the macroscopic level. By constructing a projection operator using $\Jk$, we decompose the generalized forces that drive the dynamics of the entire system into the components parallel to $\Jk$ and the random forces orthogonal to $\Jk$. Using the Mori--Zwanzig formalism, we show that the slow dynamics of $\Jk$ displays a linear response with response functions determined by the time-correlations of the random forces. For a thermal system with conservative interactions, such fluctuation--dissipation relation leads to the standard Green--Kubo relation. Here we extend the Mori--Zwanzig formalism for nonequilibrium systems involving active and dissipative interactions. By carefully evaluating the generalized forces, we prove the equilibrium-like Green--Kubo relation. 



\vspace{10mm}

\noindent\textbf{Momentum transfer} \enspace In conventional hydrodynamic theory, the Navier--Stokes equations describe the momentum transfer in a fluid.
In the same spirit, we study the evolution of momentum flux in our system. 
We use $V$ to denote the volume of the system, $N$ for the number of particles, and $m$ for their mass. The $i$th particle has position ${\bf r}_i (t)$ and velocity $\V_i(t)$.
We derive our theory in the reciprocal space by investigating the wavevector-dependent momentum current density:
\begin{equation}
   \J_{\K} (t) \triangleq \frac{1}{V} \sum_{i}^{N} m\V_i(t) \e^{-\ii \KoR_i(t)}.
\end{equation}
Taking the time derivative on the both sides, we can find an equation for the evolution of $\Jk (t)$:
\begin{equation}\label{eq:dt-J}
    \begin{split}
        \dot{\J}_{\K} (t) 
        &= \frac{1}{V} \sum_{i}^{N} \left[-\ii \K \cdot m\V_i(t)\V_i(t) + m\dot{\V}_i(t)\right] \e^{-\ii \KoR_i(t)} \\
        &= \frac{1}{V} \sum_{i}^{N} \left[-\ii \K \cdot m\V_i(t)\V_i(t) + \F_i(t)\right] \e^{-\ii \KoR_i(t)},
    \end{split}
\end{equation}
where $\F_i(t)$ is the total force on particle $i$, and we have used the fact that $m\dot{\V}_i(t)  = \F_i$. Here we employ the convention of using a double vector $\textbf{X}\textbf{Y}$ to represent a matrix with elements $X_a Y_b$.
In the following steps~[Eq.~(\ref{eq:fsum}-\ref{wavevector_dependent_stress})], we assume that the net force is the sum of reciprocal two body interactions $\F_i = \sum_{j \neq i}^{N-1} \F_{ij}$.
Here we assume that the interaction is reciprocal, but later on we will also derive the case of wet active systems that involve non-reciprocal hydrodynamic interactions.
For now, given $\F_{ij} = -\F_{ji}$,  the second term in Eq.~(\ref{eq:dt-J}) can be written as:
\begin{align} \label{eq:fsum}
\begin{split}
    \sum_i^{N} \F_i \e^{-\ii \KoR_i} 
    &= \sum_{i}^{N} \left[\sum_{j \neq i}^{N-1} \F_{ij} \right] \e^{-\ii \KoR_i} \\
    &= \frac{1}{2} \sum_{ij, \; i \neq j}^{N(N-1)} \left[\F_{ij} \e^{-\ii \KoR_i} + \F_{ji} \e^{-\ii \KoR_j}\right] \\
    &= \frac{1}{2} \sum_{ij, \; i \neq j}^{N(N-1)} \F_{ij} \left[1- \e^{\ii \KoR_{ij}}\right] \e^{-\ii \KoR_i} \\
    &= -\ii \K \cdot \frac{1}{2}\sum_{ij, \; i \neq j}^{N(N-1)} \F_{ij}\R_{ij} O_{\K, ij}\e^{-\ii \KoR_i},
\end{split}
\end{align}
with
\begin{align} 
\enspace O_{\K, ij} &\triangleq \frac{1-\e^{\ii \KoR_{ij}}}{-\ii \KoR_{ij}} = 1 + \frac{\ii \KoR_{ij}}{2} + \mathcal{O}(k^2),
\end{align}
where $\R_{ij} = \R_i - \R_j$ denotes the interparticle vector.
Given the form of Eq.~(\ref{eq:fsum}),  Eq.~(\ref{eq:dt-J}) can be summarized as
\begin{equation}\label{eq:dt-J2}
    \dot{\J}_{\K}(t) = \ii \K \cdot \s_{\K} (t),
\end{equation}
where $\s_\K (t)$ is the wavevector-dependent stress,
\begin{equation}\label{wavevector_dependent_stress}
    \\[-6pt]
    \s_{\K} \triangleq -\frac{1}{V}\sum_{i}^{N} \left[ m\V_i\V_i + \frac{1}{2}\sum_{j \neq i}^{N-1} \F_{ij}\R_{ij} O_{\K, ij} \right]\e^{-\ii \KoR_i}. \\[8pt]
\end{equation}
Eq.~(\ref{wavevector_dependent_stress}) reduces to the Irvine--Kirkwood formula Eq.~(\ref{eq:irvstress}) in the hydrodynamic limit, i.e. $\lim_{\textbf{k} \to \textbf{0}}O_{ij}(\textbf{k}) = 1$.

\vspace{10mm}

\noindent\textbf{Projection operator} \enspace 
Any instantaneous state of a dynamical system can be represented as a single point in its phase space. 
For a classical particle system like ours, the conventional phase space with coordinates $\G = (\textbf{p}^{N}, \textbf{q}^{N})$ composed of particle momentum $\textbf{p}_i = m\V_i$ and position $\textbf{q} = \R_i$ is typically used. 
Any observable of the system, for instance the aforementioned momentum current density $\J_{\K}$, is a function defined on the phase space. 
These phase-space functions form a Hilbert space, which we denote as $\mathscr{H}(\G)$. 

The steady state of a system corresponds to a stationary distribution $f_0(\G)$ of points in the phase space. Using this distribution, we define the following inner product on the space $\mathscr{H}(\G)$:
\begin{equation}
\begin{split}
    \longspace\left(A, B\right) &=  \left<A(\G)B^*(\G)\right>_0 \\
    &\triangleq \int d\G A(\G)B^*(\G)f_0(\G),
\end{split}
\end{equation}
where $A(\bm{\G})$ and $B(\bm{\G})$ are two arbitrary phase-space functions, $^*$ denotes complex conjugate, and $\left<\enspace\right>_0$ denotes the ensemble average over $f_0(\bm{\Gamma})$. With this inner product, we can perform projections among the phase-space functions near the steady state. In particular, we are interested in the projection towards $\J_{\K}$, which will be used to split slow hydrodynamics from fast fluctuations.

The operator $\J_\K (\G) = (\hat{J}_{\textbf{k}, 1}(\G), \dots, \hat{J}_{\textbf{k}, D}(\G))^\text{T}$ is in fact a vector-valued function of dimension $D$. Its different components generate a subspace $\mathscr{S}_{\Jk}(\G)$, for which we can define a projection operator:
\begin{equation}
\begin{split}
    \\[-8pt]
    \Pk \textbf{X} (\bm{\Gamma}) \triangleq (\textbf{X} \otimes \Jk) \cdot (\Jk \otimes \Jk)^{-1} \cdot \Jk, \\[6pt]
\end{split}
\end{equation}
where the outer product is given by
\begin{align}
    (\textbf{A} \otimes \textbf{B})_{pq} = (A_p, B_q).
\end{align}
For an arbitrary vector function $\textbf{X}(\bm{\Gamma})$ of dimension $Q$, $\Pk$ projects each of its components $X_p(\bm{\Gamma})$ into the subspace $\mathscr{S}_{\Jk}(\G)$ and represents the result as a linear combination of $\hat{J}_{\K, q} (\G)$. Note that $(\textbf{X} \otimes \Jk)$ and $(\Jk \otimes \Jk)$ are $Q\times D$ and $D\times D$ matrices, respectively. 

In addition to $\Pk$, we also define the projection operator to the orthogonal subspace:
\begin{equation}
    \Qk = \mathds{1} - \Pk.
\end{equation}
The operators $\Pk$ and $\Qk$ satisfies the following relations:
\begin{equation}
\begin{split}
    \Pk\Pk = \Pk&, \enspace\enspace \Qk\Qk = \Qk, \\
    \Pk\Qk &= \Qk\Pk = 0.
\end{split}
\end{equation}

\vspace{10mm}

\noindent\textbf{Mori-Zwanzig formalism} \enspace In addition to Eq.~(\ref{eq:dt-J}), the evolution of $\dot{\J}_\K(t)$ may be expressed in terms of the Liouvillian equation:
\begin{equation}\label{eq:Liou}
\begin{split}
    \dot{\J}_\K(t) = \iL \Jk (t)
\end{split}
\end{equation}
where 
\begin{equation}
    \ii \mathcal{L} \triangleq \dot{\G} \cdot \frac{\partial}{\partial \G}
\end{equation}
denotes the Liouville operator. We will apply the well-known Mori--Zwanzig formalism to derive the Green--Kubo relation presented in the main text. We proceed by decomposing $\J_\K$ in the following manner:
\begin{equation}
\begin{split} \label{eq:Mori}
    \dot{\J}_\K(t)  = \Fp(t) &\;+\;  \Fo(t)\\
    &\;- \int_{0}^{t}\textbf{K}(\tau) \cdot \Jk (t-\tau)\;d\tau,
\end{split}
\end{equation}
where
\begin{align}
   \Fp(t) &\triangleq \e^{\iLt} \Pk \iL \Jk (\G), \\[4pt]
   \Fo(t) &\triangleq \e^{\Qk \iLt} \Qk \iL \Jk (\G),\\
   \textbf{K}(\tau) &\triangleq (\Fo(\tau) \otimes \Fo(0))\cdot(\Jk \otimes \Jk)^{-1}.
\end{align}
This decomposition splits the generalized force $\iL \Jk (t)$ into contributions parallel 
$\Fp(\G, t)$ and orthogonal $\Fo(\G, t)$ to the subspace $\mathscr{S}_{\Jk}(\G)$. The former drives the systems inside the subspace, whereas the latter acts as a random noise occasionally kicking the system out of the subspace. As a consequence, the system can sustain the nonequilibrium steady state by gently fluctuating around it. The kernel $\textbf{K}(\tau)$ characterizes the linear response of the fluid towards external disturbances on the momentum current density $\Jk$. This response coefficient is also associated with the time correlation of the fluctuating random force $\Fo(\G, t)$.

While the decomposition in Eq.~(\ref{eq:Mori}) is appealing to study the Green--Kubo relation, care must be taken. 
The decomposition often requires that the system dynamics be time reversible, which corresponds to a Hermitian Liouville's operator $\mathcal{L}$~\cite{Zwanzig2001}. This assumption of Hermiticity ensures a crucial step in the derivation of Eq.~(\ref{eq:Mori}):
\begin{equation} \label{eq:LJ-JL}
    (\iL \Fo(\tau) \otimes \Jk) = -(\Fo(\tau) \otimes \iL\Jk).
\end{equation}
However, Eq.~(\ref{eq:LJ-JL}) does not generally hold for a non-Hermitian $\mathcal{L}$, which arises from effects such as activity or interparticle friction. 
However, a key insight is that the relation Eq.~(\ref{eq:LJ-JL}) still holds near the nonequilibrium steady state, even for a non-Hermitian $\mathcal{L}$. At the steady state, the probability distribution does not change over time, thus
\begin{equation}
    \frac{d}{dt} f_0(\G) = \frac{\partial}{\partial \G} (\dot{\G} f_0(\G)) = 0.
\end{equation} 
Given the assumption of steady state, one can prove Eq.~(\ref{eq:LJ-JL}) elementwise using integration by parts:
\begin{equation} \label{eq:iter-part}
\begin{split}
    &(\iL \Fo(\tau) \otimes \Jk)_{pq} + (\Fo(\tau) \otimes \iL\Jk)_{pq} \\
    &= \int d\G \fo_{\K, p}(\G, \tau) \hat{J}_{\K, q}^{*}(\G) \cdot \frac{\partial}{\partial \G} (\dot{\G} f_0(\G)) = 0.
\end{split}
\end{equation}

Hence, under the assumption of steady state, the decomposition in Eq.~(\ref{eq:Mori}) holds even for non-Hermitian Liouville operators. To make use of Eq.~(\ref{eq:Mori}) for analyzing the Green Kubo relations, we must derive the explicit form of the generalized forces and the kernel $\textbf{K}$ in terms of velocity correlation functions. To do so, we will first evaluate the projection
\begin{align}\label{eq:Pk-iLJ}
    \Pk \iL \Jk = (\iL \Jk \otimes \Jk)\cdot(\Jk \otimes \Jk)^{-1}\cdot \Jk.
\end{align}
In the following two sections, we will analyze the two outer products $(\Jk \otimes \Jk)$ and $(\iL \Jk \otimes \Jk)$ involved in Eq.~\eqref{eq:Pk-iLJ}.

\vspace{10mm}

\noindent\textbf{Evaluation of $(\Jk \otimes \Jk)$} \enspace 
We will assume for simplicity an isotropic and homogeneous steady state, i.e., one in which the distribution of particle positions and velocities are independent and have no preferred direction. For such a system, we can always choose a reference frame in which no background flow exists. (This assumption is manifestly violated close to boundaries where spontaneous active flow can arise). Therefore, in absence of external perturbations, the momentum current $\Jk$ arises purely from the fluctuations of particle velocity. We have
\begin{equation}
     \left<\V\right>_0 = \textbf{0} \midspace \text{thus} \midspace \big<\Jk\big>_0 = \textbf{0},
\end{equation}
where we use the fact that velocity of a particle does not couple with its exact position.
The magnitude of the fluctuating $\Jk$ is captured by 
\begin{align}\label{eq:JJ}
    (\Jk \otimes \Jk) &= \frac{m^2}{V^2} \sum_{ij}^{N^2} \left<\V_i \V_j \e^{-\ii \KoR_{ij}}\right>_0 \\
    &= \frac{m^2}{V^2} \left[\sum_{i}^{N} \left<\V_i \V_i\right>_0 + \sum_{i}^{N}\sum_{j\neq i}^{N-1} \left<\V_i \V_j \e^{-\ii \KoR_{ij}}\right>_0 \notag\right], \notag
\end{align}
which also serves as the normalization matrix in the projection operator $\Pk$. This quantity arises from the velocity--velocity correlations.
In particular, the first term in Eq.~(\ref{eq:JJ}) captures the correlations among different velocity components of the same particle. The isotropy of the system implies:
\begin{equation}\label{eq:vvcor}
   \sum_{i}^{N} \left<\V_i\V_i\right>_0 = \frac{N\kT}{m} \mathcal{I},
\end{equation}
where $\mathcal{I}$ denotes a $D \times D$ identity matrix. We may take Eq.~(\ref{eq:vvcor}) as definition of the effective temperature. 

The second term in Eq.~(\ref{eq:JJ}) is associated with the spatial correlations among the velocities of different particles:
\begin{align}
    \sum_{i}^{N}\sum_{j\neq i}^{N-1}\left<\V_i \V_j \e^{-\ii \KoR_{ij}}\right>_0 &= \frac{N^2}{V} \int_V  \mathcal{C}_{\V\V}(\R) \e^{-\ii \KoR} d\R, 
\end{align}
where $\mathcal{C}_{\V\V}(\R)$ is the spatial velocity--velocity correlation function:
\begin{align}
    \mathcal{C}_{\V\V}(\R) &= \left<\V(0)\V({\R})\right>_0 \notag \\ &\triangleq \frac{V}{N^2}\sum_i^N \sum_{j \neq i}^{N-1} \left<\V_i\V_j \delta(\R - \R_{ij})\right>_0,
\end{align}
which is a $D \times D$ matrix.
Eq.~(\ref{eq:JJ}) thus can be written as
\begin{align}\label{eq:JJ2}
    (\Jk \otimes \Jk) &= \frac{m^2N}{V^2}\left[\frac{\kT}{m}\mathcal{I} + n\,\hat{\mathcal{C}}_{\V\V}(\K)\right],
\end{align}
where $\hat{\mathcal{C}}_{\V\V}(\K) = \int_V  \mathcal{C}_{\V\V}(\R) \e^{-\ii \KoR} d\R$ denotes the Fourier transform of $\mathcal{C}_{\V\V}(\R)$. 
\newline

In standard fluids, the velocity--velocity correlation function $\mathcal{C}_{\V\V}(\R)$ vanishes at finite $\R$, because positions and velocities are uncorrelated.
However, this is not generally the case for non-equilibrium fluid, including the chiral active fluid presented in the main text. 
In the following, we will show that when the $\mathcal{C}_{\V\V}(\R)$ decays fast enough (at least as a power-law $r^{-D}$, where $D$ is the dimension of the system), then the only effect of nonzero velocity--velocity correlations is to renormalize the value of the effective temperature in the Green--Kubo relation. This effect mainly hinges upon the existence of isotropic correlations at small distances, which we find are very small in the system analyzed in the main text. 
\newline

\begin{figure*}[ht]
\includegraphics[width=0.96\textwidth]{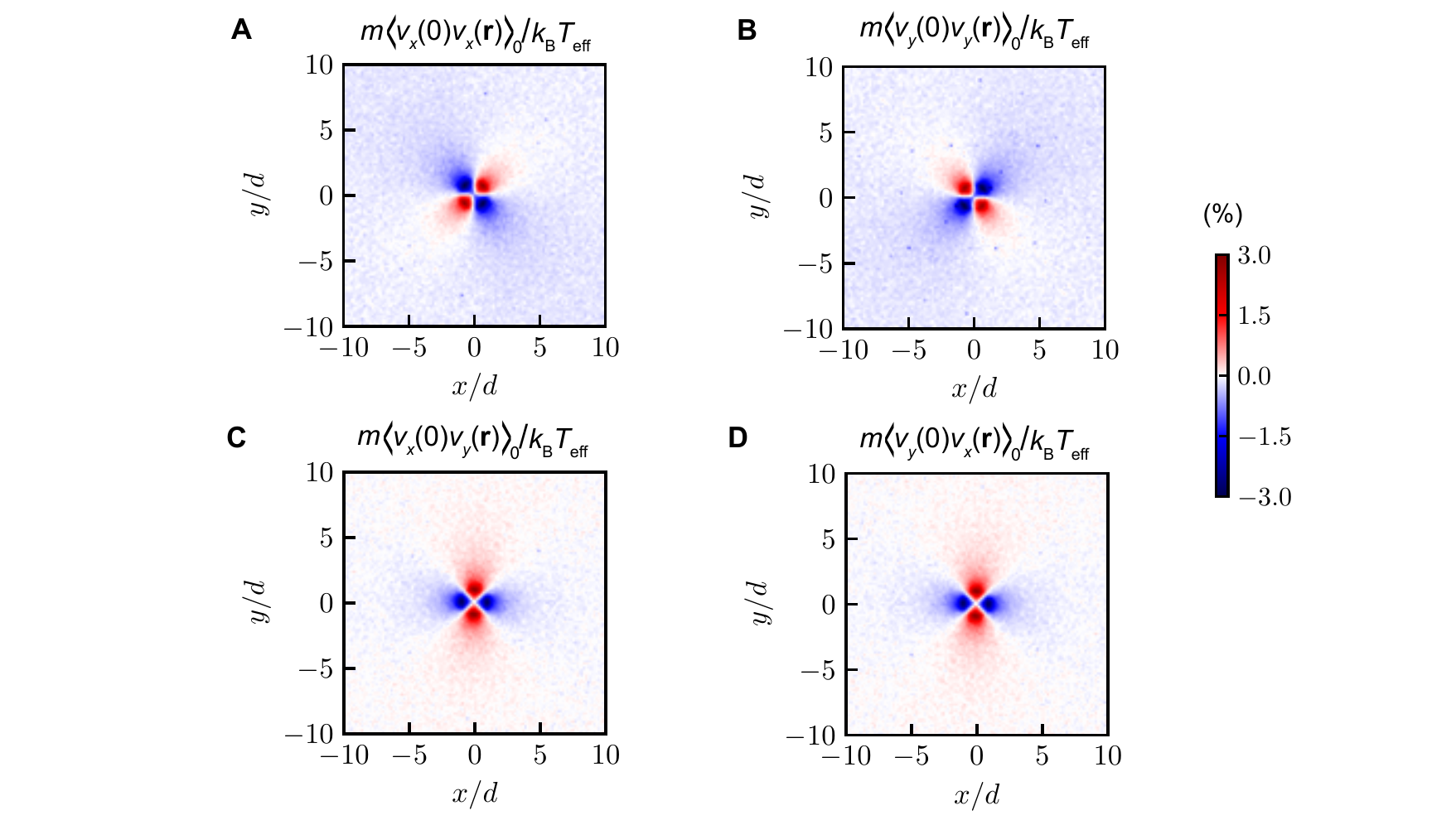}
\caption{\label{fig:rho_vv} \textbf{Velocity--velocity spatial correlations.} The correlation functions are measured at the steady state at $\Omega = 26.7/\Delta t$. Since only a weak, contact interparticle friction is employed in our simulations, the resultant velocity--velocity correlations $\mathcal{C}_{\V\V}(\R) = \big<\V(0)\V(\R)\big>_0$ are not only small ($< 5\%$ of $\kT/m$) but also short-ranged. 
Due to the conservation of momentum for the entire system, the diagonal terms  $\big<v_x(0)v_x(\R)\big>_0$ and $\big<v_y(0)v_y(\R)\big>_0$ are negative in the far field, see the blue background in 
(\textbf{A}) and (\textbf{B}). The negative correlation gives rise to a non-vanishing diagonal $\mathcal{C}_{\V\V, 0}$. The off-diagonal terms are symmetric, $\big<v_x(0)v_y(\R)\big>_0 = \big<v_y(0)v_x(\R)\big>_0$, see the identical patterns shown by (\textbf{C}) and (\textbf{D}).}
\end{figure*}

The velocity--velocity correlation $\mathcal{C}_{\V\V}(\R)$ is typically smooth at long interparticle distance $r \to \infty$.
We assume it can be decomposed as
\begin{align}\label{eq:assume}
    \mathcal{C}_{\V\V}(\R) &= \mathcal{C}_{\V\V, 0} + \mathcal{C}_{\V\V}^\text{near}(\R) + \mathcal{C}_{\V\V}^\text{far}(\R) \shortspace
\end{align}
with
\begin{align*}
    \mathcal{C}_{\V\V}^\text{near}(\R) &= \mathcal{C}_{\V\V}(\R)-\mathcal{C}_{\V\V, 0}, &  \mathcal{C}_{\V\V}^\text{far}(\R) &= 0, &  r \leq r_\text{c}, \\[6pt]
    \mathcal{C}_{\V\V}^\text{near}(\R) &\to 0, &   \mathcal{C}_{\V\V}^\text{far}(\R) &\sim \mathcal{A}r^{-D-\alpha} , &  r > r_\text{c.}\\[-6pt]
\end{align*}
$\mathcal{C}_{\V\V, 0}$ is a constant matrix independent of $\R$, capturing a background velocity--velocity coupling required by the conservation of momentum of the entire system (Fig.~\ref{fig:rho_vv}A-B). 
$\mathcal{C}_{\V\V}^\text{near}(\R)$ denotes a near-field core that is bounded near the origin with a finite boundary  $r_\text{c}$, beyond which $\mathcal{C}_{\V\V}^\text{near}(\R)$ decays faster than power law. $\mathcal{C}_{\V\V}^\text{far}(\R)$ denotes a far-field tail. We allow the matrix prefactor $\mathcal{A}$ to be any matrix. For instance, $\mathcal{A} = 0$ when there is no long-range correlation. We multiply the matrix prefactor with a generic power-law decay $r^{-D-\alpha}$. \newline

At any nonvanishing $k = |\K| > 0$, the constant matrix $\mathcal{C}_{\V\V, 0}$ does not contribute to the Fourier transform of $\mathcal{C}_{\V\V}(\R)$. Now let us evaluate the Fourier transforms of  $\mathcal{C}_{\V\V}^\text{near}(\R)$ and $\mathcal{C}_{\V\V}^\text{far}(\R)$ in the hydrodynamic limit $k \to 0$. Since $\mathcal{C}_{\V\V}^\text{near}(\R)$ is bounded within $r_\text{c}$, given that $kr_\text{c} \to 0$, 
\begin{align}
    \hat{\mathcal{C}}^\text{near}_{\V\V}(\K) &= \int_{r \leq r_\text{c}} \mathcal{C}_{\V\V}^\text{near}(\R) \e^{-\ii \KoR} d\R + \int_{r > r_\text{c}} \mathcal{C}_{\V\V}^\text{near}(\R) \e^{-\ii \KoR} d\R \notag \\
    &\approx\int_{r \leq r_\text{c}} (\mathcal{C}_{\V\V}(\R) - \mathcal{C}_{\V\V, 0}) d\R + 0 \notag\\[6pt]
    &= V_\text{c} \, \bar{\mathcal{C}}_{\V\V}^\text{near} .
\end{align}
where $V_\text{c}$ denotes the volume of the near-field region $r \leq r_\text{c}$ and $\bar{\mathcal{C}}_{\V\V}^\text{near}$ denotes the average of $\mathcal{C}_{\V\V}(\R) - \mathcal{C}_{\V\V, 0}$ in that region. We want to point out that $\bar{\mathcal{C}}_{\V\V}^\text{near}$ is a matrix independent of both position $\R$ and wavevector $\K$. The isotropy of the system requires $\bar{\mathcal{C}}_{\V\V}^\text{near} = a\delta_{pq} + b\epsilon_{pq}$, where $\delta_{pq}$ and $\epsilon_{pq}$ are the Kronecker delta and Levi-Civita tensors, respectively. Let us take a closer look at $\mathcal{C}_{\V\V}(\R) = \left<\V(0)\V(\R)\right>_0$. The isotropy of the system allows us to rotate the coordinate system by $180^\circ$ but still observe the same physics. Therefore, 
\begin{align}
    \left<v_p(0)v_q(\R)\right>_0 = \left<v_p(0)v_q(-\R)\right>_0.
\end{align}
Note that here the rotation gives rise to substitutions $v_p \to -v_p$, $v_q \to -v_q$, and $\R \to -\R$. The translational invariance then gives  
\begin{align}
    \left<v_p(0)v_q(-\R)\right>_0 = \left<v_p(\R)v_q(0)\right>_0.
\end{align}
Since here $v_p$ and $v_q$ are scalar components of $\V$, they commute. Thus we have
\begin{align}
    \left<v_p(0)v_q(\R)\right>_0 = \left<v_q(0)v_p(\R)\right>_0. \;
\end{align}
As an illustration, we confirm $\left<v_x(0)v_y(\R)\right>_0 = \left<v_y(0)v_x(\R)\right>_0$ for our system in Fig.~\ref{fig:rho_vv}C-D.
Since $\left<v_x(0)v_y(\R)\right>_0 = \left<v_y(0)v_x(\R)\right>_0$, the matrix $\mathcal{C}_{\V\V}(\R)$ is symmetric, so is  $\mathcal{C}_{\V\V_,0} = \mathcal{C}_{\V\V}(\R \to \infty)$.
As a consequence, $\bar{\mathcal{C}}_{\V\V}^\text{near}$ is proportional to the identity matrix: 
\begin{align}
    \bar{\mathcal{C}}_{\V\V}^\text{near} =  \bar{c}_{\V\V}^\text{near} \mathcal{I},
\end{align}
where $\bar{c}_{\V\V}^\text{near}$ is a scalar constant that can be extracted from the near-field correlations of particle velocity. 

Regarding $\mathcal{C}_{\V\V}^\text{far}(\R)$, we have 
\begin{align}
    \hat{\mathcal{C}}^\text{far}_{\V\V}(\K) \sim \int_V \mathcal{A} r^{-D-\alpha} \e^{-\ii \KoR} d\R = \mathcal{O}(k^{\alpha}).
\end{align}
If $\alpha > 0$, this term would always vanish in the hydrodynamic limit. Under this assumption, 
\begin{align}
    \hat{\mathcal{C}}_{\V\V}(\K) &= \hat{\mathcal{C}}^\text{near}_{\V\V}(\K) + \hat{\mathcal{C}}^\text{far}_{\V\V}(\K) \notag\\[6pt]
    &= \bar{c}_{\V\V}^\text{near}V_\text{c} \mathcal{I} + \mathcal{O}(k^{\alpha})
\end{align}
when $\K \to 0$.

To summarize, the assumption that $\mathcal{C}_{\V\V}(\R)$ decays faster than $r^{-D}$ in the far-field limit implies that 
\begin{align}\label{eq:JJ3}
    (\Jk \otimes \Jk) \approx n m c_\textbf{JJ}\mathcal{I},
\end{align}
where
\begin{align}\label{eq:cJJ}
     c_\textbf{JJ} = \frac{\kT + n m \bar{c}_{\V\V}^\text{near} V_\text{c}}{V}.
\end{align}

\vspace{20mm}

\noindent\textbf{Evaluation of $(\iL \Jk \otimes \Jk)$} \enspace  To determine this outer product, we first revisit the generalized force $\iL \Jk$. The master equation Eqs.~(\ref{eq:dt-J}) can be rewritten as
\begin{align}\label{eq:iLJ-decompose}
     \iL \Jk &= \ii \K \cdot \Sk = \ii \K \cdot  \left(\Skin + \Spos \right) + \Fd, \;\;
\end{align}
where we have decomposed the stress into three parts:
\begin{align}
    \Skin &\triangleq -\frac{1}{V}\sum_{i}^{N} m\V_i\V_i \, \e^{-\ii \KoR_i} \\
     \Spos &\triangleq-\frac{1}{2V}\sum_{ij}^{N^2}\F_{ij}^{\,\text{pos}}\, \R_{ij}O_{\K, ij}\e^{-\ii \KoR_i},\\
     \Fd &\triangleq -\frac{1}{V}\sum_{ij}^{N^2}\Upsilon_{ij}\V_{ij} \, \e^{-\ii \KoR_i}, \label{eq:dis0}
\end{align}
where $\Skin$ denotes the kinetic stress, $\Spos$ denotes the virial stress only involving the position-dependent interactions $\F_{ij}^{\,\text{pos}}$,  and $\Fd$ captures the velocity dependent forces. We assume this velocity dependent term to be of the form $\F_{ij}^{\,\text{vel}} = -\Upsilon_{ij}\V_{ij}$. Regarding the coefficient matrix, we assume the general form $\Upsilon_{ij} = \gamma(r_{ij})\mathcal{I}$, which ensures the energy transfer rate $\varepsilon = \V_{ij}^\text{T}\Upsilon_{ij}\V_{ij}$ to be both rotation- and translation-invariant, compatible with the isotropy of the system. In our case, $\Upsilon_{ij} = \gamma H(d - r_{ij}) \mathcal{I}$, where $H(x)$ is the Heaviside step function. The general form $\Upsilon_{ij} = \gamma(r_{ij})\mathcal{I}$ could also apply to long-range dissipative interactions even with a power-law behavior.  

Using the above decomposition of $\iL \Jk$, we can calculate the product $(\iL \Jk \otimes \Jk)$ term by term. Given that $\left<\V\right> = 0$, the terms involving odd power of $\V$ vanish: 
\begin{align}
     &\big(\ii \K \cdot \Skin \otimes \Jk\big) =\ii \K \cdot \sum_{ij}^{N^2}{A}_{\K, ij}\left<\V_i\V_i\V_j\right>_0 =0, \label{eq:p1} \\
     &\big(\ii \K \cdot \Spos \otimes \Jk\big) =\ii \K \cdot \sum_{i}^{N}{B}_{\K, i}\left<\V_i\right>_0 =0, \label{eq:p2}
\end{align}
where ${A}_{\K, ij}$ and ${B}_{\K, i}$ are two coefficients independent of particle velocity. This steady-state property is consistent with the thermal equilibrium of a conservative system where the generalized force $\iL \Jk = \ii \K \cdot (\Skin + \Spos)$ is always orthogonal to $\Jk$~\cite{evans2008statistical}. However, 
the product 
\begin{align}\label{eq:dis}
     \big(\Fd \otimes \Jk\big) &= -\frac{m}{V^2} \sum_{ijl}^{N^3}\big<\gamma(r_{ij})\V_{ij} \V_l \, \e^{-\ii \KoR_{il}}\big>_0 \notag\\
     & = -\frac{m}{V^2} \sum_{ijl}^{N^3}\big<\gamma(r_{ij})\V_{i}\V_l \, \e^{-\ii \KoR_{il}}\big>_0 \notag \\
     &\enspace\enspace + \frac{m}{V^2}\sum_{ijl}^{N^3}\big<\gamma(r_{ij})\V_{j}\V_l \, \e^{-\ii \KoR_{il}}\big>_0
\end{align}
is quadratic in $\V$ and thus cannot be ignored.

To evaluate the final term in Eq.~(\ref{eq:dis}), we have to construct the three particle probability, denoted as $p(\G_i, \G_j, \G_l)$. Since we assume a homogeneous steady state, the particles are uniformly distributed and their velocities and positions are independent variables. Thus for any given particle $i$, the single-body probability distribution $p(\G_i)$ reads
\begin{align}
    p(\G_i) = p(\R_i)\, p(\V_i)= \frac{1}{V} p(\V_i),
\end{align}
which satisfies
\begin{align*}
    \int p(\G_i)d\G_i &= 1, \\
    \int \V_i\, p(\G_i)d\G_i &= \left<\V_i\right>_0 = 0.
\end{align*}
For any given particle pair $ij$, the two-body probability distribution can be decomposed into
\begin{align}
    p(\G_i, \G_j) = p(\G_i)\, p(\G_j) + g(\G_i, \G_j),
\end{align}
where $g(\G_i, \G_j)$ encodes all the pairwise correlations. Given particle indexing should not affect the physics, $p(\G_i, \G_j)$ has to be invariant under $i \leftrightarrow j$, hence so does $g(\G_i, \G_j)$. 
The following relations follow directly from the above definitions: 
\begin{align*}
    \iint &p(\G_i, \G_j) d\G_i d\G_j = 1, \\
    \iint \V_i \,&p(\G_i, \G_j) d\G_i d\G_j = \left<\V_i\right>_0 = 0, \\
    \iint \V_i \V_j \e^{-\ii \KoR_{ij}} \, &p(\G_i, \G_j) d\G_i d\G_j = 
     \left<\V_i \V_j \e^{-\ii \KoR_{ij}}\right>_0,
\end{align*}
Hence, $g(\G_i, \G_j)$ should satisfy
\begin{align*}
    \iint &g(\G_i, \G_j) d\G_i d\G_j = 0, \\
    \iint \V_i \, &g(\G_i, \G_j) d\G_i d\G_j = 0, \\
    \iint \V_i \V_j \e^{-\ii \KoR_{ij}} \, &g(\G_i, \G_j) d\G_i d\G_j = 
     \left<\V_i \V_j \e^{-\ii \KoR_{ij}}\right>_0.
\end{align*}
Note that this definition of $p(\G_i, \G_j)$ also applies to the case of $i = j$. 

For any given particle triplet $ijl$, we can decompose the probability distribution into:
\begin{align}\label{eq:tri}
    p(\G_i, \G_j, \G_l) &= p(\G_i)\, p(\G_j)\, p(\G_l) + \Big[p(\G_i)\, g(\G_j, \G_l) \notag\\
    &\enspace
    \enspace + p(\G_j)\, g(\G_i, \G_l) + p(\G_l)\, g(\G_i, \G_j) \Big] \notag \\[3pt]
    & \enspace\enspace + \Tilde{g}(\G_i, \G_j, \G_l),
\end{align}
where $\Tilde{g}(\G_i, \G_j, \G_l)$ encodes the three-body correlations. 
Here we assume $\Tilde{g}(\G_i, \G_j, \G_l) \approx 0$.
To validate this decomposition of $p(\G_i, \G_j, \G_l)$, one can verify that 
\begin{align*}
    \iiint &p(\G_i, \G_j, \G_l) d\G_i d\G_j d\G_l = 1, \\
    \iiint \V_i \, &p(\G_i, \G_j, \G_l) d\G_i d\G_j d\G_l = 0, \\
    \iiint \V_i \V_j \e^{-\ii \KoR_{ij}} \, &p(\G_i, \G_j, \G_l)
    d\G_i d\G_j d\G_l = \left<\V_i\V_j\e^{-\ii \KoR_{ij}}\right>_0.
\end{align*}

Now let us evaluate the product $(\Fd \otimes \Jk)$. The first term in Eq.~\eqref{eq:dis} involves 
\begin{align} \label{eq:dis-p1}
    \notag\\[-6pt] 
    &\big<\gamma(r_{ij})\V_{i}\V_l \, \e^{-\ii \KoR_{il}}\big>_0 \notag\\[3pt]
    &= \iiint \gamma(r_{ij}) \V_i \V_l \e^{-\ii \KoR_{il}} p(\G_i, \G_j, \G_l) d\G_i d\G_j d\G_l \notag\\
    &= \iint \Big[\int \gamma(r_{ij}) p(\G_j) d\G_j\Big] 
    \V_i \V_l \e^{-\ii \KoR_{il}} g(\G_i, \G_l) d\G_i d\G_l \notag\\
    &= \iint \Big[\int_V \frac{1}{V}\gamma(r_{ij})  d\R_j\Big] 
    \V_i \V_l \e^{-\ii \KoR_{il}} g(\G_i, \G_l) d\G_i d\G_l \notag\\
    &= \iint \Big[\frac{1}{V}\int_V \gamma(r_{ji}) d\R_{ji}\Big] 
    \V_i \V_l \e^{-\ii \KoR_{il}} g(\G_i, \G_l) d\G_i d\G_l \notag\\
    &=  \frac{1}{V}\int_V \gamma(r) d\R \iint \V_i \V_l \e^{-\ii \KoR_{il}} g(\G_i, \G_l) d\G_i d\G_l \notag \\
    &= \frac{1}{V}\,\hat{\gamma}(0) \left<\V_i\V_l\e^{-\ii \KoR_{il}}\right>_0 \\[-6pt] \notag
\end{align}
where $\hat{\gamma}(\K) = \int_V \gamma(r) \e^{-\ii \KoR}d\R$ denotes the Fourier transform of $\gamma(r)$ and we have used the fact that $\left< \V \right>_0 = 0$. 
The second term in Eq.~\eqref{eq:dis} involves
\begin{align} \label{eq:dis-p2}
     \notag \\[-6pt]    
    &\big<\gamma(r_{ij})\V_{j}\V_l \, \e^{-\ii \KoR_{il}}\big>_0 \notag\\[3pt]
    &= \iiint \gamma(r_{ij}) \V_j \V_l \e^{-\ii \KoR_{il}} p(\G_i, \G_j, \G_l) d\G_i d\G_j d\G_l \notag\\[4pt]
    &= \iint \Big[\int \gamma(r_{ij}) \e^{-\ii \KoR_{ij}} p(\G_i)d\G_i \Big] \notag\\
    &\longspace\longspace\longspace \cdot\V_j \V_l \e^{-\ii \KoR_{jl}} g(\G_j, \G_l)  d\G_j d\G_l \notag\\[8pt]
    &= \frac{1}{V} \int_V \gamma(r) \e^{-\ii \KoR} d\R \iint \V_j \V_l \e^{-\ii \KoR_{jl}} g(\G_j, \G_l)  d\G_j d\G_l \notag\\
    &= \frac{1}{V}\, \hat{\gamma}(\K)\left<\V_j\V_l\e^{-\ii \KoR_{jl}}\right>_0.
    \\[-6pt] \notag
\end{align}
Plugging Eqs.~\eqref{eq:dis-p1} and \eqref{eq:dis-p2} into Eq.~\eqref{eq:dis}, we can derive the product
\begin{align}\label{eq:p3}
     \big(\Fd \otimes \Jk\big)
     &= -\frac{m}{V^2}  \sum_{ijl}^{N^3} \frac{1}{V}\,\hat{\gamma}(\K = 0) \left<\V_i\V_l\e^{-\ii \KoR_{il}}\right>_0 \notag\\
     &\enspace\enspace +\frac{m}{V^2} \sum_{ijl}^{N^3} \frac{1}{V}\, \hat{\gamma}(\K)\left<\V_j\V_l\e^{-\ii \KoR_{jl}}\right>_0\notag \\
     &= -\frac{n(\hat{\gamma}(0) - \hat{\gamma}(k))}{m}\cdot\frac{m^2}{V^2} \sum_{jl}^{N^2}\left<\V_j\V_l\e^{-\ii \KoR_{jl}}\right>_0 \notag \\
     &= -\gamma_{\Jk}\big(\Jk \otimes \Jk\big) \\[-6pt] \notag
\end{align}
where $\gamma_{\Jk} = n(\hat{\gamma}(0) - \hat{\gamma}(k))/m$ denotes the wavevector-dependent effective damping parameter for $\Jk$. In the hydrodynamic limit $\K \to 0$, 
\begin{align}\label{eq:gk}
    \gamma_{\Jk} = \frac{n}{m}\int_V \gamma(r) (1-\e^{-\ii \KoR}) d\R= \mathcal{O}(k).
\end{align}
In fact, strictly at $\K = 0$, $\gamma_{\Jk} = 0$. This is consistent with the momentum conversation of the entire system where no background friction is applied. 
To summarize, the results derived in Eqs.~\eqref{eq:p1}, \eqref{eq:p2}, and \eqref{eq:p3}, imply
\begin{align}
    (\iL \Jk \otimes \Jk) = - \gamma_{\Jk} (\Jk \otimes \Jk).
\end{align}
Therefore, the projection operator
\begin{align}
    \Pk \iL \Jk(\G) &= (\iL \Jk \otimes \Jk) \cdot (\Jk \otimes \Jk)^{-1} \cdot \Jk \notag
\end{align}
becomes
\begin{align}
    \Pk \iL \Jk(\G) = -\gamma_{\Jk} \Jk. 
\end{align}

\vspace{15mm}

\noindent\textbf{Generalized forces} \enspace 
Now we can evaluate the generalized forces $\Fp$ and $ \Fo$. At time $t = 0$, they are
\begin{align}\label{eq:F0}
    \Fp(0) &= \Pk \iL \Jk(\G) \,= - \gamma_{\Jk}  \Jk(0), \\
    \Fo(0) &= \Qk \iL \Jk(\G) =  \gamma_{\Jk}  \Jk(0) + \ii \K \cdot \Sk(0).
\end{align}
As the system evolves, 
\begin{align}
    \Fp(t) &= \e^{\iLt}\Fp(0) = -\gamma_{\Jk}  \Jk(t), \\[3pt]
    \Fo(t) &= \e^{\Qk \iLt}\Fo(0)  \notag\\
    &= \left[\e^{\iLt} - \int_0^t \e^{\iL (t-\tau)}\, \Pk \iL \, \e^{\Qk \iL \tau}\;d\tau \right]\Fo(0) \notag\\
    &= \e^{\iLt}\Fo(0) - \int_0^t \e^{\iL (t-\tau)}\, \Pk \iL \Fo(\tau)\;d\tau \notag\\
    &= \gamma_{\Jk} \Jk(t) + \ii \K \cdot \Sk(t) \label{eq:Fot}\\
    & - \int_0^t \big(\Fo(\tau)\otimes\iL \Jk(0)\big) \cdot (\Jk \otimes \Jk)^{-1} \cdot  \Jk(\tau)\, d\tau. \notag
\end{align}
We note that the parallel force $\Fp(t)$ provides an effective linear damping.
Furthermore, the propagator $\e^{\Qk \iLt}$ for the orthogonal force $\Fo$ can be expressed in terms of the standard propagator $\e^{\iLt}$ using the Dyson decomposition shown in the second step of Eq.~(\ref{eq:Fot}).  Here we argue that the integral term in the final form of Eq.~(\ref{eq:Fot}) only provides an $\mathcal{O}(k^2)$ correction, namely
\begin{equation}\label{eq:Fot2}
     \Fo(t) = \gamma_{\Jk} \Jk(t) + \ii \K \cdot \Sk(t) + \mathcal{O}(k^2).
\end{equation}
This argument can be verified by directly analyzing the product $\big(\Fo(\tau)\otimes\iL \Jk(0)\big)$, which decomposes into the following terms:
\begin{align*}
     &\big(\ii \K \cdot \Sreg(\tau) \otimes \ii\K \cdot \Sreg(0)\big) = \mathcal{O}(k^2), \\[2pt]
     &\big(\ii \K \cdot \Skin(\tau) \otimes \Fd(0)\big) = \ii \K\cdot \sum_{ij}^{N^2}C_{\K, ij}\left<\V_i^{\tau}\V_i^\tau\V_j^0\right>_0 =0, \\
     &\big(\Fd(\tau) \otimes \ii \K \cdot \Skin(0)\big) = \ii \K\cdot \sum_{ij}^{N^2}D_{\K, ij}\left<\V_i^0\V_i^0\V_j^\tau\right>_0 =0, \\
     &\big(\ii \K \cdot \Spos(\tau) \otimes \Fd(0)\big) = \ii \K\cdot \sum_{i}^{N}E_{\K, i}\left<\V_i^0\right>_0 = 0, \\
     &\big(\Fd(\tau) \otimes \ii \K \cdot \Spos(0)\big) = \ii \K\cdot \sum_{i}^{N}F_{\K, i}\left<\V_i^\tau\right>_0 = 0, \\
     &\big(\Fd(\tau) \otimes \Fd(0)\big) = \gamma^2_{\Jk}\big(\Jk(\tau) \otimes  \Jk(0) \big), \\[6pt]
     &\big(\gamma_{\Jk}\Jk(\tau) \otimes \Fd(0)\big) = -\gamma^2_{\Jk}\big(\Jk(\tau) \otimes  \Jk(0) \big), \\[6pt]
     &\text{with} \shortspace \big(\Jk(\tau) \otimes  \Jk(0) \big) = \frac{m^2}{V^2} \sum_{ij}^{N^2}\left<\V_i^\tau\V_j^0\e^{-\ii \K\cdot(\R_i^\tau - \R_j^0)}\right>_0,
\end{align*}
where the contribution $\Sreg \triangleq \Skin + \Spos$ denotes the part of the stress which excludes the velocity-dependent forces, $(C_{\K, ij}, D_{\K, ij}, E_{\K, i}, F_{\K, i})$ are velocity-independent coefficients, and time order is represented as superscripts. By summing the preceding terms, we find $\big(\Fo(\tau)\otimes\iL \Jk(0)\big) = \mathcal{O}(k^2)$, which is negligible in the hydrodynamic limit $\K \to 0$.

Using Eq.~(\ref{eq:Fot2}), we can further evaluate the linear-response coefficient matrix  $\textbf{K}(\tau)$ up to $\mathcal{O}(k^2)$:
\begin{align*}
    &(\Fo(\tau) \otimes \Fo(0)) \\
    &= \gamma^2_{\Jk}\big(\Jk(\tau) \otimes  \Jk(0) \big) + \gamma_{\Jk}\big(\Jk(\tau) \otimes  \ii \K \cdot \Sk(0)\big)\\
    &\; + \gamma_{\Jk}\big(\ii \K \cdot \Sk(\tau) \otimes  \Jk(0)\big) +
   \big(\ii \K \cdot \Sk(\tau) \otimes \ii \K \cdot \Sk(0)\big)\\
   &= \gamma^2_{\Jk}{m^2}\big(\Jk(\tau) \otimes  \Jk(0) \big) + \gamma_{\Jk}\big(\Jk(\tau) \otimes  \Fd(0)\big)\\
    &\; + \gamma_{\Jk}\big(\Fd(\tau) \otimes  \Jk(0)\big) +
   \big(\ii \K \cdot \Sk(\tau) \otimes \ii \K \cdot \Sk(0)\big)\\
   &= \big(\ii \K \cdot \Sk(\tau) \otimes \ii \K \cdot \Sk(0)\big) - \gamma^2_{\Jk}\big(\Jk(\tau) \otimes  \Jk(0) \big).
\end{align*}
Thus we have
\begin{align} \label{eq:K}
    \textbf{K}(\tau) &= \big(\Fo(\tau) \otimes \Fo(0)\big) \cdot (\Jk \otimes \Jk)^{-1} \notag\\
    &=  - \gamma^2_{\Jk}\big(\Jk(\tau) \otimes  \Jk(0) \big)\cdot (\Jk \otimes \Jk)^{-1}  \\
    &\shortspace\, + \big(\ii \K \cdot \Sk(\tau) \otimes \ii \K \cdot \Sk(0)\big) \cdot (\Jk \otimes \Jk)^{-1}.\notag 
\end{align}
In the above derivation, we considered the decomposition of the stress $\ii \K \cdot \Sk = \ii \K \cdot(\Skin + \Spos) + \Fd$ and the fact that only the velocity dependent term $\Fd$ can couple with the momentum current density $\Jk$. 

\vspace{15mm}

\noindent\textbf{Green--Kubo formula} \enspace
For a system in a nonequilibrium steady state, the noise term $\Fo$ vanishes in Eq.~(\ref{eq:Mori}) under an ensemble average. Nonetheless, the $\Fo$ term is still essential since it enters into the definition of $\textbf{K}$, which will ultimately provide the linear response. 
Thus in an average sense, the generalized Green--Kubo formula reads
\begin{align}\label{eq:Kubo}
    &\dot{\J}_\K(t) = \Fp(t) - \int_{0}^{t}d\tau \;\textbf{K}(\tau) \cdot \Jk (t-\tau). 
\end{align}

By performing the Laplace transform of both sides of Eq.~(\ref{eq:Kubo}), we obtain
\begin{align}\label{eq:Kubo-s}
    &s\Tilde{\textbf{J}}_\K(s) - \Jk(0) = \Tilde{\textbf{F}}^\parallel_{\K}(s) - \Tilde{\textbf{K}}(s) \cdot \Tilde{\textbf{J}} (s). 
\end{align}
Here we show that in the hydrodynamic limit, $\Tilde{\textbf{F}}^\parallel_{\K}(s)$ ultimately cancels with contributions from $\textbf{K}(s)\cdot\Tilde{\textbf{J}}_\K(s)$ arising from the first term in Eq.~(\ref{eq:K}). Let us first evaluate $\varrho_{\J\J}(t) = \big(\Jk(t) \otimes \Jk(0) \big)$ involved in the first term in Eq.~(\ref{eq:K}). The quantity $\varrho_{\J\J}(t)$ obeys the following master equation:
\begin{align}\label{eq:rhojj-ini}
    \dot{\varrho}_{\J\J}(t) &= \big(\dot{\J}_{\K}(t) \otimes \Jk(0) \big) = \big(\iL \Jk (t) \otimes \Jk(0) \big) \notag\\
    &= \big(\Fd(t)\otimes \Jk(0)\big)= -\gamma_{\Jk}\big(\Jk(t)\otimes \Jk(0)\big) \notag \\
    &= -\gamma_{\Jk}\varrho_{\J\J}(t)
\end{align}
After a Laplace transform, 
\begin{align}
    s\Tilde{\varrho}_{\J\J}(s) - \varrho_{\J\J}(0) = -\gamma_{\Jk}\Tilde{\varrho}_{\J\J}(s).
\end{align}
Thus we have 
\begin{align}
    \Tilde{\varrho}_{\J\J}(s) &= \frac{1}{s + \gamma_{\Jk}}\varrho_{\J\J}(0).
\end{align}
Hydrodynamics corresponds to the slowest dynamics in the system. Therefore, in the hydrodynamic limit, $t \to \infty$ and $s \to 0$. Thus, 
\begin{align}\label{eq:rhojj-end}
    \Tilde{\varrho}_{\J\J}(s)
    \approx \frac{1}{\gamma_{\Jk}}\varrho_{\J\J}(0) = \frac{1}{\gamma_{\Jk}}(\Jk \otimes \Jk).
\end{align}
Therefore, the Laplace transform of the first half of $\textbf{K}(t)$ becomes
\begin{align}\label{eq:K1st}
    \Tilde{\textbf{K}}^{\text{1st}}(s) &= -\gamma^2_{\Jk}\Tilde{\varrho}_{\J\J}(s) \cdot (\Jk \otimes \Jk)^{-1} \\
    &= -\gamma_{\Jk}(\Jk \otimes \Jk) \cdot (\Jk \otimes \Jk)^{-1} = -\gamma_{\Jk}\mathcal{I}. \notag
\end{align}
Thus we have 
\begin{align}
    \Tilde{\textbf{F}}^\parallel_{\K}(s) - \Tilde{\textbf{K}}^{\text{1st}}(s)\cdot\Tilde{\textbf{J}}_\K(s)= -\gamma_{\Jk}\Jk(s) + \gamma_{\Jk}\Jk(s) = \textbf{0}. \notag
\end{align}
As anticipated, $\Tilde{\textbf{F}}^\parallel_{\K}(s)$ drops out in the hydrodynamic limit.
After Laplace transform, the second half of $\textbf{K}(t)$ reads
\begin{align}\label{eq:K2nd}
    \Tilde{\textbf{K}}^{\text{2nd}}(s) &=
    \big(\ii \K \cdot \Tilde{\bm{\sigma}}_{\K}(s) \otimes \ii \K \cdot \Sk(0)\big) \cdot (\Jk \otimes \Jk)^{-1}. 
\end{align}
Here we keep terms up to $\mathcal{O}(k^2)$. Thus only the leading order of $(\Jk \otimes \Jk)^{-1}$ needs to be considered. Given Eq.~(\ref{eq:JJ3}), we have
\begin{align}\label{eq:K2nd}
    \Tilde{\textbf{K}}^{\text{2nd}}(s) \approx
    \frac{1}{n m c_\textbf{JJ}} \big(\ii \K \cdot \Tilde{\bm{\sigma}}_{\K}(s) \otimes \ii \K \cdot \Sk(0)\big),
\end{align}
where $c_\textbf{JJ}$ was defined in Eq.~\eqref{eq:cJJ}.
Now we can rewrite the Laplace transform of the generalized Green--Kubo formula Eq.~(\ref{eq:Kubo-s}) as
\begin{align}\label{eq:Kubo-s2}
    &s\Tilde{\textbf{J}}_\K(s) - \Jk(0) = -\Tilde{\textbf{K}}^{\text{2nd}}(s)\cdot\Tilde{\textbf{J}}_{\K}(s) \notag\\
    &= -\frac{1}{n m c_\textbf{JJ}} \big(\ii \K \cdot \Tilde{\bm{\sigma}}_{\K}(s) \otimes \ii \K \cdot \Sk(0)\big)\cdot \Tilde{\textbf{J}}_{\K}(s).
\end{align}

\vspace{10mm}

\noindent\textbf{Viscosity tensor} \enspace
Let us convert this equation to tensorial notation:
\begin{align}
    &\big(\ii \K \cdot \Tilde{\bm{\sigma}}_{\K}(s) \otimes \ii \K \cdot \Sk(0)\big) \cdot \Tilde{\textbf{J}}_{\K}(s) \notag\\
    &= -\ii k_b \left<\Tilde{\sigma}_{\K, ab}(s) \hat{\sigma}^*_{\K, cd}(0)\right>_0\ii k_d\Tilde{J}_{\K, c}(s),
\end{align}
where Einstein summation is applied to the indices $a$, $b$, $c$, and $d$. Since $\ii k_n$ corresponds to the Fourier transform of $\partial_n$, the quantity
$\ii k_q\hat{J}_{\K, p}$ is proportional to the wavevector-dependent strain rate $\hat{\dot{e}}_{\K, cd}$:
\begin{equation*}
\begin{split}
     \hat{\dot{e}}_{\K, cd} 
     &=\int_{V} \partial_d u_c(r) \e^{-\ii \KoR} d\R = \ii k_d \int_{V}  u_c(r) \e^{-\ii \KoR} d\R \\
     &= \ii k_d \int_{V} \left[\frac{1}{N} \sum_{i}^{N} v_{c, i} \delta(\R-\R_i) \right] \e^{-\ii \KoR} d\R \\
     &= \ii k_d \frac{1}{N} \sum_{i}^{N} v_{c, i} \e^{-\ii \KoR_i} = \frac{\ii k_d \jkc}{n m},
\end{split}
\end{equation*}
where $u_c(\R) = \sum_{i}^{N} v_{c, i}\delta(\R-\R_i)/N$ is the local flow field, and $\partial_d u_{c}(\R)$ is the real-space strain rate. Thus, we may rewrite Eq.~(\ref{eq:Kubo-s2}) in the following form:
\begin{align} \label{eq:KuboTen}
    &\shortspace s\Tilde{J}_{\K,a}(s) - \hat{J}_{\K, a}(0) = -\ii k_b\Tilde{\eta}_{\K, abcd}(s)\Tilde{\dot{e}}_{\K, cd},
\end{align}
with 
\begin{align}
     \;\;\Tilde{\eta}_{\K, abcd}(s) \triangleq \frac{1}{c_\textbf{JJ}}\left<\Tilde{\sigma}_{\K, ab}(s) \hat{\sigma}^*_{\K, cd}(0)\right>_0. 
\end{align}
In comparison, the Laplace transform of the master equation Eq.~(\ref{eq:dt-J2}) is
\begin{align} \label{eq:MasterLaplace}
    s\Tilde{J}_{\K, a}(s) &- \hat{J}_{\K, a}(0) = -\ii k_b
    \Tilde{\sigma}_{\K, ab} (s).
\end{align}
Combining Eq.~(\ref{eq:KuboTen}) with Eq.~(\ref{eq:MasterLaplace}) yields the following constitutive equation:
\begin{align}\label{eq:lin}
    \Tilde{\sigma}_{\K, ab} (s) = \Tilde{
    \eta}_{\K, abcd}(s) \,\Tilde{\dot{e}}_{\K, cd}(s).
\end{align}
Let us take the hydrodynamic limit $k \to 0$ and $s \to 0$.
In this limit, the system experiences a uniform, constant strain-rate $\dot{e}_{cd}$ and stress $\sigma_{ab}$ given by
\begin{align}
      \sigma_{ab}&\triangleq \lim_{\substack{\K \to \textbf{0} \\ s \to 0}} \Tilde{\sigma}_{\K, ab} (s), \\[6pt] 
        \dot{e}_{cd}& \triangleq \lim_{\substack{\K \to \textbf{0} \\ s \to 0}} \Tilde{\dot{e}}_{\K, cd}(s).
\end{align}
The constitutive equation Eq.~(\ref{eq:lin}) then reduces to
\begin{align}\label{eq:linear}
    \sigma_{ab} = \eta_{abcd} \dot{e}_{cd}. 
\end{align}
where the viscosity coefficients satisfy the standard Green--Kubo relation:
\begin{align}
    \eta_{abcd} &= \lim_{\substack{\K \to \textbf{0} \\ s \to 0}} \Tilde{
    \eta}_{\K, abcd}(s) \notag \\
    &= \frac{1}{c_\textbf{JJ}} \int_{0}^{\infty} \left<\sigma_{ab}(t)\sigma_{cd}(0)\right>_0 dt.
\end{align}

\vspace{10mm}

\noindent\textbf{Viscosity matrix for 2D isotropic fluids} \enspace In the main text, we use a graphical matrix representation of the viscosity tensor based on the symmetries of a 2D isotropic fluid (see Ref.~\cite{scheibner2019odd}).
Let us convert the Cartesian components of the viscosity tensor (denoted by subscript car) into this representation (denoted by subscript mat).
The components of the stress and strain rate in both representations may be written as vectors
\begin{align}
    \bm{\sigma}_\text{mat} &= \begin{pmatrix} \protect\p \\ \protect\tor \\ \protect\so \\ \protect\st \end{pmatrix}, &
    \dot{\textbf{e}}_\text{mat} &= \begin{pmatrix} \protect\dtV \\ \protect\dtR \\ \protect\dtSo \\ \protect\dtSt \end{pmatrix}, \\[12pt]
    \bm{\sigma}_\text{car} &= \begin{pmatrix} \sigma_{xx} \\ \sigma_{xy} \\ \sigma_{yx} \\ \sigma_{yy}\end{pmatrix}, &
    \dot{\textbf{e}}_\text{car} &= \dot{\begin{pmatrix} e_{xx} \\ e_{xy} \\ e_{yx} \\ e_{yy} \end{pmatrix}}
\end{align}
that are obtained from each other through the linear relation
\begin{align}
    \bm{\sigma}_\text{mat} &= \frac12 \mathcal{Q}\,\bm{\sigma}_\text{car}, & \dot{\textbf{e}}_\text{mat} &= \mathcal{Q}\,\dot{\textbf{e}}_\text{car},
\end{align}
where
\begin{align}
    \mathcal{Q} = \begin{pmatrix} 1 & 0 & 0 & 1 \\ 0 & -1 & 1 & 0 \\ 1 & 0 & 0 & -1 \\ 0 & 1 & 1 & 0 \end{pmatrix}.
\end{align}

In terms of the vectors of Cartesian components defined above, the tensorial equation Eq.~(\ref{eq:linear}) reads
\begin{align}\label{eq:tensor}
    \bm{\sigma}_\text{car} = \bm{\eta}_\text{car} \, \dot{\textbf{e}}_\text{car}, \;\;\;\;
\end{align}
with
\begin{align}
    \bm{\eta}_\text{car} = \frac{1}{c_\textbf{JJ}} &\int_{0}^{\infty} dt \left<\bm{\sigma}_\text{car}(t)\bm{\sigma}_\text{car}^\text{T}(0)\right>_0.
\end{align}
Since $\mathcal{Q}^\text{T}\mathcal{Q} = 2\mathcal{I}$, we can multiply both sides of Eq.~(\ref{eq:tensor}) by $\mathcal{Q}/2$ and insert $\mathcal{Q}^\text{T}\mathcal{Q}/2$ to get
\begin{align*}
    \frac12\mathcal{Q}\bm{\sigma}_\text{car} = \frac12 \mathcal{Q}\,\bm{\eta}_\text{car} \Big(\frac12\mathcal{Q}^{T}\mathcal{Q}\Big)\,\dot{\textbf{e}}_\text{car}
\end{align*}
namely
\begin{align*}
    \bm{\sigma}_\text{mat} = \Big(\frac14 \mathcal{Q}\bm{\eta}_\text{car}\mathcal{Q}^\text{T} \Big) \dot{\textbf{e}}_\text{mat}.
\end{align*}
Hence, the viscosity matrix may be expressed as:
\begin{align} \label{eq:etamat}
    \bm{\eta}_\text{mat} &= \frac{1}{4}\mathcal{Q}\bm{\eta}_\text{car}\mathcal{Q}^\text{T} \notag\\
    &= \frac{1}{c_\textbf{JJ}} \int_{0}^{\infty}  \left<(\frac12\mathcal{Q}\bm{\sigma}_\text{car}(t))(\frac12\mathcal{Q}\bm{\sigma}_\text{car}(0))^\text{T}\right>_0 dt \notag\\
    &= \frac{1}{c_\textbf{JJ}} \int_{0}^{\infty}  \left<\bm{\sigma}_\text{mat}(t)\bm{\sigma}_\text{mat}^\text{T}(0)\right>_0 dt 
\end{align}
Now let us introduce the quantity: 
\begin{align}\label{eq:Bvv}
    B_{\V\V} &= n m \bar{c}_{\V\V}^\text{near}V_\text{c} \notag\\
    &= \frac{1}{2}nm \int_{r \leq r_\text{c}} \Big[\left<\V(0)\cdot\V(\R)\right>_0 - \left<\V(0)\cdot\V(\infty)\right>_0\Big]d\R \notag \\
    &= \frac{1}{2}nm\lim_{\K \to 0} \int_{V} \left<\V(0)\cdot\V(\R)\right>_0 \e^{-\ii \KoR}d\R.  \notag\\
    &= \frac{1}{2}nm \, \hat{c}_{\V\V}(\K \to 0)
\end{align}  
Now if we use $A$ instead of $V$ to emphasize the 2D area, Eq.~(\ref{eq:etamat}) may be written in the final form used in the main text:
\begin{equation}
	\label{green_kubo}
	\eta_{\alpha \beta} = \frac{A}{\kT + B_{\V\V}} \, \int_0^{\infty} \left< \sigma_{\alpha}(t) \sigma_{\beta}(0) \right> dt.
\end{equation}

In a system with conservative interactions, the particle velocities have no spatial correlation and hence the correction $B_{\V \V}$ vanishes and the proportionality constant $(\kT + B_{\V\V})$ reduces to the standard thermal energy $\kT$, which characterizes the magnitude of velocity fluctuations of individual particles. In a system with velocity dependent interactions, the velocity of a particle is correlated with that of its neighbors. Such normalization factor in fact encodes the magnitude of the collective velocity fluctuations of an individual particle and its neighbors.

\begin{figure}[h]
\includegraphics[width=0.52\textwidth]{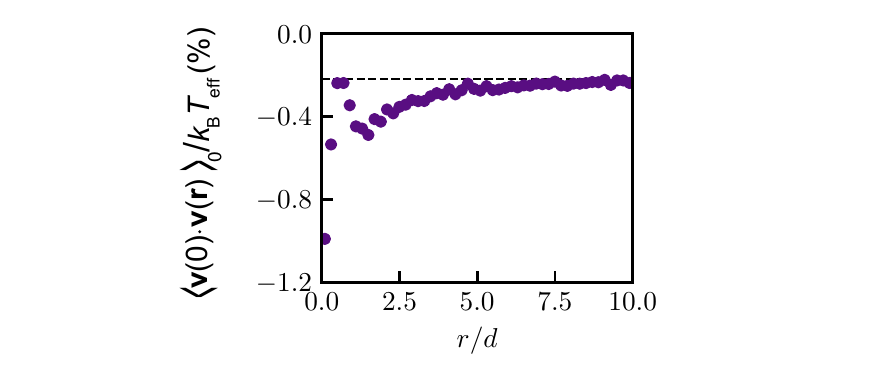}
\caption{\label{fig:vv-r} \textbf{Velocity--velocity correlation function.} We quantify the velocity--velocity correlation function $\left<\V(0)\V(\R)\right>_0$ at $\Omega = 26.7/\Delta t$. In the far field, the correlation function converges to  $\left<\V(0)\V(\infty)\right>_0 = 0.12\%$, which is marked by the dashed line. Using Eq.~\eqref{eq:Bvv}, we can estimate the correction term $B_{\V\V}/\kT = 8.3\%$.}
\end{figure}

In our chiral active fluid, we can measure the velocity--velocity correlation function $\left<\V(0)\cdot\V(\R)\right>_0$, see Fig.~\ref{fig:vv-r}. Using Eq.~\eqref{eq:Bvv}, we find the correction term $B_{\V\V}/\kT = 8.3\%$. By renormalizing $T_\text{eff}$ to $T_\text{eff}^* = T_\text{eff} + B_{\V\V}/k_\text{B}$, we achieve excellent agreement between the simulation measurement of the viscosities and the Kubo predictions, as shown in Fig.~3C.

\vspace{10mm}
\noindent\textbf{Extension to hydrodynamic interactions} \enspace The above derivation of the Green--Kubo relation assumes that the velocity dependent interactions are reciprocal: $\F_{ij}^\text{\,v} = \gamma(r_{ij})\V_{ij} = - \F_{ji}^\text{\,v}$, which primarily applies to dry active systems. 
In a wet active system, hydrodynamic interactions between particles are mediated by a liquid environment and can often be non-reciprocal. 
For our analysis, we assume that the hydrodynamic force generated by particle $j$ on particle $i$ is well approximated by the linear relationship
\begin{align} \label{eq:fh}
    \F^\text{\,h}_{ij} = - \Upsilon(\R_{ij}) \cdot \V_j,
\end{align}
where $\Upsilon(\R_{ij})$ are a symmetric drag coefficient matrix depending on the interparticle vector $\R_{ij}$. 
One can show that as long as $\V_i \neq -\V_j$, $\F^\text{\,h}_{ij} \neq -\F^\text{\,h}_{ji}$.
\newline

Let us take a 3D suspension of spherical colloids as an example (see e.g. Ref.~\cite{dileonardo2008hydro2d} for a discussion on 2D hydrodynamic interactions). In general, the hydrodynamic interaction is truly a many-body effect, which involves direct coupling among all the particles~\cite{ermak1978brownian}:
\begin{align}
    \F^\text{\,h}_{p} = -\sum_{q}^{3N} \zeta_{pq} v_j
\end{align}
where indices $p$ and $q$ run over the $3N$ particle coordinates $(x_1, y_1, z_1, x_2, y_2, z_2,...)$. $\zeta_{pq}$ is a friction tensor that depends on the configuration of the entire system and satisfies the relation:
\begin{align}
    \sum_{q}^{3N} \zeta_{pq} D_{qs} = \sum_{q}^{3N} D_{pq}\zeta_{qs} = \kb T\delta_{ps}
\end{align}
where $D_{pq}$ is the configuration-dependent diffusion tensor. 
The diffusion tensor contains two parts: (i) when $p$ and $q$ are coordinates of the same particle, $D_{pq}$ are the corresponding elements of the Stokes drag coefficient matrix
\begin{align}
    \mathcal{D}^{\text{(i)}} = \frac{\kb T}{6\pi\mu a} \,\mathcal{I}_{3 \times 3}
\end{align}
where $\mu$ is the viscosity of the fluid, $a$ is the particle radius, and $\mathcal{I}_{3 \times 3}$ denotes a $3 \times 3$ identity matrix.
(ii) when $p$ and $q$ are coordinates of two different particles $i$ and $j$, $D_{pq}$ are the corresponding elements $\mathcal{D}^{\text{(ii)}\, {a b}}_{i j}$ of the Oseen tensor (i.e., $p=(i,a)$, $q=(j,b)$ where $i,j$ index the particles and  $a,b$ the Cartesian coordinates of the particles)
\begin{align}
    \mathcal{D}^\text{(ii)}_{i j} = \frac{\kb T}{8\pi\mu r_{ij}} \,\Big[\mathcal{I}_{3 \times 3} + \frac{\R_{ij}\R_{ij}}{r_{ij}^2}\Big].
\end{align}
The first and second parts give the diagonal and off-diagonal components of the diffusion matrix $D$, denoted by:
\begin{align}
    D = \frac{\kb T}{6\pi \mu a} \, \mathcal{I}_{3N \times 3N} + D^\text{(ii)}.
\end{align}
Note that $D^\text{(ii)}$ scales inversely linear with particle distance, $D^\text{(ii)} \sim r_{ij}^{-1}$. In a dilute system where the average particle distance $\bar{r}_{ij} \gg a$, $D^\text{(ii)}$ can be treated as a small perturbation.
Thus we can estimate the friction matrix as 
\begin{align} \label{eq:dragApprox}
    \zeta \approx \kb T\, D^{-1} = 6\pi \mu a \Big[\mathcal{I}_{3N \times 3N} - \frac{6\pi \mu a}{\kb T} D^\text{(ii)}\Big].
\end{align}
The approximation in Eq.~(\ref{eq:dragApprox}) reduces the many-body interaction into a pairwise hydrodynamic interaction:
\begin{align}
    \F^\text{\,h}_{ij} = -\Upsilon_{ij} \V_j
\end{align}
with the drag coefficient matrix 
\begin{align}
     \Upsilon_{ij} = 6\pi \mu a \, \mathcal{I} \delta_{ij} - \frac{9\pi \mu a^2}{2r_{ij}} \left[\mathcal{I}_{3\times3} + \frac{\R_{ij}\R_{ij}}{r_{ij}^2}\right].
\end{align}
Note that $\Upsilon_{ij}$ is a symmetric matrix depending on the interparticle distance.

Now let us continue the derivation of the Green-Kubo relations with the general linear ansatz for the hydrodynamic interaction $\F^\text{\,h}_{ij} = - \Upsilon(\R_{ij}) \cdot \V_j$.  
Since this interaction is non-reciprocal $\F^\text{\,h}_{ij} \neq -\F^\text{\,h}_{ji}$, it cannot be included in the calculation of Irvine--Kirkwood stress. Instead, we use the decomposition of the generalized force $\iL \Jk$ similar to Eq.~\eqref{eq:iLJ-decompose}
\begin{align}
    \iL \Jk = \ii \K \cdot \Sreg + \Fd
\end{align}
where the stress $\Sreg$ includes the kinetic part as well as the virial part contributed by position dependent stresses only, which we assume to be reciprocal. The term $\Fd$ summarizes the contributions of the hydrodynamic interaction:
\begin{align}
    \Fd \triangleq -\frac{1}{V}\sum_{ij}^{N^2}\Upsilon(\R_{ij})\V_{j} \, \e^{-\ii \KoR_i}.
\end{align}
We will now proceed with our previous derivation while highlighting the steps potentially affected by $\Fd$. 

First, let us re-derive $\Fp$, $\Fo$ and $\textbf{K}$ in the Mori--Zwanzig by evaluating all the outer products involving $\Fd$. 
To evaluate the equal-time product $\big(\Fd \otimes \Jk\big)$, we first follow the steps in Eq.~\eqref{eq:dis-p2} to derive the relation:
\begin{align*}
    \big<\Upsilon(\R_{ij})\V_{j}\V_l \, \e^{-\ii \KoR_{il}}\big>_0  = \frac{1}{V}\, \hat{\Upsilon}(\K)\left<\V_j\V_l\e^{-\ii \KoR_{jl}}\right>_0,
\end{align*}
where $\hat{\Upsilon}(\K) = \int_V \Upsilon(\R) \e^{-\ii \KoR}d\R$ denotes the Fourier transform of the matrix $\Upsilon(\R)$. We find  
\begin{align}
    \big(\Fd \otimes \Jk\big) &= -\frac{m}{V^2} \sum_{ijl}^{N^3}\big<\gamma(r_{ij})\V_{j} \V_l \, \e^{-\ii \KoR_{il}}\big>_0 \notag\\
    &= -\frac{m}{V^2}\sum_{ijl}^{N^3}\frac{1}{V}\, \hat{\Upsilon}(\K)\left<\V_j\V_l\e^{-\ii \KoR_{jl}}\right>_0 \notag \\
    &= -\frac{n\hat{\Upsilon}(\K)}{m} \cdot \frac{m^2}{V^2} \sum_{jl}^{N^2} \left<\V_j\V_l\e^{-\ii \KoR_{jl}}\right>_0 \notag \\
    &= -\gamma_{\Jk}\big(\Jk \otimes \Jk\big),
\end{align}
where the effective damping coefficient $\gamma_{\Jk} = n\hat{\Upsilon}(\K)/m$ now is a matrix. Since $\Upsilon(\R)$ is a symmetric matrix, $\gamma_{\Jk}$ is also symmetric.
We then derive the time-correlated products
\begin{align*}
    &\big(\ii \K \cdot \Sreg(\tau) \otimes \ii\K \cdot \Sreg(0)\big) = \mathcal{O}(k^2), \\[2pt]
     &\big(\ii \K \cdot \Skin(\tau) \otimes \Fd(0)\big) = \ii \K\cdot \sum_{ij}^{N^2}C_{\K, ij}\left<\V_i^{\tau}\V_i^\tau\V_j^0\right>_0 =0, \\
     &\big(\Fd(\tau) \otimes \ii \K \cdot \Skin(0)\big) = \ii \K\cdot \sum_{ij}^{N^2}D_{\K, ij}\left<\V_i^0\V_i^0\V_j^\tau\right>_0 =0, \\
     &\big(\ii \K \cdot \Spos(\tau) \otimes \Fd(0)\big) = \ii \K\cdot \sum_{i}^{N}E_{\K, i}\left<\V_i^0\right>_0 = 0, \\
     &\big(\Fd(\tau) \otimes \ii \K \cdot \Spos(0)\big) = \ii \K\cdot \sum_{i}^{N}F_{\K, i}\left<\V_i^\tau\right>_0 = 0, \\[3pt]
     &\big(\Fd(\tau) \otimes \Fd(0)\big) = \gamma_{\Jk}\big(\Jk(\tau) \otimes  \Jk(0) \big)\,\gamma_{\Jk}, \\[12pt]
     &\big(\Fd(\tau) \otimes \Jk(0) \big) = -\gamma_{\Jk}\big(\Jk(\tau) \otimes  \Jk(0) \big)\,\gamma_{\Jk}. \\[12pt]
     &\big(\gamma_{\Jk}\Jk(\tau) \otimes \Fd(0)\big) = -\gamma_{\Jk}\big(\Jk(\tau) \otimes  \Jk(0) \big)\,\gamma_{\Jk}. \\[12pt]
     &\big(\gamma_{\Jk}\Jk(\tau) \otimes \gamma_{\Jk}\Jk(0) \big) = \gamma_{\Jk}\big(\Jk(\tau) \otimes  \Jk(0) \big)\,\gamma_{\Jk}. \\[-6pt]
\end{align*}
These products allow us to derive the generalized forces in the Mori--Zwanzig formalism:
\begin{align}
    \Fp(t) &= -\gamma_{\Jk} \Jk(t), \\[6pt]
    \Fo(t) &= \ii \K \cdot \Sreg(t) + \Fd(t) \\ 
           &\longspace\midspace\;\, + \gamma_{\Jk} \Jk(t)+ \mathcal{O}(k^2), \notag
\end{align}
as well as the response function 
\begin{align}
    \textbf{K}(\tau) &= \big(\Fo(\tau) \otimes \Fo(0)\big) \cdot (\Jk \otimes \Jk)^{-1} \\[6pt]
    &=  \big(\ii \K \cdot \Sreg(\tau) \otimes \ii \K \cdot \Sreg(0)\big) \cdot (\Jk \otimes \Jk)^{-1}.\notag 
\end{align}

With the newly derived $\Fp(t)$ and $\textbf{K}(\tau)$, the generalized Green--Kubo relation becomes
\begin{align}\label{eq:Kubo-hydro}
    &s\Tilde{\textbf{J}}_\K(s) - \Jk(0) = \Tilde{\textbf{F}}^\parallel_{\K}(s) - \Tilde{\textbf{K}}(s) \cdot \Tilde{\textbf{J}} (s) \\[9pt]
    &= - \frac{\big(\ii \K \cdot \Tilde{\bm{\sigma}}_{\K}^\text{reg}(s) \otimes \ii \K \cdot \Sreg(0)\big)\cdot \Tilde{\textbf{J}}_{\K}(s)}{n m c_\textbf{JJ}} -\gamma_{\Jk}\Tilde{\textbf{J}}_\K(s). \notag 
\end{align}

Following the section ``viscosity tensor," we find that the term involving the product $\big(\ii \K \cdot \Tilde{\bm{\sigma}}_{\K}^\text{reg}(s) \otimes \ii \K \cdot \Sreg(0)\big)$ in Eq.~\eqref{eq:Kubo-hydro} gives rise to a viscous coefficient tensor satisfying the Green--Kubo relation
\begin{align}
    \eta_{abcd}^{(0)} = \frac{1}{c_\textbf{JJ}} \int_0^\infty \left<\sigma^\text{reg}_{ab}(t)\sigma^\text{reg}_{cd}(0)\right>_0 dt.
\end{align}
\vspace{1mm}

Regarding the term $-\gamma_{\Jk}\Jk(s)$ in Eq.~\eqref{eq:Kubo-hydro}, we note that  $\gamma_{\Jk}$ is a function of wavevector $\K$, and thus can lead to corrections in the Green--Kubo formula. 
We perform a Taylor series expansion on $\gamma_{\Jk}$ in the vicinity of $k=0$.
Due to isotropy, the series expansion of $\gamma_{\Jk}$ takes the general form
\begin{align} \label{eq:Tay1}
    \begin{split}
    \gamma_{\Jk} &= \gamma_0\mathcal{I} + \gamma_1 k \, \mathcal{I} + \big(\gamma_2^\text{A} k^2 \,\mathcal{I} +  \gamma_2^\text{B}\, \K\K \big) + \mathcal{O}(k^3)
    \end{split}
\end{align}
where $\mathcal{I}$ is a $D\times D$ identity matrix.  Here all the coefficients are scalars that do not depend on either $\R$ or $\K$. 
In addition, the definition of $\gamma_{\Jk}$ implies
\begin{align} \label{eq:Tay2}
\begin{split}
    \gamma_{\Jk} &= \frac{n\hat{\Upsilon}(\K)}{m} = \frac{n}{m} \int_V \Upsilon(\R)\, \e^{-\ii \KoR} d\R \\
    &= \frac{n}{m} \int_V \Upsilon(\R)\, \Big[1- (\ii \KoR) + (\ii \KoR)^2/2 + \mathcal{O}(k^3)\Big] d\R.
\end{split}
\end{align}
Comparing Eq.~(\ref{eq:Tay1}) with Eq.(\ref{eq:Tay2}), we find
\begin{align}
    \gamma_0\mathcal{I} &= \frac{n}{m} \int_V \Upsilon(\R)\, d\R,\\
    \gamma_1 k \, \mathcal{I} &= -\frac{n}{m} \int_V (\ii \KoR)\,\Upsilon(\R)\, d\R, \label{eq:k1}\\
    \gamma_2^\text{A}\, k^2 \,\mathcal{I} +  \gamma_2^\text{B}\, \K\K & = \frac{n}{2m} \int_V (\ii \KoR)^2\,\Upsilon(\R)\, d\R. \label{eq:gamma2}
\end{align}
Since $k = \sqrt{\K\cdot\K}$ is an operation not allowed in Eq.~\eqref{eq:k1}, the corresponding coefficient has to vanish, $\gamma_1 = 0$. Thus, we can summarize the effective damping coefficient matrix as
\begin{align}
    \gamma_{\Jk} = \gamma_0\mathcal{I} +  \gamma_2^\text{A}\, k^2 \,\mathcal{I} +  \gamma_2^\text{B}\, \K\K.
\end{align}

Now let us perform an inverse Laplace transform of $-\gamma_{\Jk}\Tilde{\textbf{J}}_\K(s)$ in Eq.~\eqref{eq:Kubo-hydro}:
\begin{align}\label{eq:gamma-J}
    -\gamma_{\Jk} \Jk(t) = -\gamma_0 \Jk(t) - \gamma_2^\text{A} k^2 \Jk(t) - \gamma_2^\text{B} \K\K\,\Jk(t). 
\end{align}
The first term in Eq.~\eqref{eq:gamma-J} may be written as:
\begin{align*}
    -\gamma_0 \Jk &= -\frac{\gamma_0}{V} \sum_i^N m\V_i \e^{-\ii \KoR_i} \\
    &= -nm\gamma_0 \int_V \Big[\frac{1}{N} \sum_i^N \V_i \delta(\R-\R_i)\Big] \e^{-\ii \KoR} d\R \notag\\
    &= -nm\gamma_0 \int_V \U(\R) \e^{-\ii \KoR} d\R = -nm\gamma_0 \, \hat{\U}(\K), \notag
\end{align*}
where $\U(\R) = \sum_i^N \V_i \delta(\R-\R_i)/N$ is the local flow field. 
This term correspond to a background friction
\begin{align}
    -\gamma_0 \Jk \to  -\gamma_{f} \U(\R),
\end{align}
with frictional coefficient
\begin{align}
    \gamma_{f} = nm\gamma_0.
\end{align}
\vspace{1mm}

The second term in Eq.~\eqref{eq:gamma-J} may be written as
\begin{align} \label{eq:Term2}
\begin{split}
    -\gamma_2^\text{A} k^2 \Jk &= -\gamma_2^\text{A} \,k_b \, k_b \,\jka = \gamma_2^\text{A}\, (\ii k_b) \,(\ii k_b \,\jka) \notag\\[6pt]
    &= (\ii k_b)\, (nm\gamma_2^\text{A}\, \dot{\hat{e}}_{\K, ab} )\notag = \ii \K \cdot (nm\gamma_2^\text{A}\, \hat{\dot{\textbf{e}}}_{\K} ) \notag,
\end{split}
\end{align}
where $\hat{\dot{\textbf{e}}}_{\K}$ is the Fourier transform of the real-space strain rate $\dot{\textbf{e}} = \bm\nabla \U$. 
Using the correspondence $\ii \K \to \bm\nabla$, the right hand side of Eq.~(\ref{eq:Term2}) becomes the divergence of the stress under Fourier transformation:
\begin{align}
    -\gamma_2^\text{A} k^2 \Jk \to \bm\nabla\cdot\sigma^{(1)}_{ab}
\end{align}
The right-hand side of Eq.~(\ref{eq:Term2}) gives a linear viscous response:
\begin{align}
    \sigma^{(1)}_{ab} = \eta^{(1)}_{abcd} \dot{e}_{cd}
\end{align}
where the viscosity tensor reads
\begin{align}
    \eta^{(1)}_{abcd} = nm\gamma_2^\text{A}\delta_{ac}\delta_{bd},
\end{align}
\vspace{1mm}

The third term of Eq.~\eqref{eq:gamma-J} may be written as
\begin{align*}
    -\gamma_2^\text{B}  \K\K \Jk &= -\gamma_2^\text{B} \,k_a \, k_b \,\jkb = \gamma_2^\text{B} (\ii k_a) \, (\ii k_b \,\jkb)\notag \\[6pt]
    &= \gamma_2^\text{B} (\ii \K) \, (\ii \K \cdot \Jk) = nm\gamma_2^\text{B} (\ii \K) \, (\ii \K \cdot \hat{\U}(\K)) \notag
\end{align*}
This corresponds to a linear viscous response towards compression
\begin{align}
    -\gamma_2^\text{B}\K\K \Jk \to \xi^{(1)} \bm\nabla (\bm\nabla \cdot \U(\R)),
\end{align}
with bulk viscosity
\begin{align}\label{eq:xi1}
    \xi^{(1)} = nm\gamma_2^\text{B}.
\end{align}
\vspace{1mm}

To summarize, in a wet active fluid involving hydrodynamic interactions, the viscosity tensor contains two parts: (i) $\eta^{(0)}_{abcd}$ associated with reciprocal interactions, still satisfying the Green--Kubo relation Eq.~\eqref{eq:Kubo-hydro}; (ii) Corrections $\eta^{(1)}_{abcd}$ and $\xi^{(1)}$ due to non-reciprocal hydrodynamic interactions. Note that the correction terms can be derived from the hydrodynamic interaction $\Upsilon(\R)$ by using Eq.~\eqref{eq:gamma2}. Furthermore, since $\Upsilon(\R)$ is symmetric, the correction terms do not affect the anti-symmetric components of the viscosity tensor, for instance $\eta^\text{o}$.

\vspace{10mm}

\noindent\textbf{Discussion} \enspace 
In our derivation, we utilize the assumption that the system is symmetric under inversion  $\R \leftrightarrow -\R$ and isotropic. In 2D, such systems, including the chiral active fluids studied in this work, may still violate parity. 
We note, however, that chiral fluids in 3D are necessarily anisotropic. Hence, our conclusions may require modification when applied, for example, to a 3D chiral active fluid with spinners all sharing the same rotation axis. Nonetheless, for active 3D fluids obeying isotropy and inversion symmetry, our results are valid. 
\newline

Furthermore, the derivation assumes the existence of a stable steady state, which allows us to take the hydrodynamic limit $k \to 0$ and $t \to \infty$. 
This is a crucial prerequisite for using the Mori--Zwanzig formalism to derive a generalized linear response.  
The limit $k \to 0$ isolates the long-wavelength hydrodynamic modes as the slow variables. 
The limit $t \to \infty$ provides separation of  timescales so that one can treat the orthogonal forces $\Fo$ as a fast-fluctuating noise and ignore it after averaging over the initial conditions. 
For nonequilibrium dynamics where $\Fo$ and $\textbf{K}$ have comparable timescale, $\Fo$ needs to be considered as well and typically gives rise to a nonlinear response of the system~\cite{Zwanzig2001}.
\newline

To summarize, 
the scope of our derivation can be described by four categories of systems:
(i) Without velocity-dependent interactions (or very weak), the standard Green--Kubo relation holds; (ii) With velocity-dependent interactions but no spinning, an equilibrium-like Green--Kubo relation with a renormalized temperature $T_\text{eff} + B_{\V\V}/k_\text{B}$ still holds; 
(iii) For 2D chiral active fluids, the same Kubo relation holds as long as $|\left<\V(0)\V(\R)\right>_0| < \mathcal{O}(r^{-D})$. The condition on $\left<\V(0)\V(\R)\right>_0$ can be easily evaluated in either simulations or experiments of active fluids in which the motions of individual particles are traceable; (iv) For 3D chiral active fluids with long-range hydrodynamic interactions, the Kubo relation is not guaranteed.

\onecolumngrid
\clearpage
\newpage
\twocolumngrid

\section{Langevin equation of the stress}
\noindent\textbf{Effective Langevin equation of the shear stresses} \enspace 
Figure 3B shows that the shear stress $\textbf{S} = (\protect\so, \protect\st)^\text{T}$ evolves as a 2D random walker with a tendency towards rotation (see Supplementary Mov.~S2). This inspires us to propose a phenomenological model using linear Langevin equations:
\begin{equation} \label{eq:lgv}
\begin{split}
    \frac{d}{dt}\begin{bmatrix} s_1(t) \\[3pt] s_2(t)\end{bmatrix} = -\begin{bmatrix} a & b \\ -b & a\end{bmatrix} \begin{bmatrix} s_1(t) \\[3pt] s_2(t)\end{bmatrix} + C_\textbf{R}\begin{bmatrix} w_1(t) \\[3pt] w_2(t)\end{bmatrix}, 
\end{split}
\end{equation}
where $s_1 = \protect\so$, $s_2 = \protect\st$, $a$ characterizes the relaxation of the fluctuating stress $\textbf{S}$ towards $(0, 0)^\text{T}$ due to shear viscosity $\eta$, $b$ characterizes the chiral response involving odd viscosity $\eta^\text{o}$, and $w_1(t)$ and $w_2(t)$ are two independent white noises. 

Let us first consider two normalized correlation functions:
\begin{align}
    \rho_{11}(t) &= \frac{\left<s_1(t)s_1(0)\right>}{\big<s_1^2(0)\big>},  \\[6pt] 
    \rho_{21}(t) &= \frac{\left<s_2(t)s_1(0)\right>}{\big<s_1^2(0)\big>}.
\end{align}
According to Eq.~(\ref{eq:lgv}), the two correlation functions should obey:
\begin{equation}
    \frac{d}{dt}\begin{bmatrix} \rho_{11}(t) \\[3pt] \rho_{21}(t)\end{bmatrix} = -\begin{bmatrix} a & b \\[3pt] -b & a\end{bmatrix} \begin{bmatrix} \rho_{11}(t) \\[3pt] \rho_{21}(t)\end{bmatrix}.
\end{equation}
Given the initial condition $\rho_{11}(0) = 1$ and $\rho_{21}(0) = 0$, one can derive
\begin{align} \label{eq:scorr11}
    \rho_{11}(t) &= \frac{\left<s_1(t)\,s_1(0)\right>}{\big<s_1^2(0)\big>} = \e^{-at}\text{cos}(bt) \\[6pt]
    \label{eq:scorr21}
    \rho_{21}(t) &= \frac{\left<s_2(t)\,s_1(0)\right>}{\big<s_1^2(0)\big>} = \e^{-at}\text{sin}(bt).
\end{align}

Parameters $a$ and $b$ can be fixed by evaluating the Green--Kubo relation, which we have derived in Sec.~\ref{sec:Kubo}:
\begin{align*}
    \eta &= \frac{A\big<s_1^2(0)\big>}{\kT} \int_0^{\infty} \rho_{11}(t) dt
    = \frac{A \big<s_1^2(0)\big>}{\kT} \frac{a}{a^2 + b^2} \\[6pt]
    -\eta^\text{o} &= \frac{A\big<s_1^2(0)\big>}{\kT} \int_0^{\infty} \rho_{12}(t)dt = \frac{A \big<s_1^2(0)\big>}{\kT} \frac{b}{a^2 + b^2}.
\end{align*}
Solving the above equations, we have
\begin{align}
    a &= \frac{\hat{\eta}}{\hat{\eta}^2 + \hat{\eta}^{\text{o}2}}, & b &= -\frac{\hat{\eta}^\text{o}}{\hat{\eta}^2 + \hat{\eta}^{\text{o}2}}
\end{align}
with the normalized shear and odd viscosities
\begin{align}
    \hat{\eta} &= \frac{\kT}{A\big<s_1^2(0)\big>}\eta,& \shortspace \hat{\eta}^\text{o} &=\frac{\kT}{A\big<s_1^2(0)\big>}\eta^\text{o}.\shortspace
\end{align}
To achieve the steady-state fluctuations, we have to choose a proper noise magnitude $C_\textbf{R}$ so that the following quantity vanishes:
\begin{align}
    &\big<s_1^2(\Delta t)\big> - \big<s_1^2(0)\big> \notag\\[3pt]
    &= \big<\left[s_1(0) -as_1(0)\Delta t - bs_2(0)\Delta t + C_\textbf{R} dw_1\right]^2\big> - \big<s_1^2(0)\big> \notag\\[3pt]
    &= -2a\big<s_1^2(0)\big>\Delta t + C_\textbf{R}^2 \left<dw_1^2\right> \notag\\[3pt]
    &= -2a\big<s_1^2(0)\big>\Delta t + 2C_\textbf{R}^2 \Delta t,
\end{align}
where we use the relation $\left<s_1(0)s_2(0)\right> = \left<s_2(0)s_1(0)\right> = 0$ and only keep the leading order $\mathcal{O}(\Delta t)$.
This implies $C_\textbf{R} = \sqrt{a\big<s_1^2(0)\big>} = \big<s_1^2(0)\big> \sqrt{A/\kT \cdot \eta/(\eta^2 + \eta^{\text{o}2})}$.
Therefore, Eq.~(\ref{eq:lgv}) becomes the Langevin equation that we provided in the main text:
\begin{align*}
    \frac{d}{dt}\begin{bmatrix} s_1(t) \\[6pt] s_2(t)\end{bmatrix} = -C_{\bm{\eta}}\begin{bmatrix} \eta & \eta^\text{o} \\[6pt] -\eta^\text{o} & \eta\end{bmatrix}^{-1} \begin{bmatrix} s_1(t) \\[6pt] s_2(t)\end{bmatrix} + C_\textbf{R}\begin{bmatrix} w_1(t) \\[6pt] w_2(t)\end{bmatrix}
\end{align*}
where $C_{\bm{\eta}} =  \big<s_1^2(0)\big>_0\, A/k_\text{B}T_\text{eff}$. 

\begin{figure*}[ht]
\includegraphics[width=0.96\textwidth]{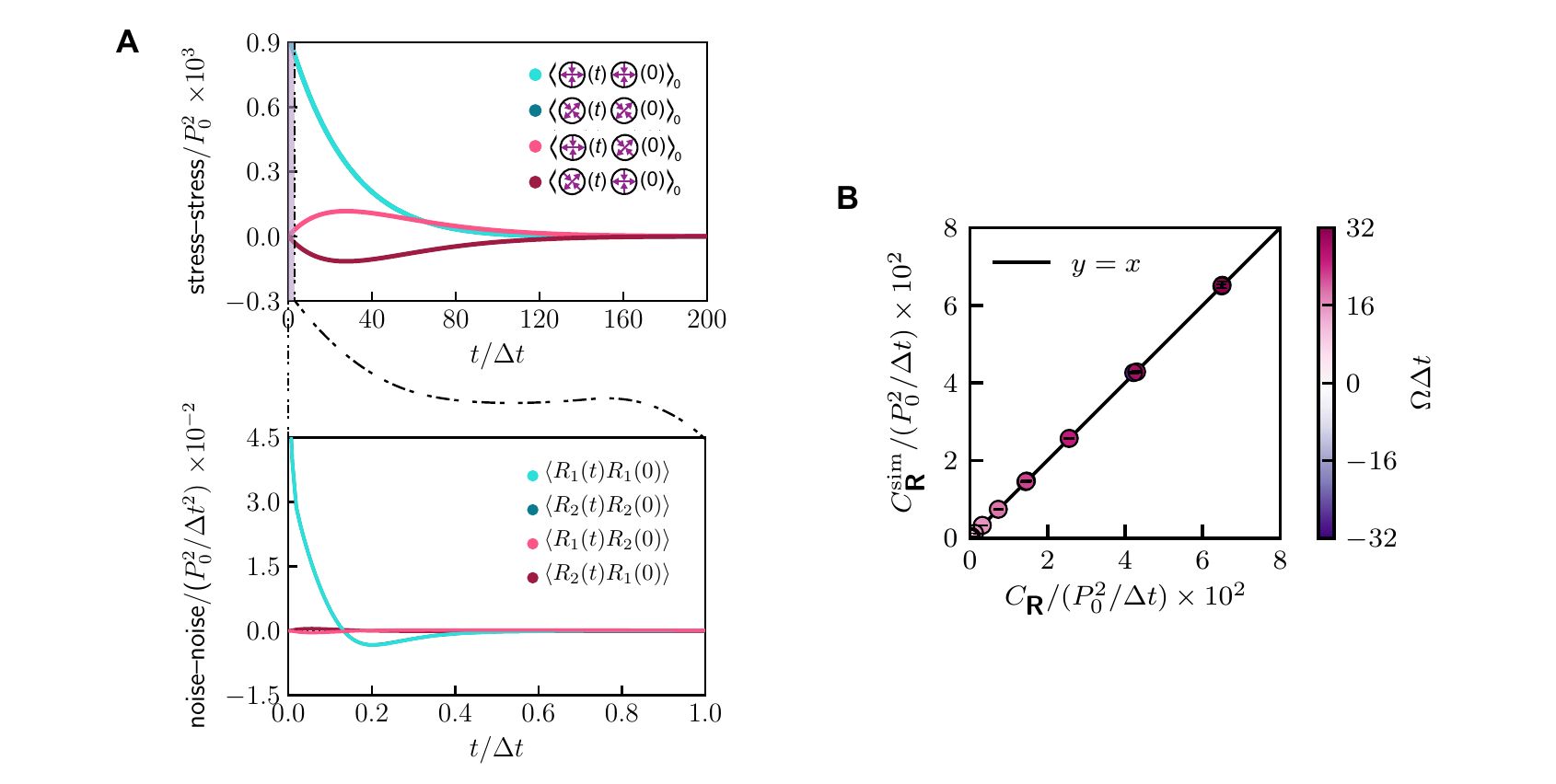}
\caption{\label{fig:noise} \textbf{Validating the Langevin equation.} We perform molecular dynamic simulations at the steady state and record the time evolution of the shear stresses $\textbf{S} = (s_1, s_2)^\text{T}$. We further extract the noise term $\textbf{R} = (R_1, R_2)^\text{T}$ in the Langevin dynamics of the stress using Eq.~\eqref{eq:res}. \textbf{A.} Time-correlation functions of \textbf{S} versus that of  \textbf{R}. First, \textbf{R} decorrelates much faster than  \textbf{S}, with correlation time $\tau_\textbf{R} < \tau_\textbf{S}/200$. Second, the different components of \textbf{R} are barely correlated with each other. Thus the noise term \textbf{R} in actual simulation can be treated as an independent white noise. \textbf{B.} Comparison between the theoretical and measured magnitude of the noise. The noise magnitude measured from the simulations using Eq.~\eqref{eq:R2} agrees well with the theoretical value $C_\textbf{R}$. These evidences provide a strong validation of our Langevin equation.}
\end{figure*}

To validate this effective Langevin equation, we perform molecular dynamics simulations at the steady state and evaluate the following residue terms
\begin{align}\label{eq:res}
   \begin{bmatrix}R_1(t) \\[6pt] R_2(t)\end{bmatrix} = \frac{d}{dt}\begin{bmatrix} s_1(t) \\[6pt] s_2(t)\end{bmatrix} + C_{\bm{\eta}}\begin{bmatrix} \eta & \eta^\text{o} \\[6pt] -\eta^\text{o} & \eta\end{bmatrix}^{-1} \begin{bmatrix} s_1(t) \\[6pt] s_2(t)\end{bmatrix}.
\end{align}
These residue terms decorrelate much faster than the stress (Fig.~\ref{fig:noise}A). 
Thus, they indeed can be treated as independent sources of white noises. 
We further measure the magnitude of the noises as
\begin{equation} \label{eq:R2}
    C_\textbf{R}^\text{sim} = \int_0^\infty \left<R_1(0)R_1(t)\right> dt.
\end{equation}
Remarkably, we find that this measured value $C_\textbf{R}^\text{sim}$ agrees well with what we have derived theoretically $C_\textbf{R} = \big<s_1^2(0)\big> \sqrt{A/\kT \cdot \eta/(\eta^2 + \eta^{\text{o}2})}$.
These findings strongly suggests the validity of our Langevin theory for the stress.

\vspace{10mm}
\noindent \textbf{Frequency dependence} \enspace Our Langevin equation allows us to derive the stress--stress correlation functions, see Eqs.~(\ref{eq:scorr11}--\ref{eq:scorr21}). From the stress-stress correlation functions, we can further predict the frequency dependence of the viscous coefficients by using the Green--Kubo formula:
\begin{equation}
\begin{split}
    \eta_{\alpha\beta}(f) = \frac{A}{\kT} \int_0^\infty \left<\sigma_\alpha(t)\sigma_\beta(0)\right> \e^{-\ii 2\pi ft}dt,
\end{split}
\end{equation}
where $f$ denotes the deformation frequency. 

Here we find the analytical form of the frequency-dependent shear and odd viscosities:
\begin{align}
    &\eta(f) = \frac{A}{\kT} \int_0^\infty \left<s_1(t)\,s_1(0)\right> \e^{-\ii 2\pi ft}dt \notag\\
    &= \frac{A\big<s_1^2(0)\big>}{\kT} \int_0^\infty \e^{-at} \text{cos}(bt) \e^{-\ii 2\pi ft}dt \notag\\
    &= C_{\bm{\eta}}
    \frac{a+\ii 2 \pi f}{b^2 + (a+\ii 2 \pi f)^2}  \\[6pt]
    &= \frac{\eta^2 + \eta^{\text{o}2}}{2}\Big[\frac{1}{(1+\ii 2 \pi f t_\text{c})\eta - \ii \eta^\text{o}} + \frac{1}{(1+\ii 2 \pi f t_\text{c})\eta + \ii \eta^\text{o}}\Big] \notag
\end{align}
\begin{align}
    &\eta^{\text{o}}(f) = -\frac{A}{\kT} \int_0^\infty \left<s_2(t)\,s_1(0)\right> \e^{2\pi ft}dt \notag\\
    &= -\frac{A\big<s_1^2(0)\big>}{\kT} \int_0^\infty \e^{-at} \text{sin}(bt) \e^{-\ii 2\pi ft}dt \notag \\
    &= -C_{\bm{\eta}}
    \cdot \frac{b}{b^2 + (a+\ii 2 \pi f)^2} \\[6pt]
    &= \frac{\eta^2 + \eta^{\text{o}2}}{2\ii}\Big[\frac{1}{(1+\ii 2 \pi f t_\text{c})\eta - \ii \eta^\text{o}}  - \frac{1}{(1+\ii 2 \pi f t_\text{c})\eta + \ii \eta^\text{o}}\Big] \notag
\end{align}
where $t_\text{c} = (\eta^2 + \eta^{\text{o}2})/\eta\, C_{\bm\eta}$ is the characteristic time.
In the main text, we have demonstrated the excellent agreement between this theoretical prediction and the simulation measurement (Fig.~3D-E).

This frequency dependence has a kinetic origin. The characteristic frequency is set by the inverse of the tumbling time $\Delta t_\text{tumble}$ required by a particle to randomize its velocity through interparticle collisions. 
Using the simulation data on the particle interactions, we can measure the average velocity reduction in its original (incoming) moving direction after each collision
\begin{align}
    \overline{\Delta v} = \left<\Delta v^\text{s}_x\right>_0,
\end{align}
where $\Delta v_x$ is the $x$-component of the symmetrized velocity change $\Delta \V^\text{s}$ (see Fig.~\ref{fig:odd-viscosity-dV}).
To completely eliminate the correlation between its current and original velocity, a particle needs to collide $N_\text{col}$ times, where 
\begin{align}
    N_\text{col} \approx \bar{v}/\overline{\Delta v}.
\end{align}
The average waiting time for a new collision is
\begin{align}
    t_\text{wait} = \Delta t + \Delta t_\text{col},
\end{align}
where $\Delta t$ is the collision duration and $\Delta t_\text{col}$ is the travelling time between two adjacent collisions.
Therefore, we can estimate the tumbling time as
\begin{align}
    t_\text{tumble} = (\Delta t + \Delta t_\text{col}) \cdot \bar{v}/\overline{\Delta v}.
\end{align}
Consistently, as shown in Fig.~3D-E of the main text, both the characteristic frequency of $\eta(f)$ and that of $\eta^\text{o}(f)$ are of order $t_\text{tumble}^{-1}$. 

\onecolumngrid
\clearpage
\newpage
\twocolumngrid

\section{Hydrodynamics} \label{secHydro}

Here we verify that the transport coefficients obtained from microscopic measurements and statistical mechanics calculations do capture hydrodynamic phenomena at the macroscopic level. To do so, we compare large-scale molecular dynamics simulations of a shock wave as well as a steady-state flow with the corresponding  predictions from the hydrodynamic theory:
\begin{equation}\label{eq:hydro}
    \begin{cases}
    D_t \rho = -\rho \nabla \cdot \U, \\[9pt]
    \rho D_t \U = \div{\bm{\sigma}_\text{ss}} + \xi \, \bm\nabla \, (\div{\bm{u}}) +\, \eta \, \Delta \bm{u} \\[3pt]
    \longspace\,
    +\, \eta^\text{o} \, \mathcal{R} \cdot \Delta\bm{u} - \, \bm\nabla \times (\Gamma(\rho) \bm\nabla \times \bm{u}), 
   \end{cases}
\end{equation}
with the steady-state stress 
\begin{equation}
    \bm{\sigma}_\text{ss} = -P \, \mathcal{I} + \tau \, \mathcal{R},
\end{equation}
which is set by the equations of the states, pressure $P = c^2\rho$ ($c = \sqrt{\kT/m}$ is the speed of sound) and anti-symmetric stress $\tau = \Gamma(\rho)\Omega$.
All the parameters $\eta$, $\eta^\text{o}$,  $\xi$, $\Gamma$, and $c$ are obtained from microscopic measurements. 
Since all the particles are forced to rotate at a constant speed $\Omega$, the rotation field $\Omega(\R)$ is not included in this hydrodynamic theory but treated as an adjustable parameter.
Compared to the general form of the Navier--Stokes equation Eq.~\eqref{explicit_navier_stokes}, we have ignored the terms involving compression--rotation viscosities $\eta^\text{A}$ and $\eta^\text{B}$ that are negligible compared to the other viscous coefficients (see Fig.~2C in the main text and Fig.~\ref{fig:rotB-viscosity}). But we allow the rotation viscosity $\Gamma(\rho)$ to depend on density. Note that the mass density $\rho = mn$. 

\vspace{10mm}

\noindent \textbf{Shock wave} \enspace
In the main text, Fig.~4 shows the simulation of a shock wave generated by a fast moving piston. The system has initial particle number density $n_0 = 0.125d^{-2}$. Here we solve the hydrodynamic equation Eq.~\eqref{eq:hydro} numerically in the co-moving frame of the shock. 
For simplicity, we use the coefficients $\eta$, $\eta^\text{o}$,  $\xi$, and $c$, which were previously measured at a different density $n_\text{m} = 2n_0$.
Nonetheless, the numerical solutions of $\rho(x)$, $u_x(x)$, $u_y(x)$ using these coefficients agree well with the simulation measurements, until local particle density becomes too high $n(x) > 3n_0$ (see Fig.~4B-D in the main text).

\begin{figure}[h]
\includegraphics[width=0.52\textwidth]{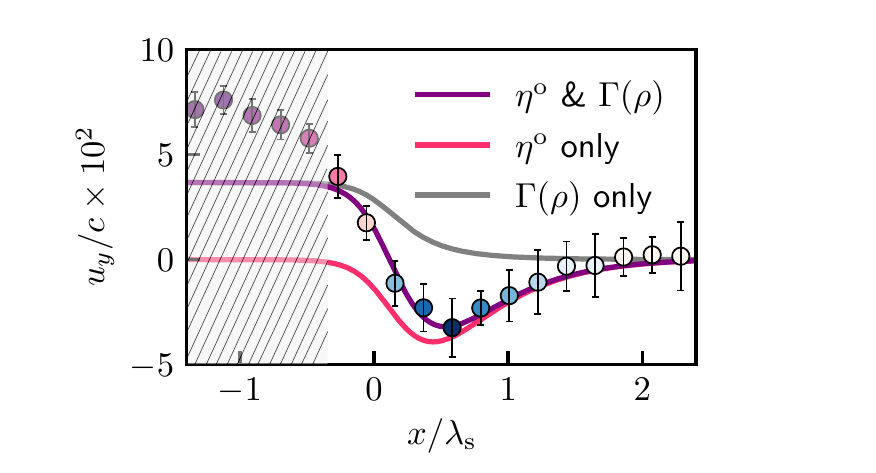}
\caption{\label{fig:shock} \textbf{Transverse flow in a shock.} Here we investigate the role of odd viscosity $\eta^\text{o}$ and density-dependent rotation viscosity $\Gamma(\rho)$ in causing the shear flow perpendicular to the shock. We solve Eq.~\eqref{eq:hydro} in three different cases: (i) both the coefficients are included; (ii) only $\eta^\text{o}$ is included;  (iii) only $\Gamma(\rho)$ is included. Note that the molecular dynamic simulations (dotted data) are performed at initial particle number density $n_0 = 0.125 d^{-2}$. The parameters for the hydrodynamic theory was previously measured at density $n_\text{m} = 2n_0$. When local density becomes too dense $n(x) > 3n_0$, such hydrodynamic prediction breaks down (see the shaded region).}
\end{figure}

A common shock wave only displays longitudinal modes. However, our shock wave is accompanied by a shear flow in the transverse direction near the wavefront (Fig.~4A, Supplementary Mov.~S3). This transverse shear flow originates from the $\eta^\text{o}$ and $\Gamma(\rho)$ terms, which are disallowed in traditional, achiral fluids. 
To investigate the roles of $\eta^\text{o}$ and $\Gamma(\rho)$, we numerically solve Eq.~\eqref{eq:hydro} in three different cases: (i) both the coefficients are included; (ii) only $\eta^\text{o}$ is included; (iii) only $\Gamma(\rho)$ is included.
As illustrated in Fig.~\ref{fig:shock}, $\eta^\text{o}$ gives rise to a dip near the wavefront (red line), whereas $\Gamma(\rho)$ remains monotonic even further away from the wavefront (grey line). 
The locations of these features can be understood by dimensional analysis. 
In the Navier--Stokes equation, odd viscosity appears in the term $\eta^\text{o} \Delta \U$ that involves a second spatial derivative, whereas rotation viscosity appears in the term $\bm\nabla \cdot (\Gamma(\rho)(\Omega - \omega) \,\mathcal{R})$ that involves a first spatial derivative. 
Therefore, the impact of $\eta^\text{o}$ shows up in a smaller length scale, explaining why the resulting dip is closer to the wavefront.

Considering $|u_y| \ll |u_x|$ (see Fig.~4C-D in main text), we assume that the transverse shear flow barely affects the longitudinal propagation of the shock. Thus, we can use the standard viscid Burgers' equation to estimate the width of the shock~\cite{burgers1948mathematical}:
\begin{align}
    \frac{\partial}{\partial t} u_x + u_x \frac{\partial}{\partial x} u_x = \nu \frac{\partial^2}{\partial x^2} u_x,
\end{align}
where $\nu = \eta/n_0m$ is the kinematic viscosity.
For the shock generated by a piston moving at speed $U$, the analytical solution of the Burgers' equation is 
\begin{align}
    u_x(x, t) = \frac{U}2 - \frac{U}2 \text{tanh}\Big[\frac{(x - Ut/2)\,U}{4\nu}\Big].
\end{align}
This suggests the width of the shock is
\begin{align}
    \lambda_\text{s} = \frac{4\nu}{U},
\end{align}
consistent with what we observe in Fig.~\ref{fig:shock}.

\vspace{25mm}

\begin{figure}
\includegraphics[width=0.5\textwidth]{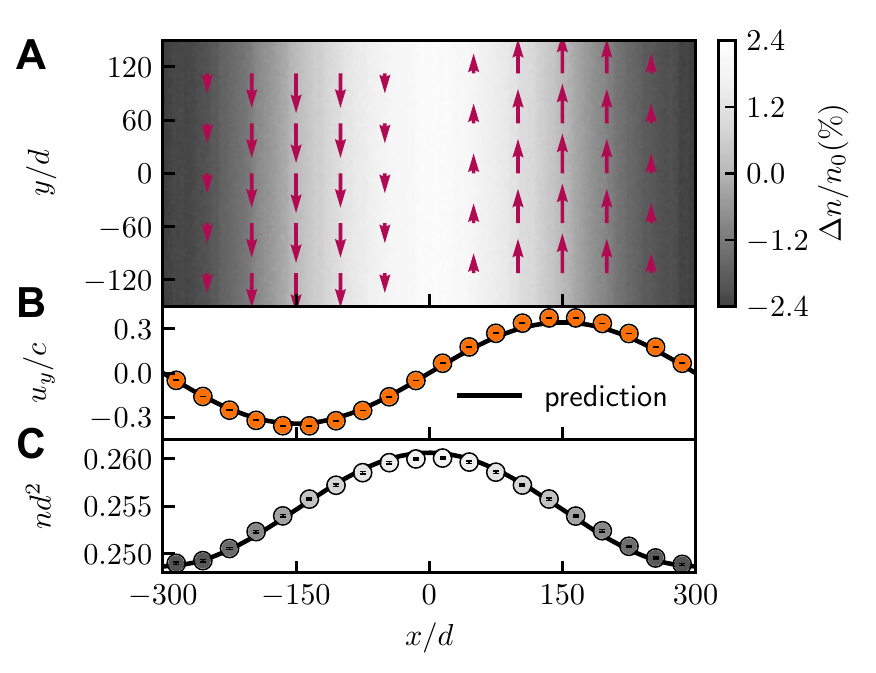}
\caption{\label{fig:shear} \textbf{Compression mode in a steady-state shear flow.} \textbf{A.} 2D density profile obtained from molecular dynamic simulation. 
In the simulation, a force field $\textbf{F} = (0,\,F_0 \text{sin}(2\pi x/L_x))$ of magnitude $F_0 = 2.6 \times 10^{-3} md^2/\Delta t^2$ (see red arrows) is applied to the particles, in order to drive a $y$-direction steady-state shear flow. The system is periodic on both $x$ and $y$-directions, having size $L_x = 600d$ and $L_y = 300d$. The global density is $n_0 = 0.25d^{-2}$ . The time-averaged density profile $\Delta n(\R)/n_0$ , where $\Delta n(\R) = n(\R) - n_0$, is color-coded in grey scale. The shear flow leads to a density modulation in the transverse direction. \textbf{B.} Flow profile $u_y(x)$. \textbf{C.} Density profile $n(x)$. The analytical solution Eq.~(\ref{eq:steady-shear1}-\ref{eq:steady-shear2}) derived from the hydrodynamic theory (solid line) matches with the simulation measurement (points).}
\end{figure}

\noindent \textbf{Steady-state shear flow} \enspace 
The simulation of the shock shows that a compression wave can lead to shear flow in the transverse direction via odd viscosity $\eta^\text{o}$. 
Here we investigate whether a shear flow can in turn also lead to compression via $\eta^\text{o}$.
To do so, we first perform molecular dynamics simulations of a steady-state shear flow.
A force field $\textbf{F} = (0,\, F_0 \text{sin}(2\pi x/L_x))$ (red arrows in Fig.~\ref{fig:shear}A) is applied to the particles to drive a vertical shear flow.

We find that this shear flow leads to a density modulation in the $x$-direction (Fig.~\ref{fig:shear}A).
However, this density modulation is rather small, with magnitude up to only $2.5\%$.
Thus we can ignore the secondary effects, i.e. the correction to the shear flow due to $\Gamma(n)$.
In this approximation, we can reduce the Navier--Stokes equations Eq.~\eqref{eq:hydro} to two stationary equations for $u_y(x)$ and $n(x)$: 
\begin{align}
    -mc^2 \frac{\partial}{\partial x} n(x) + \eta^\text{o} \frac{\partial^2}{\partial x^2} u_y(x) &= 0,\\
    \eta \frac{\partial^2}{\partial x^2} u_y(x) + nF_y(x) &= 0.
\end{align}
Given that $F_y(x) = F_0 \text{sin}(2\pi x/L_x)$, the analytical solution of the above equations read
\begin{align}
    u_y(x) &= \frac{nF_0L_x^2}{4\pi^2\eta} \text{sin}(2\pi x/L_x), \label{eq:steady-shear1}\\
    n(x) &= n_0+ \frac{nF_0L_x}{2\pi mc^2} \cdot \frac{\eta^\text{o}}{\eta}\cdot \text{cos}(2\pi x/L_x). \label{eq:steady-shear2}
\end{align}
As illustrated in Fig.~\ref{fig:shear}B-C, this analytical solution agrees well with the simulation measurement, further validating the hydrodynamic theory.
Given Eq.~\eqref{eq:steady-shear2}, this steady-state shear provides an additional probe to directly measure odd viscosity.

\clearpage
\newpage
\twocolumngrid

\section{Supplementary movies}

\noindent MOV.~S1. \textbf{Interparticle collision.} Two particles both spinning at speed $\Omega = 9.5 / \Delta t$ are set to collide with a relative velocity $v_\text{rel} = 0.63 d/\Delta t$. Although the particles are perfectly aligned to undergo a head-to-head collision, due to the presence of self-spinning and interparticle friction, they gain transverse motion after the collision.

\vspace{10mm}

\noindent MOV.~S2. \textbf{Chiral Brownian motion in shear-stress space.} To illustrate the steady-state fluctuations, we plot the shear stresses $\protect\so$ and $\protect\st$ of an unperturbed system against each other. In the system, all the particles spin counter-clockwise at speed  $\Omega = 26.7/\Delta t$ and have global density $n_0 = 0.25 d^{-2}$. Over time, the stress vector $\textbf{S} = (\protect\so, \protect\st)^\text{T}$ exhibits a chiral Brownian motion, confined near the origin $(0, 0)^\text{T}$ and having a tendency towards a clockwise rotation. To illustrate such chiral rotation, we plot the normalized stress $\hat{\textbf{S}} = \textbf{S}/||\textbf{S}||$ and mark its polar angle $\theta$. The winding number $n_\text{w}(t) = [\theta(t) - \theta(0)]/2\pi$ clearly shows the tendency of the stress vector $\textbf{S}$ to perform a clockwise ration. 

\vspace{10mm}

\noindent MOV.~S3. \textbf{Shock wave.} A piston moving at speed $v = 1.4 c$, where $c = 1.4 d/\Delta t$ is the speed of sound, generates a shock wave propagating from left to right in our chiral active fluid. Here the particles spin counter-clockwise at speed $\Omega = 26.7/\Delta t$ and have initial global density $n_0 = 0.125 d^{-2}$. To demonstrate the resultant shear flow in the transverse direction, we color-code the fluid according to the y-component of the flow velocity $u_y$. To characterize the shock wave, we further show the density profile $\rho(x)$ as well as the flow profile $u_x(x)$ and $u_y(x)$.

\onecolumngrid
\clearpage
\newpage
\twocolumngrid

\bibliography{active_spinners.bib}

\begin{thebibliography}{66}%
\makeatletter
\providecommand \@ifxundefined [1]{%
 \@ifx{#1\undefined}
}%
\providecommand \@ifnum [1]{%
 \ifnum #1\expandafter \@firstoftwo
 \else \expandafter \@secondoftwo
 \fi
}%
\providecommand \@ifx [1]{%
 \ifx #1\expandafter \@firstoftwo
 \else \expandafter \@secondoftwo
 \fi
}%
\providecommand \natexlab [1]{#1}%
\providecommand \enquote  [1]{``#1''}%
\providecommand \bibnamefont  [1]{#1}%
\providecommand \bibfnamefont [1]{#1}%
\providecommand \citenamefont [1]{#1}%
\providecommand \href@noop [0]{\@secondoftwo}%
\providecommand \href [0]{\begingroup \@sanitize@url \@href}%
\providecommand \@href[1]{\@@startlink{#1}\@@href}%
\providecommand \@@href[1]{\endgroup#1\@@endlink}%
\providecommand \@sanitize@url [0]{\catcode `\\12\catcode `\$12\catcode
  `\&12\catcode `\#12\catcode `\^12\catcode `\_12\catcode `\%12\relax}%
\providecommand \@@startlink[1]{}%
\providecommand \@@endlink[0]{}%
\providecommand \url  [0]{\begingroup\@sanitize@url \@url }%
\providecommand \@url [1]{\endgroup\@href {#1}{\urlprefix }}%
\providecommand \urlprefix  [0]{URL }%
\providecommand \Eprint [0]{\href }%
\providecommand \doibase [0]{http://dx.doi.org/}%
\providecommand \selectlanguage [0]{\@gobble}%
\providecommand \bibinfo  [0]{\@secondoftwo}%
\providecommand \bibfield  [0]{\@secondoftwo}%
\providecommand \translation [1]{[#1]}%
\providecommand \BibitemOpen [0]{}%
\providecommand \bibitemStop [0]{}%
\providecommand \bibitemNoStop [0]{.\EOS\space}%
\providecommand \EOS [0]{\spacefactor3000\relax}%
\providecommand \BibitemShut  [1]{\csname bibitem#1\endcsname}%
\let\auto@bib@innerbib\@empty
\bibitem [{\citenamefont {Kubo}(1966)}]{kubo1966fluctuation}%
  \BibitemOpen
  \bibfield  {author} {\bibinfo {author} {\bibfnamefont {R.}~\bibnamefont
  {Kubo}},\ }\href@noop {} {\bibfield  {journal} {\bibinfo  {journal} {Reports
  on progress in physics}\ }\textbf {\bibinfo {volume} {29}},\ \bibinfo {pages}
  {255} (\bibinfo {year} {1966})}\BibitemShut {NoStop}%
\bibitem [{\citenamefont {Kurchan}(2005)}]{Kurchan2005}%
  \BibitemOpen
  \bibfield  {author} {\bibinfo {author} {\bibfnamefont {J.}~\bibnamefont
  {Kurchan}},\ }\href {\doibase 10.1038/nature03278} {\bibfield  {journal}
  {\bibinfo  {journal} {Nature}\ }\textbf {\bibinfo {volume} {433}},\ \bibinfo
  {pages} {222} (\bibinfo {year} {2005})}\BibitemShut {NoStop}%
\bibitem [{\citenamefont {Ciliberto}\ \emph {et~al.}(2010)\citenamefont
  {Ciliberto}, \citenamefont {Joubaud},\ and\ \citenamefont
  {Petrosyan}}]{Ciliberto2010}%
  \BibitemOpen
  \bibfield  {author} {\bibinfo {author} {\bibfnamefont {S.}~\bibnamefont
  {Ciliberto}}, \bibinfo {author} {\bibfnamefont {S.}~\bibnamefont {Joubaud}},
  \ and\ \bibinfo {author} {\bibfnamefont {A.}~\bibnamefont {Petrosyan}},\
  }\href {\doibase 10.1088/1742-5468/2010/12/p12003} {\bibfield  {journal}
  {\bibinfo  {journal} {Journal of Statistical Mechanics: Theory and
  Experiment}\ }\textbf {\bibinfo {volume} {2010}},\ \bibinfo {pages} {P12003}
  (\bibinfo {year} {2010})}\BibitemShut {NoStop}%
\bibitem [{\citenamefont {Cugliandolo}(2011)}]{Cugliandolo2011}%
  \BibitemOpen
  \bibfield  {author} {\bibinfo {author} {\bibfnamefont {L.~F.}\ \bibnamefont
  {Cugliandolo}},\ }\href {\doibase 10.1088/1751-8113/44/48/483001} {\bibfield
  {journal} {\bibinfo  {journal} {Journal of Physics A: Mathematical and
  Theoretical}\ }\textbf {\bibinfo {volume} {44}},\ \bibinfo {pages} {483001}
  (\bibinfo {year} {2011})}\BibitemShut {NoStop}%
\bibitem [{\citenamefont {Seifert}(2012)}]{Seifert2012}%
  \BibitemOpen
  \bibfield  {author} {\bibinfo {author} {\bibfnamefont {U.}~\bibnamefont
  {Seifert}},\ }\href {\doibase 10.1088/0034-4885/75/12/126001} {\bibfield
  {journal} {\bibinfo  {journal} {Reports on Progress in Physics}\ }\textbf
  {\bibinfo {volume} {75}},\ \bibinfo {pages} {126001} (\bibinfo {year}
  {2012})}\BibitemShut {NoStop}%
\bibitem [{\citenamefont {Makse}\ and\ \citenamefont
  {Kurchan}(2002)}]{Makse2002}%
  \BibitemOpen
  \bibfield  {author} {\bibinfo {author} {\bibfnamefont {H.~A.}\ \bibnamefont
  {Makse}}\ and\ \bibinfo {author} {\bibfnamefont {J.}~\bibnamefont
  {Kurchan}},\ }\href {\doibase 10.1038/415614a} {\bibfield  {journal}
  {\bibinfo  {journal} {Nature}\ }\textbf {\bibinfo {volume} {415}},\ \bibinfo
  {pages} {614} (\bibinfo {year} {2002})}\BibitemShut {NoStop}%
\bibitem [{\citenamefont {D'Anna}\ \emph {et~al.}(2003)\citenamefont {D'Anna},
  \citenamefont {Mayor}, \citenamefont {Barrat}, \citenamefont {Loreto},\ and\
  \citenamefont {Nori}}]{DAnna2003}%
  \BibitemOpen
  \bibfield  {author} {\bibinfo {author} {\bibfnamefont {G.}~\bibnamefont
  {D'Anna}}, \bibinfo {author} {\bibfnamefont {P.}~\bibnamefont {Mayor}},
  \bibinfo {author} {\bibfnamefont {A.}~\bibnamefont {Barrat}}, \bibinfo
  {author} {\bibfnamefont {V.}~\bibnamefont {Loreto}}, \ and\ \bibinfo {author}
  {\bibfnamefont {F.}~\bibnamefont {Nori}},\ }\href {\doibase
  10.1038/nature01867} {\bibfield  {journal} {\bibinfo  {journal} {Nature}\
  }\textbf {\bibinfo {volume} {424}},\ \bibinfo {pages} {909} (\bibinfo {year}
  {2003})}\BibitemShut {NoStop}%
\bibitem [{\citenamefont {Ojha}\ \emph {et~al.}(2004)\citenamefont {Ojha},
  \citenamefont {Lemieux}, \citenamefont {Dixon}, \citenamefont {Liu},\ and\
  \citenamefont {Durian}}]{ojha2004statistical}%
  \BibitemOpen
  \bibfield  {author} {\bibinfo {author} {\bibfnamefont {R.}~\bibnamefont
  {Ojha}}, \bibinfo {author} {\bibfnamefont {P.-A.}\ \bibnamefont {Lemieux}},
  \bibinfo {author} {\bibfnamefont {P.}~\bibnamefont {Dixon}}, \bibinfo
  {author} {\bibfnamefont {A.}~\bibnamefont {Liu}}, \ and\ \bibinfo {author}
  {\bibfnamefont {D.}~\bibnamefont {Durian}},\ }\href@noop {} {\bibfield
  {journal} {\bibinfo  {journal} {Nature}\ }\textbf {\bibinfo {volume} {427}},\
  \bibinfo {pages} {521} (\bibinfo {year} {2004})}\BibitemShut {NoStop}%
\bibitem [{\citenamefont {Drescher}\ \emph {et~al.}(2009)\citenamefont
  {Drescher}, \citenamefont {Leptos}, \citenamefont {Tuval}, \citenamefont
  {Ishikawa}, \citenamefont {Pedley},\ and\ \citenamefont
  {Goldstein}}]{drescher2009dancing}%
  \BibitemOpen
  \bibfield  {author} {\bibinfo {author} {\bibfnamefont {K.}~\bibnamefont
  {Drescher}}, \bibinfo {author} {\bibfnamefont {K.~C.}\ \bibnamefont
  {Leptos}}, \bibinfo {author} {\bibfnamefont {I.}~\bibnamefont {Tuval}},
  \bibinfo {author} {\bibfnamefont {T.}~\bibnamefont {Ishikawa}}, \bibinfo
  {author} {\bibfnamefont {T.~J.}\ \bibnamefont {Pedley}}, \ and\ \bibinfo
  {author} {\bibfnamefont {R.~E.}\ \bibnamefont {Goldstein}},\ }\href@noop {}
  {\bibfield  {journal} {\bibinfo  {journal} {Physical Review Letters}\
  }\textbf {\bibinfo {volume} {102}},\ \bibinfo {pages} {168101} (\bibinfo
  {year} {2009})}\BibitemShut {NoStop}%
\bibitem [{\citenamefont {F{\"u}rthauer}\ \emph {et~al.}(2012)\citenamefont
  {F{\"u}rthauer}, \citenamefont {Strempel}, \citenamefont {Grill},\ and\
  \citenamefont {J{\"u}licher}}]{furthauer2012active}%
  \BibitemOpen
  \bibfield  {author} {\bibinfo {author} {\bibfnamefont {S.}~\bibnamefont
  {F{\"u}rthauer}}, \bibinfo {author} {\bibfnamefont {M.}~\bibnamefont
  {Strempel}}, \bibinfo {author} {\bibfnamefont {S.~W.}\ \bibnamefont {Grill}},
  \ and\ \bibinfo {author} {\bibfnamefont {F.}~\bibnamefont {J{\"u}licher}},\
  }\href@noop {} {\bibfield  {journal} {\bibinfo  {journal} {The European
  physical journal E}\ }\textbf {\bibinfo {volume} {35}},\ \bibinfo {pages} {1}
  (\bibinfo {year} {2012})}\BibitemShut {NoStop}%
\bibitem [{\citenamefont {Nguyen}\ \emph {et~al.}(2014)\citenamefont {Nguyen},
  \citenamefont {Klotsa}, \citenamefont {Engel},\ and\ \citenamefont
  {Glotzer}}]{nguyen2014emergent}%
  \BibitemOpen
  \bibfield  {author} {\bibinfo {author} {\bibfnamefont {N.~H.}\ \bibnamefont
  {Nguyen}}, \bibinfo {author} {\bibfnamefont {D.}~\bibnamefont {Klotsa}},
  \bibinfo {author} {\bibfnamefont {M.}~\bibnamefont {Engel}}, \ and\ \bibinfo
  {author} {\bibfnamefont {S.~C.}\ \bibnamefont {Glotzer}},\ }\href@noop {}
  {\bibfield  {journal} {\bibinfo  {journal} {Physical review letters}\
  }\textbf {\bibinfo {volume} {112}},\ \bibinfo {pages} {075701} (\bibinfo
  {year} {2014})}\BibitemShut {NoStop}%
\bibitem [{\citenamefont {Petroff}\ \emph {et~al.}(2015)\citenamefont
  {Petroff}, \citenamefont {Wu},\ and\ \citenamefont
  {Libchaber}}]{petroff2015fast}%
  \BibitemOpen
  \bibfield  {author} {\bibinfo {author} {\bibfnamefont {A.~P.}\ \bibnamefont
  {Petroff}}, \bibinfo {author} {\bibfnamefont {X.-L.}\ \bibnamefont {Wu}}, \
  and\ \bibinfo {author} {\bibfnamefont {A.}~\bibnamefont {Libchaber}},\
  }\href@noop {} {\bibfield  {journal} {\bibinfo  {journal} {Physical review
  letters}\ }\textbf {\bibinfo {volume} {114}},\ \bibinfo {pages} {158102}
  (\bibinfo {year} {2015})}\BibitemShut {NoStop}%
\bibitem [{\citenamefont {Kokot}\ and\ \citenamefont
  {Snezhko}(2018)}]{kokot2018manipulation}%
  \BibitemOpen
  \bibfield  {author} {\bibinfo {author} {\bibfnamefont {G.}~\bibnamefont
  {Kokot}}\ and\ \bibinfo {author} {\bibfnamefont {A.}~\bibnamefont
  {Snezhko}},\ }\href@noop {} {\bibfield  {journal} {\bibinfo  {journal}
  {Nature communications}\ }\textbf {\bibinfo {volume} {9}},\ \bibinfo {pages}
  {2344} (\bibinfo {year} {2018})}\BibitemShut {NoStop}%
\bibitem [{\citenamefont {Tsai}\ \emph {et~al.}(2005)\citenamefont {Tsai},
  \citenamefont {Ye}, \citenamefont {Rodriguez}, \citenamefont {Gollub},\ and\
  \citenamefont {Lubensky}}]{tsai2005chiral}%
  \BibitemOpen
  \bibfield  {author} {\bibinfo {author} {\bibfnamefont {J.-C.}\ \bibnamefont
  {Tsai}}, \bibinfo {author} {\bibfnamefont {F.}~\bibnamefont {Ye}}, \bibinfo
  {author} {\bibfnamefont {J.}~\bibnamefont {Rodriguez}}, \bibinfo {author}
  {\bibfnamefont {J.~P.}\ \bibnamefont {Gollub}}, \ and\ \bibinfo {author}
  {\bibfnamefont {T.}~\bibnamefont {Lubensky}},\ }\href@noop {} {\bibfield
  {journal} {\bibinfo  {journal} {Physical review letters}\ }\textbf {\bibinfo
  {volume} {94}},\ \bibinfo {pages} {214301} (\bibinfo {year}
  {2005})}\BibitemShut {NoStop}%
\bibitem [{\citenamefont {Scholz}\ \emph {et~al.}(2018)\citenamefont {Scholz},
  \citenamefont {Engel},\ and\ \citenamefont
  {P{\"o}schel}}]{scholz2018rotating}%
  \BibitemOpen
  \bibfield  {author} {\bibinfo {author} {\bibfnamefont {C.}~\bibnamefont
  {Scholz}}, \bibinfo {author} {\bibfnamefont {M.}~\bibnamefont {Engel}}, \
  and\ \bibinfo {author} {\bibfnamefont {T.}~\bibnamefont {P{\"o}schel}},\
  }\href@noop {} {\bibfield  {journal} {\bibinfo  {journal} {Nature
  communications}\ }\textbf {\bibinfo {volume} {9}},\ \bibinfo {pages} {931}
  (\bibinfo {year} {2018})}\BibitemShut {NoStop}%
\bibitem [{\citenamefont {Avron}(1998)}]{avron1998odd}%
  \BibitemOpen
  \bibfield  {author} {\bibinfo {author} {\bibfnamefont {J.}~\bibnamefont
  {Avron}},\ }\href@noop {} {\bibfield  {journal} {\bibinfo  {journal} {Journal
  of statistical physics}\ }\textbf {\bibinfo {volume} {92}},\ \bibinfo {pages}
  {543} (\bibinfo {year} {1998})}\BibitemShut {NoStop}%
\bibitem [{\citenamefont {De~Groot}\ and\ \citenamefont
  {Mazur}(2013{\natexlab{a}})}]{de2013non}%
  \BibitemOpen
  \bibfield  {author} {\bibinfo {author} {\bibfnamefont {S.~R.}\ \bibnamefont
  {De~Groot}}\ and\ \bibinfo {author} {\bibfnamefont {P.}~\bibnamefont
  {Mazur}},\ }\href@noop {} {\emph {\bibinfo {title} {Non-equilibrium
  thermodynamics}}}\ (\bibinfo  {publisher} {Courier Corporation},\ \bibinfo
  {year} {2013})\BibitemShut {NoStop}%
\bibitem [{\citenamefont {Banerjee}\ \emph {et~al.}(2017)\citenamefont
  {Banerjee}, \citenamefont {Souslov}, \citenamefont {Abanov},\ and\
  \citenamefont {Vitelli}}]{banerjee2017odd}%
  \BibitemOpen
  \bibfield  {author} {\bibinfo {author} {\bibfnamefont {D.}~\bibnamefont
  {Banerjee}}, \bibinfo {author} {\bibfnamefont {A.}~\bibnamefont {Souslov}},
  \bibinfo {author} {\bibfnamefont {A.~G.}\ \bibnamefont {Abanov}}, \ and\
  \bibinfo {author} {\bibfnamefont {V.}~\bibnamefont {Vitelli}},\ }\href@noop
  {} {\bibfield  {journal} {\bibinfo  {journal} {Nature communications}\
  }\textbf {\bibinfo {volume} {8}},\ \bibinfo {pages} {1573} (\bibinfo {year}
  {2017})}\BibitemShut {NoStop}%
\bibitem [{\citenamefont {Souslov}\ \emph {et~al.}(2019)\citenamefont
  {Souslov}, \citenamefont {Dasbiswas}, \citenamefont {Fruchart}, \citenamefont
  {Vaikuntanathan},\ and\ \citenamefont {Vitelli}}]{souslov2019topological}%
  \BibitemOpen
  \bibfield  {author} {\bibinfo {author} {\bibfnamefont {A.}~\bibnamefont
  {Souslov}}, \bibinfo {author} {\bibfnamefont {K.}~\bibnamefont {Dasbiswas}},
  \bibinfo {author} {\bibfnamefont {M.}~\bibnamefont {Fruchart}}, \bibinfo
  {author} {\bibfnamefont {S.}~\bibnamefont {Vaikuntanathan}}, \ and\ \bibinfo
  {author} {\bibfnamefont {V.}~\bibnamefont {Vitelli}},\ }\href@noop {}
  {\bibfield  {journal} {\bibinfo  {journal} {Physical review letters}\
  }\textbf {\bibinfo {volume} {122}},\ \bibinfo {pages} {128001} (\bibinfo
  {year} {2019})}\BibitemShut {NoStop}%
\bibitem [{\citenamefont {Liao}\ \emph {et~al.}(2019)\citenamefont {Liao},
  \citenamefont {Han}, \citenamefont {Fruchart}, \citenamefont {Vitelli},\ and\
  \citenamefont {Vaikuntanathan}}]{liao2019mechanism}%
  \BibitemOpen
  \bibfield  {author} {\bibinfo {author} {\bibfnamefont {Z.}~\bibnamefont
  {Liao}}, \bibinfo {author} {\bibfnamefont {M.}~\bibnamefont {Han}}, \bibinfo
  {author} {\bibfnamefont {M.}~\bibnamefont {Fruchart}}, \bibinfo {author}
  {\bibfnamefont {V.}~\bibnamefont {Vitelli}}, \ and\ \bibinfo {author}
  {\bibfnamefont {S.}~\bibnamefont {Vaikuntanathan}},\ }\href {\doibase
  10.1063/1.5126962} {\bibfield  {journal} {\bibinfo  {journal} {The Journal of
  Chemical Physics}\ }\textbf {\bibinfo {volume} {151}},\ \bibinfo {pages}
  {194108} (\bibinfo {year} {2019})}\BibitemShut {NoStop}%
\bibitem [{\citenamefont {Epstein}\ and\ \citenamefont
  {Mandadapu}(2019)}]{epstein2019time}%
  \BibitemOpen
  \bibfield  {author} {\bibinfo {author} {\bibfnamefont {J.~M.}\ \bibnamefont
  {Epstein}}\ and\ \bibinfo {author} {\bibfnamefont {K.~K.}\ \bibnamefont
  {Mandadapu}},\ }\href@noop {} {\bibfield  {journal} {\bibinfo  {journal}
  {arXiv preprint arXiv:1907.10041}\ } (\bibinfo {year} {2019})}\BibitemShut
  {NoStop}%
\bibitem [{\citenamefont {Soni}\ \emph {et~al.}(2019)\citenamefont {Soni},
  \citenamefont {Bililign}, \citenamefont {Magkiriadou}, \citenamefont
  {Sacanna}, \citenamefont {Bartolo}, \citenamefont {Shelley},\ and\
  \citenamefont {Irvine}}]{soni2019odd}%
  \BibitemOpen
  \bibfield  {author} {\bibinfo {author} {\bibfnamefont {V.}~\bibnamefont
  {Soni}}, \bibinfo {author} {\bibfnamefont {E.~S.}\ \bibnamefont {Bililign}},
  \bibinfo {author} {\bibfnamefont {S.}~\bibnamefont {Magkiriadou}}, \bibinfo
  {author} {\bibfnamefont {S.}~\bibnamefont {Sacanna}}, \bibinfo {author}
  {\bibfnamefont {D.}~\bibnamefont {Bartolo}}, \bibinfo {author} {\bibfnamefont
  {M.~J.}\ \bibnamefont {Shelley}}, \ and\ \bibinfo {author} {\bibfnamefont
  {W.~T.}\ \bibnamefont {Irvine}},\ }\href@noop {} {\bibfield  {journal}
  {\bibinfo  {journal} {Nature Physics}\ ,\ \bibinfo {pages} {1}} (\bibinfo
  {year} {2019})}\BibitemShut {NoStop}%
\bibitem [{\citenamefont {Alekseev}(2016)}]{alekseev2016negative}%
  \BibitemOpen
  \bibfield  {author} {\bibinfo {author} {\bibfnamefont {P.}~\bibnamefont
  {Alekseev}},\ }\href@noop {} {\bibfield  {journal} {\bibinfo  {journal}
  {Physical review letters}\ }\textbf {\bibinfo {volume} {117}},\ \bibinfo
  {pages} {166601} (\bibinfo {year} {2016})}\BibitemShut {NoStop}%
\bibitem [{\citenamefont {Korving}\ \emph {et~al.}(1966)\citenamefont
  {Korving}, \citenamefont {Hulsman}, \citenamefont {Knaap},\ and\
  \citenamefont {Beenakker}}]{korving1966transverse}%
  \BibitemOpen
  \bibfield  {author} {\bibinfo {author} {\bibfnamefont {J.}~\bibnamefont
  {Korving}}, \bibinfo {author} {\bibfnamefont {H.}~\bibnamefont {Hulsman}},
  \bibinfo {author} {\bibfnamefont {H.}~\bibnamefont {Knaap}}, \ and\ \bibinfo
  {author} {\bibfnamefont {J.}~\bibnamefont {Beenakker}},\ }\href@noop {}
  {\bibfield  {journal} {\bibinfo  {journal} {Physics Letters}\ }\textbf
  {\bibinfo {volume} {21}},\ \bibinfo {pages} {5} (\bibinfo {year}
  {1966})}\BibitemShut {NoStop}%
\bibitem [{\citenamefont {Wiegmann}\ and\ \citenamefont
  {Abanov}(2014)}]{wiegmann2014anomalous}%
  \BibitemOpen
  \bibfield  {author} {\bibinfo {author} {\bibfnamefont {P.}~\bibnamefont
  {Wiegmann}}\ and\ \bibinfo {author} {\bibfnamefont {A.~G.}\ \bibnamefont
  {Abanov}},\ }\href@noop {} {\bibfield  {journal} {\bibinfo  {journal}
  {Physical review letters}\ }\textbf {\bibinfo {volume} {113}},\ \bibinfo
  {pages} {034501} (\bibinfo {year} {2014})}\BibitemShut {NoStop}%
\bibitem [{\citenamefont {Berdyugin}\ \emph {et~al.}(2019)\citenamefont
  {Berdyugin}, \citenamefont {Xu}, \citenamefont {Pellegrino}, \citenamefont
  {Kumar}, \citenamefont {Principi}, \citenamefont {Torre}, \citenamefont
  {Shalom}, \citenamefont {Taniguchi}, \citenamefont {Watanabe}, \citenamefont
  {Grigorieva}, \citenamefont {Polini}, \citenamefont {Geim},\ and\
  \citenamefont {Bandurin}}]{Berdyugin2019}%
  \BibitemOpen
  \bibfield  {author} {\bibinfo {author} {\bibfnamefont {A.~I.}\ \bibnamefont
  {Berdyugin}}, \bibinfo {author} {\bibfnamefont {S.~G.}\ \bibnamefont {Xu}},
  \bibinfo {author} {\bibfnamefont {F.~M.~D.}\ \bibnamefont {Pellegrino}},
  \bibinfo {author} {\bibfnamefont {R.~K.}\ \bibnamefont {Kumar}}, \bibinfo
  {author} {\bibfnamefont {A.}~\bibnamefont {Principi}}, \bibinfo {author}
  {\bibfnamefont {I.}~\bibnamefont {Torre}}, \bibinfo {author} {\bibfnamefont
  {M.~B.}\ \bibnamefont {Shalom}}, \bibinfo {author} {\bibfnamefont
  {T.}~\bibnamefont {Taniguchi}}, \bibinfo {author} {\bibfnamefont
  {K.}~\bibnamefont {Watanabe}}, \bibinfo {author} {\bibfnamefont {I.~V.}\
  \bibnamefont {Grigorieva}}, \bibinfo {author} {\bibfnamefont
  {M.}~\bibnamefont {Polini}}, \bibinfo {author} {\bibfnamefont {A.~K.}\
  \bibnamefont {Geim}}, \ and\ \bibinfo {author} {\bibfnamefont {D.~A.}\
  \bibnamefont {Bandurin}},\ }\href {\doibase 10.1126/science.aau0685}
  {\bibfield  {journal} {\bibinfo  {journal} {Science}\ ,\ \bibinfo {pages}
  {eaau0685}} (\bibinfo {year} {2019})}\BibitemShut {NoStop}%
\bibitem [{\citenamefont {Pellegrino}\ \emph {et~al.}(2017)\citenamefont
  {Pellegrino}, \citenamefont {Torre},\ and\ \citenamefont
  {Polini}}]{pellegrino2017nonlocal}%
  \BibitemOpen
  \bibfield  {author} {\bibinfo {author} {\bibfnamefont {F.~M.}\ \bibnamefont
  {Pellegrino}}, \bibinfo {author} {\bibfnamefont {I.}~\bibnamefont {Torre}}, \
  and\ \bibinfo {author} {\bibfnamefont {M.}~\bibnamefont {Polini}},\
  }\href@noop {} {\bibfield  {journal} {\bibinfo  {journal} {Physical Review
  B}\ }\textbf {\bibinfo {volume} {96}},\ \bibinfo {pages} {195401} (\bibinfo
  {year} {2017})}\BibitemShut {NoStop}%
\bibitem [{\citenamefont {Bradlyn}\ \emph {et~al.}(2012)\citenamefont
  {Bradlyn}, \citenamefont {Goldstein},\ and\ \citenamefont
  {Read}}]{bradlyn2012kubo}%
  \BibitemOpen
  \bibfield  {author} {\bibinfo {author} {\bibfnamefont {B.}~\bibnamefont
  {Bradlyn}}, \bibinfo {author} {\bibfnamefont {M.}~\bibnamefont {Goldstein}},
  \ and\ \bibinfo {author} {\bibfnamefont {N.}~\bibnamefont {Read}},\
  }\href@noop {} {\bibfield  {journal} {\bibinfo  {journal} {Physical Review
  B}\ }\textbf {\bibinfo {volume} {86}},\ \bibinfo {pages} {245309} (\bibinfo
  {year} {2012})}\BibitemShut {NoStop}%
\bibitem [{\citenamefont {Offertaler}\ and\ \citenamefont
  {Bradlyn}(2019)}]{offertaler2019viscoelastic}%
  \BibitemOpen
  \bibfield  {author} {\bibinfo {author} {\bibfnamefont {B.}~\bibnamefont
  {Offertaler}}\ and\ \bibinfo {author} {\bibfnamefont {B.}~\bibnamefont
  {Bradlyn}},\ }\href@noop {} {\bibfield  {journal} {\bibinfo  {journal}
  {Physical Review B}\ }\textbf {\bibinfo {volume} {99}},\ \bibinfo {pages}
  {035427} (\bibinfo {year} {2019})}\BibitemShut {NoStop}%
\bibitem [{\citenamefont {Son}(2019)}]{son2019chiral}%
  \BibitemOpen
  \bibfield  {author} {\bibinfo {author} {\bibfnamefont {D.~T.}\ \bibnamefont
  {Son}},\ }\href@noop {} {\bibfield  {journal} {\bibinfo  {journal} {arXiv
  preprint arXiv:1907.07187}\ } (\bibinfo {year} {2019})}\BibitemShut {NoStop}%
\bibitem [{\citenamefont {Farhadi}\ \emph {et~al.}(2018)\citenamefont
  {Farhadi}, \citenamefont {Machaca}, \citenamefont {Aird}, \citenamefont
  {Maldonado}, \citenamefont {Davis}, \citenamefont {Arratia},\ and\
  \citenamefont {Durian}}]{Farhadi2018}%
  \BibitemOpen
  \bibfield  {author} {\bibinfo {author} {\bibfnamefont {S.}~\bibnamefont
  {Farhadi}}, \bibinfo {author} {\bibfnamefont {S.}~\bibnamefont {Machaca}},
  \bibinfo {author} {\bibfnamefont {J.}~\bibnamefont {Aird}}, \bibinfo {author}
  {\bibfnamefont {B.~O.~T.}\ \bibnamefont {Maldonado}}, \bibinfo {author}
  {\bibfnamefont {S.}~\bibnamefont {Davis}}, \bibinfo {author} {\bibfnamefont
  {P.~E.}\ \bibnamefont {Arratia}}, \ and\ \bibinfo {author} {\bibfnamefont
  {D.~J.}\ \bibnamefont {Durian}},\ }\href {\doibase 10.1039/c8sm00403j}
  {\bibfield  {journal} {\bibinfo  {journal} {Soft Matter}\ }\textbf {\bibinfo
  {volume} {14}},\ \bibinfo {pages} {5588} (\bibinfo {year}
  {2018})}\BibitemShut {NoStop}%
\bibitem [{\citenamefont {Harada}\ and\ \citenamefont
  {Sasa}(2005)}]{harada2005equality}%
  \BibitemOpen
  \bibfield  {author} {\bibinfo {author} {\bibfnamefont {T.}~\bibnamefont
  {Harada}}\ and\ \bibinfo {author} {\bibfnamefont {S.-i.}\ \bibnamefont
  {Sasa}},\ }\href@noop {} {\bibfield  {journal} {\bibinfo  {journal} {Physical
  review letters}\ }\textbf {\bibinfo {volume} {95}},\ \bibinfo {pages}
  {130602} (\bibinfo {year} {2005})}\BibitemShut {NoStop}%
\bibitem [{\citenamefont {Fodor}\ \emph {et~al.}(2016)\citenamefont {Fodor},
  \citenamefont {Nardini}, \citenamefont {Cates}, \citenamefont {Tailleur},
  \citenamefont {Visco},\ and\ \citenamefont {van Wijland}}]{Fodor2016}%
  \BibitemOpen
  \bibfield  {author} {\bibinfo {author} {\bibfnamefont {{\'{E}}.}~\bibnamefont
  {Fodor}}, \bibinfo {author} {\bibfnamefont {C.}~\bibnamefont {Nardini}},
  \bibinfo {author} {\bibfnamefont {M.~E.}\ \bibnamefont {Cates}}, \bibinfo
  {author} {\bibfnamefont {J.}~\bibnamefont {Tailleur}}, \bibinfo {author}
  {\bibfnamefont {P.}~\bibnamefont {Visco}}, \ and\ \bibinfo {author}
  {\bibfnamefont {F.}~\bibnamefont {van Wijland}},\ }\href {\doibase
  10.1103/physrevlett.117.038103} {\bibfield  {journal} {\bibinfo  {journal}
  {Physical Review Letters}\ }\textbf {\bibinfo {volume} {117}},\ \bibinfo
  {pages} {038103} (\bibinfo {year} {2016})}\BibitemShut {NoStop}%
\bibitem [{\citenamefont {Shankar}\ and\ \citenamefont
  {Marchetti}(2018)}]{shankar2018hidden}%
  \BibitemOpen
  \bibfield  {author} {\bibinfo {author} {\bibfnamefont {S.}~\bibnamefont
  {Shankar}}\ and\ \bibinfo {author} {\bibfnamefont {M.~C.}\ \bibnamefont
  {Marchetti}},\ }\href@noop {} {\bibfield  {journal} {\bibinfo  {journal}
  {Physical Review E}\ }\textbf {\bibinfo {volume} {98}},\ \bibinfo {pages}
  {020604} (\bibinfo {year} {2018})}\BibitemShut {NoStop}%
\bibitem [{\citenamefont {Nardini}\ \emph {et~al.}(2017)\citenamefont
  {Nardini}, \citenamefont {Fodor}, \citenamefont {Tjhung}, \citenamefont {van
  Wijland}, \citenamefont {Tailleur},\ and\ \citenamefont
  {Cates}}]{Nardini2017}%
  \BibitemOpen
  \bibfield  {author} {\bibinfo {author} {\bibfnamefont {C.}~\bibnamefont
  {Nardini}}, \bibinfo {author} {\bibfnamefont {{\'{E}}.}~\bibnamefont
  {Fodor}}, \bibinfo {author} {\bibfnamefont {E.}~\bibnamefont {Tjhung}},
  \bibinfo {author} {\bibfnamefont {F.}~\bibnamefont {van Wijland}}, \bibinfo
  {author} {\bibfnamefont {J.}~\bibnamefont {Tailleur}}, \ and\ \bibinfo
  {author} {\bibfnamefont {M.~E.}\ \bibnamefont {Cates}},\ }\href {\doibase
  10.1103/physrevx.7.021007} {\bibfield  {journal} {\bibinfo  {journal}
  {Physical Review X}\ }\textbf {\bibinfo {volume} {7}},\ \bibinfo {pages}
  {021007} (\bibinfo {year} {2017})}\BibitemShut {NoStop}%
\bibitem [{\citenamefont {Le~Goff}\ \emph {et~al.}(2001)\citenamefont
  {Le~Goff}, \citenamefont {Amblard},\ and\ \citenamefont
  {Furst}}]{le2001motor}%
  \BibitemOpen
  \bibfield  {author} {\bibinfo {author} {\bibfnamefont {L.}~\bibnamefont
  {Le~Goff}}, \bibinfo {author} {\bibfnamefont {F.}~\bibnamefont {Amblard}}, \
  and\ \bibinfo {author} {\bibfnamefont {E.~M.}\ \bibnamefont {Furst}},\
  }\href@noop {} {\bibfield  {journal} {\bibinfo  {journal} {Physical review
  letters}\ }\textbf {\bibinfo {volume} {88}},\ \bibinfo {pages} {018101}
  (\bibinfo {year} {2001})}\BibitemShut {NoStop}%
\bibitem [{\citenamefont {Berthier}\ and\ \citenamefont
  {Kurchan}(2013)}]{berthier2013non}%
  \BibitemOpen
  \bibfield  {author} {\bibinfo {author} {\bibfnamefont {L.}~\bibnamefont
  {Berthier}}\ and\ \bibinfo {author} {\bibfnamefont {J.}~\bibnamefont
  {Kurchan}},\ }\href@noop {} {\bibfield  {journal} {\bibinfo  {journal}
  {Nature Physics}\ }\textbf {\bibinfo {volume} {9}},\ \bibinfo {pages} {310}
  (\bibinfo {year} {2013})}\BibitemShut {NoStop}%
\bibitem [{\citenamefont {Palacci}\ \emph {et~al.}(2010)\citenamefont
  {Palacci}, \citenamefont {Cottin-Bizonne}, \citenamefont {Ybert},\ and\
  \citenamefont {Bocquet}}]{palacci2010sedimentation}%
  \BibitemOpen
  \bibfield  {author} {\bibinfo {author} {\bibfnamefont {J.}~\bibnamefont
  {Palacci}}, \bibinfo {author} {\bibfnamefont {C.}~\bibnamefont
  {Cottin-Bizonne}}, \bibinfo {author} {\bibfnamefont {C.}~\bibnamefont
  {Ybert}}, \ and\ \bibinfo {author} {\bibfnamefont {L.}~\bibnamefont
  {Bocquet}},\ }\href@noop {} {\bibfield  {journal} {\bibinfo  {journal}
  {Physical Review Letters}\ }\textbf {\bibinfo {volume} {105}},\ \bibinfo
  {pages} {088304} (\bibinfo {year} {2010})}\BibitemShut {NoStop}%
\bibitem [{\citenamefont {Egolf}(2000)}]{egolf2000equilibrium}%
  \BibitemOpen
  \bibfield  {author} {\bibinfo {author} {\bibfnamefont {D.~A.}\ \bibnamefont
  {Egolf}},\ }\href@noop {} {\bibfield  {journal} {\bibinfo  {journal}
  {Science}\ }\textbf {\bibinfo {volume} {287}},\ \bibinfo {pages} {101}
  (\bibinfo {year} {2000})}\BibitemShut {NoStop}%
\bibitem [{\citenamefont {Prost}\ \emph {et~al.}(2009)\citenamefont {Prost},
  \citenamefont {Joanny},\ and\ \citenamefont
  {Parrondo}}]{prost2009generalized}%
  \BibitemOpen
  \bibfield  {author} {\bibinfo {author} {\bibfnamefont {J.}~\bibnamefont
  {Prost}}, \bibinfo {author} {\bibfnamefont {J.-F.}\ \bibnamefont {Joanny}}, \
  and\ \bibinfo {author} {\bibfnamefont {J.}~\bibnamefont {Parrondo}},\ }\href
  {\doibase 10.1103/PhysRevLett.103.090601} {\bibfield  {journal} {\bibinfo
  {journal} {Physical review letters}\ }\textbf {\bibinfo {volume} {103}},\
  \bibinfo {pages} {090601} (\bibinfo {year} {2009})}\BibitemShut {NoStop}%
\bibitem [{\citenamefont {Gomez-Solano}\ \emph {et~al.}(2009)\citenamefont
  {Gomez-Solano}, \citenamefont {Petrosyan}, \citenamefont {Ciliberto},
  \citenamefont {Chetrite},\ and\ \citenamefont
  {Gaw{\k{e}}dzki}}]{GomezSolano2009}%
  \BibitemOpen
  \bibfield  {author} {\bibinfo {author} {\bibfnamefont {J.~R.}\ \bibnamefont
  {Gomez-Solano}}, \bibinfo {author} {\bibfnamefont {A.}~\bibnamefont
  {Petrosyan}}, \bibinfo {author} {\bibfnamefont {S.}~\bibnamefont
  {Ciliberto}}, \bibinfo {author} {\bibfnamefont {R.}~\bibnamefont {Chetrite}},
  \ and\ \bibinfo {author} {\bibfnamefont {K.}~\bibnamefont {Gaw{\k{e}}dzki}},\
  }\href {\doibase 10.1103/physrevlett.103.040601} {\bibfield  {journal}
  {\bibinfo  {journal} {Physical Review Letters}\ }\textbf {\bibinfo {volume}
  {103}},\ \bibinfo {pages} {040601} (\bibinfo {year} {2009})}\BibitemShut
  {NoStop}%
\bibitem [{\citenamefont {Seifert}\ and\ \citenamefont
  {Speck}(2010)}]{Seifert2010}%
  \BibitemOpen
  \bibfield  {author} {\bibinfo {author} {\bibfnamefont {U.}~\bibnamefont
  {Seifert}}\ and\ \bibinfo {author} {\bibfnamefont {T.}~\bibnamefont
  {Speck}},\ }\href {\doibase 10.1209/0295-5075/89/10007} {\bibfield  {journal}
  {\bibinfo  {journal} {{EPL} (Europhysics Letters)}\ }\textbf {\bibinfo
  {volume} {89}},\ \bibinfo {pages} {10007} (\bibinfo {year}
  {2010})}\BibitemShut {NoStop}%
\bibitem [{\citenamefont {Cengio}\ \emph {et~al.}(2019)\citenamefont {Cengio},
  \citenamefont {Levis},\ and\ \citenamefont {Pagonabarraga}}]{Cengio2019}%
  \BibitemOpen
  \bibfield  {author} {\bibinfo {author} {\bibfnamefont {S.~D.}\ \bibnamefont
  {Cengio}}, \bibinfo {author} {\bibfnamefont {D.}~\bibnamefont {Levis}}, \
  and\ \bibinfo {author} {\bibfnamefont {I.}~\bibnamefont {Pagonabarraga}},\
  }\href@noop {} {\enquote {\bibinfo {title} {Linear response theory and
  green-kubo relations for active matter},}\ } (\bibinfo {year} {2019}),\
  \Eprint {http://arxiv.org/abs/1907.02560v1} {arXiv:1907.02560v1} \BibitemShut
  {NoStop}%
\bibitem [{\citenamefont {Sarracino}\ and\ \citenamefont
  {Vulpiani}(2019)}]{sarracino2019fluctuation}%
  \BibitemOpen
  \bibfield  {author} {\bibinfo {author} {\bibfnamefont {A.}~\bibnamefont
  {Sarracino}}\ and\ \bibinfo {author} {\bibfnamefont {A.}~\bibnamefont
  {Vulpiani}},\ }\href@noop {} {\bibfield  {journal} {\bibinfo  {journal}
  {Chaos: An Interdisciplinary Journal of Nonlinear Science}\ }\textbf
  {\bibinfo {volume} {29}},\ \bibinfo {pages} {083132} (\bibinfo {year}
  {2019})}\BibitemShut {NoStop}%
\bibitem [{\citenamefont {Han}\ \emph {et~al.}(2017)\citenamefont {Han},
  \citenamefont {Yan}, \citenamefont {Granick},\ and\ \citenamefont
  {Luijten}}]{han2017effective}%
  \BibitemOpen
  \bibfield  {author} {\bibinfo {author} {\bibfnamefont {M.}~\bibnamefont
  {Han}}, \bibinfo {author} {\bibfnamefont {J.}~\bibnamefont {Yan}}, \bibinfo
  {author} {\bibfnamefont {S.}~\bibnamefont {Granick}}, \ and\ \bibinfo
  {author} {\bibfnamefont {E.}~\bibnamefont {Luijten}},\ }\href@noop {}
  {\bibfield  {journal} {\bibinfo  {journal} {Proceedings of the National
  Academy of Sciences}\ }\textbf {\bibinfo {volume} {114}},\ \bibinfo {pages}
  {7513} (\bibinfo {year} {2017})}\BibitemShut {NoStop}%
\bibitem [{\citenamefont {Irving}\ and\ \citenamefont
  {Kirkwood}(1950)}]{irving1950statistical}%
  \BibitemOpen
  \bibfield  {author} {\bibinfo {author} {\bibfnamefont {J.}~\bibnamefont
  {Irving}}\ and\ \bibinfo {author} {\bibfnamefont {J.~G.}\ \bibnamefont
  {Kirkwood}},\ }\href@noop {} {\bibfield  {journal} {\bibinfo  {journal} {The
  Journal of chemical physics}\ }\textbf {\bibinfo {volume} {18}},\ \bibinfo
  {pages} {817} (\bibinfo {year} {1950})}\BibitemShut {NoStop}%
\bibitem [{\citenamefont {Scheibner}\ \emph {et~al.}(2019)\citenamefont
  {Scheibner}, \citenamefont {Souslov}, \citenamefont {Banerjee}, \citenamefont
  {Surowka}, \citenamefont {Irvine},\ and\ \citenamefont
  {Vitelli}}]{scheibner2019odd}%
  \BibitemOpen
  \bibfield  {author} {\bibinfo {author} {\bibfnamefont {C.}~\bibnamefont
  {Scheibner}}, \bibinfo {author} {\bibfnamefont {A.}~\bibnamefont {Souslov}},
  \bibinfo {author} {\bibfnamefont {D.}~\bibnamefont {Banerjee}}, \bibinfo
  {author} {\bibfnamefont {P.}~\bibnamefont {Surowka}}, \bibinfo {author}
  {\bibfnamefont {W.~T.~M.}\ \bibnamefont {Irvine}}, \ and\ \bibinfo {author}
  {\bibfnamefont {V.}~\bibnamefont {Vitelli}},\ }\href@noop {} {\bibfield
  {journal} {\bibinfo  {journal} {arXiv preprint arXiv:1902.07760, Nature
  Physics {\it in press}}\ } (\bibinfo {year} {2019})}\BibitemShut {NoStop}%
\bibitem [{\citenamefont {Casimir}(1945)}]{casimir1945onsager}%
  \BibitemOpen
  \bibfield  {author} {\bibinfo {author} {\bibfnamefont {H.~B.~G.}\
  \bibnamefont {Casimir}},\ }\href@noop {} {\bibfield  {journal} {\bibinfo
  {journal} {Reviews of Modern Physics}\ }\textbf {\bibinfo {volume} {17}},\
  \bibinfo {pages} {343} (\bibinfo {year} {1945})}\BibitemShut {NoStop}%
\bibitem [{\citenamefont {Chapman}\ \emph {et~al.}(1990)\citenamefont
  {Chapman}, \citenamefont {Cowling}, \citenamefont {Burnett},\ and\
  \citenamefont {Cercignani}}]{ChapmanCowling}%
  \BibitemOpen
  \bibfield  {author} {\bibinfo {author} {\bibfnamefont {S.}~\bibnamefont
  {Chapman}}, \bibinfo {author} {\bibfnamefont {T.}~\bibnamefont {Cowling}},
  \bibinfo {author} {\bibfnamefont {D.}~\bibnamefont {Burnett}}, \ and\
  \bibinfo {author} {\bibfnamefont {C.}~\bibnamefont {Cercignani}},\
  }\href@noop {} {\emph {\bibinfo {title} {The Mathematical Theory of
  Non-uniform Gases: An Account of the Kinetic Theory of Viscosity, Thermal
  Conduction and Diffusion in Gases}}},\ Cambridge Mathematical Library\
  (\bibinfo  {publisher} {Cambridge University Press},\ \bibinfo {year}
  {1990})\BibitemShut {NoStop}%
\bibitem [{Note1()}]{Note1}%
  \BibitemOpen
  \bibinfo {note} {We verified that the long-time tail associated with the
  breakdown of 2D hydrodynamics is too small to impact the viscosity prediction
  (see SI Fig.xx).}\BibitemShut {Stop}%
\bibitem [{\citenamefont {Zwanzig}(2001)}]{Zwanzig2001}%
  \BibitemOpen
  \bibfield  {author} {\bibinfo {author} {\bibfnamefont {R.}~\bibnamefont
  {Zwanzig}},\ }\href@noop {} {\emph {\bibinfo {title} {Nonequilibrium
  Statistical Mechanics}}},\ \bibinfo {edition} {3rd}\ ed.\ (\bibinfo
  {publisher} {Oxford University Press},\ \bibinfo {year} {2001})\BibitemShut
  {NoStop}%
\bibitem [{\citenamefont {Mori}(1965)}]{Mori1965}%
  \BibitemOpen
  \bibfield  {author} {\bibinfo {author} {\bibfnamefont {H.}~\bibnamefont
  {Mori}},\ }\href {\doibase 10.1143/ptp.33.423} {\bibfield  {journal}
  {\bibinfo  {journal} {Progress of Theoretical Physics}\ }\textbf {\bibinfo
  {volume} {33}},\ \bibinfo {pages} {423} (\bibinfo {year} {1965})}\BibitemShut
  {NoStop}%
\bibitem [{\citenamefont {Nakajima}(1958)}]{Nakajima1958}%
  \BibitemOpen
  \bibfield  {author} {\bibinfo {author} {\bibfnamefont {S.}~\bibnamefont
  {Nakajima}},\ }\href {\doibase 10.1143/ptp.20.948} {\bibfield  {journal}
  {\bibinfo  {journal} {Progress of Theoretical Physics}\ }\textbf {\bibinfo
  {volume} {20}},\ \bibinfo {pages} {948} (\bibinfo {year} {1958})}\BibitemShut
  {NoStop}%
\bibitem [{\citenamefont {Zwanzig}(1960)}]{Zwanzig1960}%
  \BibitemOpen
  \bibfield  {author} {\bibinfo {author} {\bibfnamefont {R.}~\bibnamefont
  {Zwanzig}},\ }\href {\doibase 10.1063/1.1731409} {\bibfield  {journal}
  {\bibinfo  {journal} {The Journal of Chemical Physics}\ }\textbf {\bibinfo
  {volume} {33}},\ \bibinfo {pages} {1338} (\bibinfo {year}
  {1960})}\BibitemShut {NoStop}%
\bibitem [{Note2()}]{Note2}%
  \BibitemOpen
  \bibinfo {note} {The linear relation between stresses and velocity gradients
  holds only for the macroscopic, averaged (or on shell) quantities not the
  fluctuating ones.}\BibitemShut {Stop}%
\bibitem [{\citenamefont {Evans}\ and\ \citenamefont
  {Morriss}(1984)}]{evans1984nonlinear}%
  \BibitemOpen
  \bibfield  {author} {\bibinfo {author} {\bibfnamefont {D.~J.}\ \bibnamefont
  {Evans}}\ and\ \bibinfo {author} {\bibfnamefont {G.}~\bibnamefont
  {Morriss}},\ }\href@noop {} {\bibfield  {journal} {\bibinfo  {journal}
  {Physical Review A}\ }\textbf {\bibinfo {volume} {30}},\ \bibinfo {pages}
  {1528} (\bibinfo {year} {1984})}\BibitemShut {NoStop}%
\bibitem [{\citenamefont {Daivis}\ and\ \citenamefont
  {Todd}(2006)}]{daivis2006simple}%
  \BibitemOpen
  \bibfield  {author} {\bibinfo {author} {\bibfnamefont {P.~J.}\ \bibnamefont
  {Daivis}}\ and\ \bibinfo {author} {\bibfnamefont {B.}~\bibnamefont {Todd}},\
  }\href@noop {} {\bibfield  {journal} {\bibinfo  {journal} {The Journal of
  chemical physics}\ }\textbf {\bibinfo {volume} {124}},\ \bibinfo {pages}
  {194103} (\bibinfo {year} {2006})}\BibitemShut {NoStop}%
\bibitem [{\citenamefont {Evans}\ and\ \citenamefont
  {Morriss}(2008)}]{evans2008statistical}%
  \BibitemOpen
  \bibfield  {author} {\bibinfo {author} {\bibfnamefont {D.~J.}\ \bibnamefont
  {Evans}}\ and\ \bibinfo {author} {\bibfnamefont {G.}~\bibnamefont
  {Morriss}},\ }\href@noop {} {\emph {\bibinfo {title} {Statistical mechanics
  of nonequilibrium liquids}}}\ (\bibinfo  {publisher} {Cambridge University
  Press},\ \bibinfo {year} {2008})\BibitemShut {NoStop}%
\bibitem [{\citenamefont {Sierou}\ and\ \citenamefont
  {Brady}(2002)}]{sierou2002rheology}%
  \BibitemOpen
  \bibfield  {author} {\bibinfo {author} {\bibfnamefont {A.}~\bibnamefont
  {Sierou}}\ and\ \bibinfo {author} {\bibfnamefont {J.}~\bibnamefont {Brady}},\
  }\href@noop {} {\bibfield  {journal} {\bibinfo  {journal} {Journal of
  Rheology}\ }\textbf {\bibinfo {volume} {46}},\ \bibinfo {pages} {1031}
  (\bibinfo {year} {2002})}\BibitemShut {NoStop}%
\bibitem [{\citenamefont {Weissenberg}(1947)}]{weissenberg1947continuum}%
  \BibitemOpen
  \bibfield  {author} {\bibinfo {author} {\bibfnamefont {K.}~\bibnamefont
  {Weissenberg}},\ }\href@noop {} {\enquote {\bibinfo {title} {A continuum
  theory of rhelogical phenomena},}\ } (\bibinfo {year} {1947})\BibitemShut
  {NoStop}%
\bibitem [{\citenamefont {Campbell}(1989)}]{campbell1989stress}%
  \BibitemOpen
  \bibfield  {author} {\bibinfo {author} {\bibfnamefont {C.~S.}\ \bibnamefont
  {Campbell}},\ }\href@noop {} {\bibfield  {journal} {\bibinfo  {journal}
  {Journal of Fluid Mechanics}\ }\textbf {\bibinfo {volume} {203}},\ \bibinfo
  {pages} {449} (\bibinfo {year} {1989})}\BibitemShut {NoStop}%
\bibitem [{\citenamefont {De~Groot}\ and\ \citenamefont
  {Mazur}(2013{\natexlab{b}})}]{DGM}%
  \BibitemOpen
  \bibfield  {author} {\bibinfo {author} {\bibfnamefont {S.~R.}\ \bibnamefont
  {De~Groot}}\ and\ \bibinfo {author} {\bibfnamefont {P.}~\bibnamefont
  {Mazur}},\ }\href@noop {} {\emph {\bibinfo {title} {Non-equilibrium
  thermodynamics}}}\ (\bibinfo  {publisher} {Courier Corporation},\ \bibinfo
  {year} {2013})\BibitemShut {NoStop}%
\bibitem [{\citenamefont {de~Groot}\ and\ \citenamefont
  {Mazur}(1954)}]{deGroot1954}%
  \BibitemOpen
  \bibfield  {author} {\bibinfo {author} {\bibfnamefont {S.~R.}\ \bibnamefont
  {de~Groot}}\ and\ \bibinfo {author} {\bibfnamefont {P.}~\bibnamefont
  {Mazur}},\ }\href {\doibase 10.1103/physrev.94.218} {\bibfield  {journal}
  {\bibinfo  {journal} {Physical Review}\ }\textbf {\bibinfo {volume} {94}},\
  \bibinfo {pages} {218} (\bibinfo {year} {1954})}\BibitemShut {NoStop}%
\bibitem [{\citenamefont {Leonardo}\ \emph {et~al.}(2008)\citenamefont
  {Leonardo}, \citenamefont {Keen}, \citenamefont {Ianni}, \citenamefont
  {Leach}, \citenamefont {Padgett},\ and\ \citenamefont
  {Ruocco}}]{dileonardo2008hydro2d}%
  \BibitemOpen
  \bibfield  {author} {\bibinfo {author} {\bibfnamefont {R.~D.}\ \bibnamefont
  {Leonardo}}, \bibinfo {author} {\bibfnamefont {S.}~\bibnamefont {Keen}},
  \bibinfo {author} {\bibfnamefont {F.}~\bibnamefont {Ianni}}, \bibinfo
  {author} {\bibfnamefont {J.}~\bibnamefont {Leach}}, \bibinfo {author}
  {\bibfnamefont {M.~J.}\ \bibnamefont {Padgett}}, \ and\ \bibinfo {author}
  {\bibfnamefont {G.}~\bibnamefont {Ruocco}},\ }\href {\doibase
  10.1103/physreve.78.031406} {\bibfield  {journal} {\bibinfo  {journal}
  {Physical Review E}\ }\textbf {\bibinfo {volume} {78}},\ \bibinfo {pages}
  {031406} (\bibinfo {year} {2008})}\BibitemShut {NoStop}%
\bibitem [{\citenamefont {Ermak}\ and\ \citenamefont
  {McCammon}(1978)}]{ermak1978brownian}%
  \BibitemOpen
  \bibfield  {author} {\bibinfo {author} {\bibfnamefont {D.~L.}\ \bibnamefont
  {Ermak}}\ and\ \bibinfo {author} {\bibfnamefont {J.~A.}\ \bibnamefont
  {McCammon}},\ }\href@noop {} {\bibfield  {journal} {\bibinfo  {journal} {The
  Journal of chemical physics}\ }\textbf {\bibinfo {volume} {69}},\ \bibinfo
  {pages} {1352} (\bibinfo {year} {1978})}\BibitemShut {NoStop}%
\bibitem [{\citenamefont {Burgers}(1948)}]{burgers1948mathematical}%
  \BibitemOpen
  \bibfield  {author} {\bibinfo {author} {\bibfnamefont {J.~M.}\ \bibnamefont
  {Burgers}},\ }in\ \href@noop {} {\emph {\bibinfo {booktitle} {Advances in
  applied mechanics}}},\ Vol.~\bibinfo {volume} {1}\ (\bibinfo  {publisher}
  {Elsevier},\ \bibinfo {year} {1948})\ pp.\ \bibinfo {pages}
  {171--199}\BibitemShut {NoStop}%
\end{thebibliography}%
\end{document}